\documentclass[journal]{IEEEtran}
\usepackage{yhmath}
\usepackage{balance}

\usepackage{cite}
\ifCLASSINFOpdf
   \usepackage[pdftex]{graphicx}
\else
   \usepackage[dvips]{graphicx}
\fi
\usepackage{amsmath}
\ifCLASSOPTIONcompsoc
  \usepackage[caption=false,font=normalsize,labelfont=sf,textfont=sf]{subfig}
\else
  \usepackage[caption=true,font=footnotesize]{subfig}
\fi
\begin{document}
\title{A New Weighting Scheme for Fan-beam and Circle Cone-beam CT Reconstructions}

\author{Wei Wang, Xiang-Gen Xia, Chuanjiang He, Zemin Ren, Jian Lu,  Tianfu Wang and  Baiying Lei
	\thanks{This work was supported partly by National Natural Science Foundation of China (Nos.12001381, 61871274 and 61801305), China Postdoctoral Science Foundation
		(2018M64081), Peacock Plan (No. KQTD2016053112051497), Shenzhen Key Basic Research Project (Nos. JCYJ20180507184647636, JCYJ20170412104656685, JCYJ20170818094109846, and JCYJ20190808155618806).}
	\thanks{Wei Wang, Tianfu Wang, and  Baiying Lei are with the School of Biomedical Engineering, Shenzhen University,
		National-Regional Key Technology Engineering Laboratory for Medical Ultrasound, Guangdong Key Laboratory for Biomedical Measurements and Ultrasound Imaging, School of Biomedical Engineering, Health Science Center, Shenzhen University, Shenzhen, China. (e-mail: wangwei@szu.edu.cn, leiby@szu.edu.cn, tfwang@szu.edu.cn). }
	\thanks{Xiang-Gen Xia is with the Department of Electrical and Computer Engineering, University of Delaware, Newark, DE 19716, USA.  (e-mail: xxia@ee.udel.edu).}
	\thanks{Chuanjiang He is with the
		College of Mathematics and Statistics, Chongqing University, Chongqing, China (e-mail: cjhe@cqu.edu.cn).}
	\thanks{Zemin Ren is with the
		College of Mathematics and Physics, Chongqing University of Science and Technology, Chongqing, China (e-mail: zeminren@cqu.edu.cn).}
	\thanks{Jian Lu is with the
		Shenzhen Key Laboratory of Advanced Machine Learning and Applications, Shenzhen University, Shenzhen, China (e-mail: jianlu@szu.edu.cn).}
}


\markboth{Journal of \LaTeX\ Class Files,~Vol.~14, No.~8, August~2015}%
{Shell \MakeLowercase{\textit{et al.}}: Bare Demo of IEEEtran.cls for IEEE Journals}



\maketitle

% in the abstract or k eywords.
\begin{abstract}
In this paper, we first present an arc based algorithm for fan-beam computed tomography (CT) reconstruction via applying Katsevich’s helical CT formula to 2D fan-beam CT reconstruction. Then, we propose a new weighting function to deal with the redundant projection data. By extending the weighted  arc based fan-beam algorithm to circle cone-beam geometry, we also obtain a new FDK-similar algorithm for  circle cone-beam CT reconstruction. Experiments show that our methods can obtain higher PSNR and SSIM compared to the Parker-weighted conventional fan-beam algorithm and the FDK algorithm for super-short-scan trajectories.
\end{abstract}


% \fi
\IEEEpeerreviewmaketitle

\section{Introduction}
\label{sec:introduction}
\IEEEPARstart{C}{omputed} tomography (CT) has been widely used in clinical diagnosis and industrial applications since its ability of providing inner vision of an object without destructing it. In classical tomography, the fan-beam CT reconstruction algorithm is fundamental since it can be heuristically extended to 3D helical cone-beam \cite{RN15}\cite{ISI:000222216800008}\cite{RN8}\cite{RN10}\cite{RN17}\cite{RN20} and circle cone-beam \cite{ISI:A1984SU73300005}\cite{RN1}\cite{RN5}\cite{RN14}\cite{ISI:000432982200006}\cite{RN19} CT reconstructions. The standard fan-beam reconstruction method is the ramp filter-based filtered backprojection (FBP) algorithm \cite{RN2}\cite{RN3}, which can be derived from the  Radon inversion formula. 
For 2D CT image reconstruction,  to exactly and stably reconstruct the whole image, it requires to measure all the line integrals of the X-rays that diverge from all directions and pass through the object. To make  a fan-beam CT that samples data on a circular trajectory meet this condition, the detector must be large enough to cover the fan-angle  of $\pm \gamma_m=\arcsin(R_m/R_o)$ to avoid truncated projections and the X-ray source must travel on a continuous arc of $\pi+2\gamma_m$ on the circle to ensure that  the line integrals of  the X-rays diverging from all directions in the 2D plane and passing through the object are measured, where $R_o$ is the radius of the scanning  trajectory and $R_m<R_o$ is the radius of the object. In the literature, the range of  $\pi+2\gamma_m$ is called as short-scan.
  In \cite{ISI:A1982NJ29700015},  Parker  proposed a weighting function to weight the short-scan fan-beam projection data before convoluting it to avoid completing the projection data of the remain angles ([$ \pi+2\gamma_m$:$2\pi$]) and implement the reconstruction algorithm efficiently. When the range of scanning angles is larger than the short-scan, redundant projection data is measured. In \cite{RN4}, Silver extended Parker’s weighting function by utilizing the virtual detectors to tackle these redundant projection data. 

In 2002, Noo et al. \cite{ISI:000177297700011} proposed another type of FBP algorithm for fan-beam CT reconstruction by decomposing the convolution of the ramp filter into a successive  convolutions of a Hilbert filter and a derivative filter. After that, many other algorithms \cite{RN5}\cite{RN6}\cite{ISI:000223896700006} based on the Hilbert transform were proposed. One advantage of these new algorithms is that they can exactly reconstruct a part of the image even though the range of the scanning angles is less than  the short scan. These new algorithms can be seen as special cases that applying the 3D exact helical cone-beam inversion formula \cite{ISI:000220649400006}\cite{ISI:000221245100004} to the 2D CT reconstructions.  
 In \cite{RN7}, You et al. derived a Hilbert transform based FBP algorithm for fan-beam full- and partial-scans, in which the backprojection does not include position-dependent weights.  In \cite{RN5} and \cite{ISI:000177297700011}, to deal with the redundant data, a continuous weight function was also proposed to weight the filtered sinograms. 

A specific view-dependent data differentiation was a common processing step in these Hilbert transform based algorithms. In \cite{RN8}\cite{RN9},  the implementation of this step  was researched in order to improve the resolution and quality of the reconstructed image. In \cite{RN10}, Zamyatin et al. used the Taylor series expansions to approximate the derivative of the Hilbert transform and proposed new algorithms for fan-beam and helical cone-beam CT reconstruction.

The fan-beam algorithms can be heuristically extended to 3D cone-beam CT reconstructions \cite{RN15}\cite{ISI:000222216800008}\cite{RN8}\cite{RN17}\cite{RN20}
\cite{ISI:A1984SU73300005}\cite{RN5}\cite{ISI:000432982200006}\cite{RN19}\cite{RN13}, which are referred to as approximating algorithms. On the other hand, there exist a lot of exact cone-beam algorithms \cite{RN14}\cite{ISI:000220649400006}\cite{ISI:000221245100004}\cite{ISI:000177412500008}\cite{RN16}\cite{RN12} in the literature. The advantages of the approximating algorithms are that they are easy to implement and flexible to modulate the ramp kernel for different clinical applications, while those of the exact algorithms are that they can reconstruct images with good resolution even when the cone angle is large. In \cite{ISI:000222216800008}\cite{RN20}\cite{RN14}\cite{ISI:000432982200006}\cite{RN19}\cite{RN13}\cite{RN12}, the weighting functions are also proposed to tackle the redundant cone-beam projection data.

In this paper, we first present an arc based algorithm for fan-beam CT reconstruction by applying Katsevich’s helical CT \cite{ISI:000177297700011} formula to 2D fan-beam CT reconstruction. Then, we propose a new weighting function to tackle the redundant projection data. By extending the arc based algorithm to the circle cone-beam geometry, we also obtain a new algorithm for circle cone-beam CT reconstruction. Our weighting function is different from the ones used in \cite{RN5}\cite{ISI:A1982NJ29700015}\cite{ISI:000177297700011}. The weighting functions in \cite{RN5}\cite{ISI:A1982NJ29700015}\cite{ISI:000177297700011} depend on the rotation angle of the X-ray source and the diverging direction of the X-ray, and are required to be continuous with respect to these arguments while our weighting function depends on the rotation angle of the X-ray source and the positions of the pixels of the reconstructed image, and has no continuity constraint. Thus, the CT images reconstructed by our method may have  higher resolutions and less artifacts. Moreover, there exists a hyper-parameter in the weighting functions of \cite{RN5} \cite{ISI:000177297700011} that greatly influences the performance of the algorithm while our weighting function has no hyper-parameter.  

The rest of the paper is organized as follows. In Section II we briefly introduce the related works. In Section III, we present an arc based algorithm for fan-beam CT reconstruction and
derive a weighting function for tackling the redundant data, and then extend the algorithm with the weighting function to circular cone-beam CT reconstruction. Numerical evaluation is
presented in Section IV, followed by  conclusion
in Section V.

\section{Related works}
In this section, we briefly describe some works related to our method.
\subsection{Noo's fan-beam formula} 
In \cite{ISI:000177297700011}, Noo et al. decomposed the  convolution of the ramp filter into a successive convolutions of the Hilbert filter and the derivative filter in the Fourier domain by observing that
\begin{equation}
|\sigma|=(1/2\pi)(i 2\pi\sigma)(-i \text{sign}(\sigma )),
\end{equation}
where $|\sigma|$, $i 2\pi\sigma$ and $-i \text{sign}(\sigma )$ are the Fourier transform of the ramp filter, the derivative filter and the Hilbert filter, respectively. By utilizing this relationship, they proposed a FBP reconstruction formula for fan-beam CT. 

Let \begin{equation}
g(\lambda, \gamma)=g(\lambda, \underline{a}(\gamma))
\end{equation}
be the measured projections, where $\lambda$ is a polar angle of the X-ray source, $\underline{a}(\lambda)=\left(R_{o} \cos \lambda, R_{o} \sin \lambda\right)$ is a position of the X-ray source, $R_o$ is the radius of the trajectory circle of the X-ray source,
 $\underline{\gamma}=\cos \gamma \underline{e}_{1}+\sin \gamma \underline{e}_{2}$ is a diverging direction of the X-ray, $\gamma \in\left[-\gamma_{m}, \gamma_{m}\right]$, $\gamma_m$ is the half fan-angle, $\underline{e}_{1}=-(\cos \lambda, \sin \lambda)$, and $\underline{e}_{2}=(-\sin \lambda,  \cos \lambda)$. Then the reconstruction formula for fan-beam CT with equi-angular detetcor reads:
\begin{equation}\label{e3}
f(\underline{x})=\frac{1}{2 \pi} \int_{\Lambda} \mathrm{d} \lambda \frac{1}{\|\underline{x}-\underline{a}(\lambda)\|}\left[w(\lambda, \phi) g_{F}(\lambda, \phi)\right]_{\phi=\phi^{*}(\lambda, \underline{x})},
\end{equation}
 where  $\Lambda$ is the X-ray source trajectory satisfying the data completeness condition \cite{ISI:000177297700011},
 \begin{equation}
 \phi^{*}(\lambda, \underline{x})=\arctan \frac{x \cdot \underline{e}_{2}}{R_{o}+\underline{x} \cdot \underline{e}_{1}}, \quad\left|\phi^{*}(\lambda, \underline{x})\right|<\pi / 2
 \end{equation}
 is the angle characterizing the ray that diverges from $\underline{a}(\lambda)$ and passes $\underline{x}$, 
 \begin{equation}\label{e5}
 g_{F}(\lambda, \phi)=\int_{-\gamma_{m}}^{\gamma_{m}}\text{d}\gamma h_{H}(\sin (\phi-\gamma))\left(\frac{\partial}{\partial \lambda}+\frac{\partial}{\partial \gamma}\right) g(\lambda, \gamma),
 \end{equation}
 \begin{equation}\label{A1}
 h_{H}(s)=-\int_{-\infty}^{+\infty} \mathrm{d} \sigma \mathrm{i} \operatorname{sgn}(\sigma) \mathrm{e}^{\mathrm{i} 2 \pi \sigma s}=\frac{1}{\pi s}
 \end{equation}
is the Hilbert filter
and 
\begin{equation}
w(\lambda, \phi)=\frac{c(\lambda)}{c(\lambda)+c(\lambda+\pi-2 \phi)}
\end{equation}
is the weighting function with
\begin{equation}
c(\lambda)=\left\{\begin{array}{ll}
\cos ^{2} \frac{\pi\left(\lambda-\lambda_{s}-d\right)}{2 d} & \text { if } \quad \lambda_{s}<\lambda<\lambda_{s}+d, \\
1 & \text { if } \quad \lambda_{s}+d<\lambda<\lambda_{e}-d, \\
\cos ^{2} \frac{\pi\left(\lambda-\lambda_{e}+d\right)}{2 d} & \text { if } \quad \lambda_{e}-d<\lambda<\lambda_{e}
\end{array}\right.
\end{equation}
and $d$ is an angular interval over which $c(\lambda)$ smoothly drops from 1 to 0. In the experiments, they set $d=10\pi/180$.

Applying the changes of variables $u=D \tan \gamma$ and $\tilde{u}=D \tan \phi$ in equations (\ref{e3}) and (\ref{e5}), they obtain the reconstruction formula for fan-beam CT with equally spaced collinear detectors:
\begin{equation}
f(\underline{x})=\frac{1}{2 \pi} \int_{\Lambda} \mathrm{d} \lambda \frac{1}{R_{o}+\underline{x} \cdot e_{1}}\left[w(\lambda, \tilde{u}) g_{F}(\lambda, \tilde{u})\right]_{\tilde{u}=\tilde{a}^{*}(\lambda, \underline{x})}
\end{equation}
with \begin{equation}
\begin{aligned}
g_{F}(\lambda, \tilde{u})=\int_{-u_{m}}^{u_{m}} \text{d}& u h_{H}(\tilde{u}-u) \frac{D}{\sqrt{D^{2}+u^{2}}}\\
&\left(\frac{\partial}{\partial \lambda}+\frac{D^{2}+u^{2}}{D} \frac{\partial}{\partial u}\right) g(\lambda, u),
\end{aligned}
\end{equation}
where $u_{m}=D \tan \gamma_{m}$,
\begin{equation}
\tilde{u}^{*}(\lambda, \underline{x})=D \tan \phi^{*}(\lambda, \underline{x})=\frac{D \underline{x} \cdot \underline{e}_{2}}{R_{o}+\underline{x} \cdot \underline{e}_{1}}
\end{equation}
is the detector location of the line connecting $\underline{x}$ to $\underline{a}(\lambda)$
and 
\begin{equation}
w(\lambda, \tilde{u})=\frac{c(\lambda)}{c(\lambda)+c(\lambda+\pi-2 \arctan (\tilde{u} / D))}.
\end{equation}

\subsection{Katsevich's helical cone-beam formula}
In \cite{ISI:000221245100004}, Katsevich proposed an exact reconstruction formula for helical cone-beam CT and  Noo et al. \cite{RN18} researched how to efficiently and accurately implement
it. The reconstruction formula can be written as 
\begin{equation}\label{e13}
f(\underline{x})=-\frac{1}{2 \pi} \int_{\lambda_{i}(\underline{x})}^{\lambda_{o}(\underline{x})} \mathrm{d} \lambda \frac{1}{\|\underline{x}-\underline{a}(\lambda)\|} g^{F}\left(\lambda, \frac{\underline{x}-\underline{a}(\lambda)}{\|\underline{x}-\underline{a}(\lambda)\|}\right)
\end{equation}
where $\lambda_i(x)$ and $\lambda_o(x)$ are the extremities of the $\pi$-line passing through $x$ with $\lambda_i(x)<\lambda_o(x)$,  
\begin{equation}
g^{F}(\lambda, \underline{\theta})=\int_{0}^{2 \pi} \mathrm{d} \gamma h_{H}(\sin \gamma) g^{\prime}(\lambda, \cos \gamma \underline{\theta}+\sin \gamma(\underline{\theta} \times \underline{m}(\lambda, \underline{\theta}))),
\end{equation}
\begin{equation}\label{e16}
g^{\prime}(\lambda, \underline{\theta})=\lim _{\varepsilon \rightarrow 0} \frac{g(\lambda+\varepsilon, \underline{\theta})-g(\lambda, \underline{\theta})}{\varepsilon},
\end{equation}
 $h_{H}(s)$ is the Hilbert filter defined in equation (\ref{A1}), $g(\lambda, \underline{\theta})$ is the measured projection data and  vector $\underline{m}(\lambda, \underline{\theta})$ is  normal to the $\kappa$-plane $\mathcal{K}(\lambda, \psi)$ of the smallest $|\psi|$ value that contains the line of direction $\underline{\theta}$ through $\underline{a}(\lambda)$.

\section{Proposed Method}
\subsection{Arc based fan-beam algorithm}
Applying equation (\ref{e13}) to the 2D fan-beam reconstruction, the $\pi$-line becomes a chord of the circle of the scanning trajectory, the $\kappa$-plane $\mathcal{K}(\lambda, \psi)$ coincides with the image plane to reconstruct and so $\underline{\theta} \times \underline{m}(\lambda, \underline{\theta})=\underline\theta^{\bot}$.

Let $\underline{a}(\lambda)=\left(R_{o} \cos \lambda, R_{o} \sin \lambda\right)$ be the position of the X-ray source on the trajectory circle of radius $R_o$ and
\begin{equation}
g(\lambda, \underline{\theta})=\int_{0}^{+\infty} \mathrm{d} t f(\underline{a}(\lambda)+t \underline{\theta})
\end{equation}
be the measured projection data, where $\underline{\theta}\in S^1$ is a diverging direction of the X-ray and $S^1$ is the unit circle in the 2D plane. Then, according to equation (\ref{e13}), the reconstruction formula for fan-beam CT can be written as
\begin{equation}\label{e18}
f(\underline{x})=-\frac{1}{2 \pi} \int_{\text{chord}(\underline{x})} \mathrm{d} \lambda \frac{1}{\|\underline{x}-\underline{a}(\lambda)\|} g^{F}\left(\lambda, \frac{\underline{x}-\underline{a}(\lambda)}{\|\underline{x}-\underline{a}(\lambda)\|}\right),
\end{equation}
where $|\underline x|<R_m<R_o$, $R_m$ is the radius of the test object,
\begin{equation}\label{e19}
g^{F}(\lambda, \underline{\theta})=\int_{0}^{2 \pi} \mathrm{d} \bar\gamma h_{H}(\sin \bar\gamma) g^{\prime}(\lambda, \cos \bar\gamma \underline{\theta}+\sin \bar\gamma\underline{\theta}^{\bot}),
\end{equation}
 $h_{H}(s)$ and $g^{\prime}(\lambda, \underline{\theta})$ are defined by equations (\ref{A1}) and (\ref{e16}), respectively,  $\text{chord}(\underline{x})$ is a chord passing through $\underline{x}$ and dividing the  trajectory circle into two arcs, and the integral $\text{d}\lambda$ is done on any one of the two arcs. 

\begin{figure}[h]
	\includegraphics[scale=0.5]{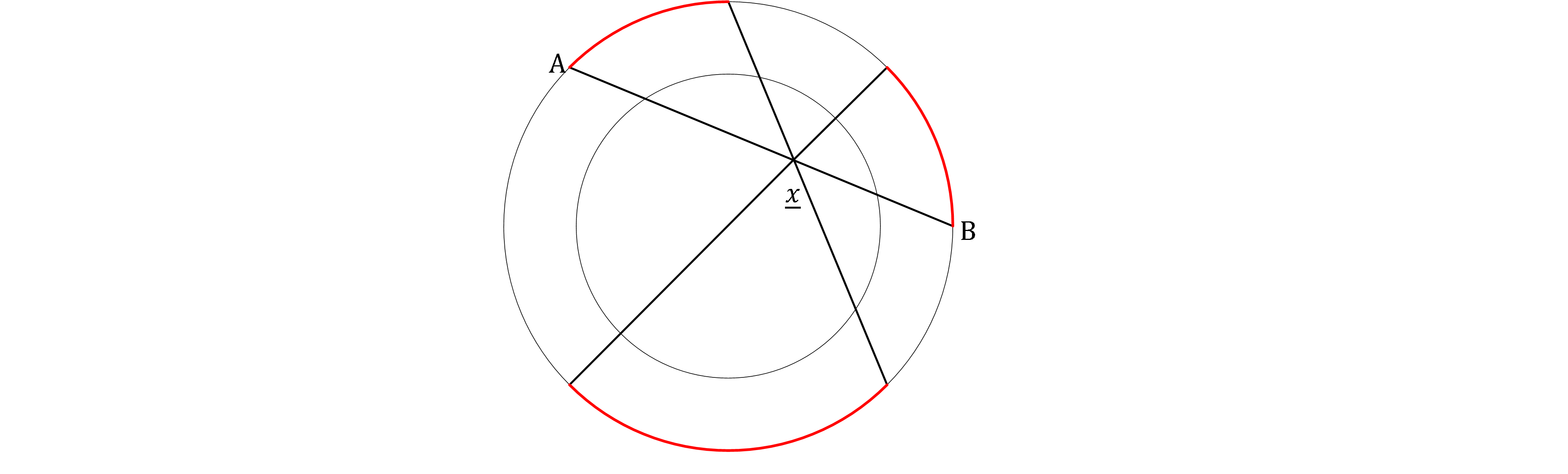}\\			
	\caption{An example of the X-ray source trajectory. The projection data measured along  the three red arcs can  be used to reconstruct the intensity of $\underline x$.}
	\label{fig1}
\end{figure}
From equation (\ref{e18}), we can argue that the intensity of any point $\underline{x}$ with $|\underline x|<R_m<R_o$ in the circle can be reconstructed if there exists a chord passing through $\underline{x}$ such that all the projections $g(\lambda, \underline{\theta})$ on the corresponding arcs of the chord passing through the neighborhood of $\underline{x}$ are  measured.  Note that $g(\lambda, \underline{\theta})$ can be measured  from another direction since $g(\lambda, \underline{\theta})=g(\lambda_1, -\underline{\theta})$, where $\underline\theta=\frac{\underline{a}(\lambda_1)-\underline{a}(\lambda)}{|\underline{a}(\lambda_1)-\underline{a}(\lambda)|}$. See Fig. \ref{fig1} for example.

For any fixed $\underline{x}$ in the trajectory circle,  there may exist many available chords  passing through $\underline{x}$ and so the integral arcs can be different. Moreover, when the range of scanning arc is larger than the short scan, redundant projection data needs to be processed. A simple method to tackle these redundant projection data is to calculate equation (\ref{e18}) for all available chords and average them. However, averaging them is equivalent to filtering the reconstructed image by a low-pass filter and so may reduce the resolution of the reconstructed image.  In this paper we only consider two types of chords that respectively pass through  the two endpoints of the scanning arcs.

Therefore, we propose the  following formula for fan-beam reconstruction:
\begin{equation}\label{e20}
f(\underline{x})=-\frac{1}{2 \pi} \int_{\lambda_0}^{\lambda_P} \mathrm{d} \lambda \frac{\varpi(\underline{x},\lambda)}{\|\underline{x}-\underline{a}(\lambda)\|} g^{F}\left(\lambda, \frac{\underline{x}-\underline{a}(\lambda)}{\|\underline{x}-\underline{a}(\lambda)\|}\right),
\end{equation}
where $\lambda_0<\lambda_P$ correspond to the two endpoints of the scanning arcs,
\begin{equation}\label{e21}
\varpi(\underline{x},\lambda)=\frac{1}{2}(\varpi_1(\underline{x},\lambda)+\varpi_2(\underline{x},\lambda))
\end{equation}
 is a weighting function, and $\varpi_1(\underline{x},\lambda)$ and $\varpi_2(\underline{x},\lambda)$ are   defined by
\begin{equation}\label{e22}
\varpi_1(\underline{x},\lambda)=\left\{\begin{array}{l}
1,~~~~~~\text{if}~~ \underline{a}(\lambda)\in \wideparen{\underline{a}(\lambda_0),\underline{a}(\lambda(\underline{x},\lambda_0))},\\
0~~~~~~~\text{else},
\end{array}\right.
\end{equation}
\begin{equation}\label{e23}
\varpi_2(\underline{x},\lambda)=\left\{\begin{array}{l}
1,~~~~~~\text{if}~~ \underline{a}(\lambda)\in \wideparen{\underline{a}(\lambda(\underline{x},\lambda_P)),\underline{a}(\lambda_P)},\\
0~~~~~~~\text{else}, 
\end{array}\right.
\end{equation}
where $\underline{a}(\lambda(\underline{x},\lambda_0))$ is another endpoint of the chord passing through $\underline{a}(\lambda_0)$ and $x$,  $\underline{a}(\lambda(\underline{x},\lambda_P))$ is another endpoint of the chord passing through $\underline{a}(\lambda_P)$ and $x$, and $\wideparen{\underline{a}(\lambda_0),\underline{a}(\lambda(\underline{x},\lambda_0))}$ denotes the arc starting from $\underline{a}(\lambda_0)$ and ending at $\underline{a}(\lambda(\underline{x},\lambda_0))$.

\subsubsection{Implementation for the curved-line detector}
In this subsection,  we describe how to implement  equation (\ref{e20})  when the fan-beam
projections are measured by using an equi-angular curved-line detector. 

\begin{figure}[h]
	\includegraphics[scale=0.5]{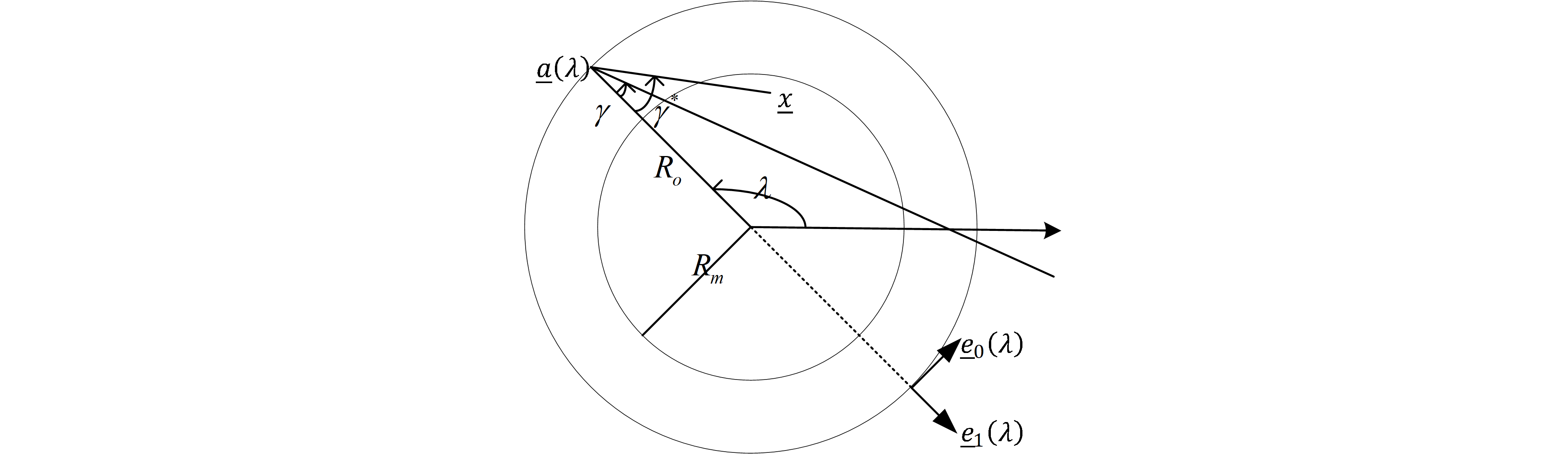}\\			
	\caption{Geometry of data acquisition by using an equi-angular curved detector.}
	\label{fig2}
\end{figure}
Let $\underline{a}(\lambda)=\left(R_{o} \cos \lambda, R_{o} \sin \lambda\right)$ be the position of the X-ray source, $\underline{\theta}(\lambda,\gamma)=\cos\gamma\underline{e}_0(\lambda)+\sin\gamma\underline{e}_1(\lambda)$ be the diverging direction, $g_c(\lambda,\gamma)=g(\lambda,\underline{\theta}(\lambda,\gamma))$ be the measured projection data using the curved-line detector, where
$\gamma$ is the sampling coordinate of the fan-angle,
 $\underline{e}_0(\lambda)=(-\sin(\lambda),\cos(\lambda))$, $\underline{e}_1(\lambda)=(-\cos(\lambda),-\sin(\lambda))$ (See Fig. \ref{fig2}). By the chain rule \cite{ISI:000177297700011}, equation (\ref{e16}) can be implemented as
\begin{equation}\label{e24}
g_1(\lambda,\gamma):=g^{\prime}(\lambda, \underline{\theta}(\lambda,\gamma))=\frac{\partial g_c(\lambda,\gamma)}{\partial \lambda}+\frac{\partial g_c(\lambda,\gamma)}{\partial \gamma}.
\end{equation}

Applying the changes of variable $\bar \gamma=\gamma^{\prime}-\gamma$, equation (\ref{e19}) can be rewritten as 
\begin{equation}\label{e25}
\begin{aligned}
 g_2(\lambda,\gamma):=&g^{F}(\lambda, \underline{\theta}(\lambda,\gamma))\\
 =&-\int_{0}^{2 \pi} \mathrm{d} \gamma^{\prime} h_{H}(\sin( \gamma-\gamma^{\prime})) g_1(\lambda, \gamma^{\prime}).
\end{aligned}
\end{equation}

Let $\gamma^{*}(\underline{x},\lambda)=\arctan\left(\frac{\underline{x} \cdot \underline{e}_{0}(\lambda)}{R_o+\underline{x} \cdot \underline{e}_{1}(\lambda)}\right)$ be the fan angle of the measured X-ray that diverges from $\underline{a}(\lambda)$ and passes through $\underline{x}$. Then, equation (\ref{e20}) can be rewritten as 
\begin{equation}\label{e26}
f(\underline{x})=-\frac{1}{2 \pi}  \int_{\lambda_0}^{\lambda_P} \mathrm{d} \lambda \frac{\varpi(\underline{x},\lambda)}{\|\underline{x}-\underline{a}(\lambda)\|} g_2\left(\lambda,\gamma^{*} \right).
\end{equation}

Note that the main differences of equation (\ref{e26}) and equation (\ref{e3}) are the weighting functions $w(\lambda, \phi)$ and $w(\underline{x},\lambda)$.

\subsubsection{Implementation for straight-line detector}
In this subsection,  we show how to implement  the algorithm  for the fan-beam
projections  measured by using an equi-space straight-line detector. 

\begin{figure}[h]
	\includegraphics[scale=0.5]{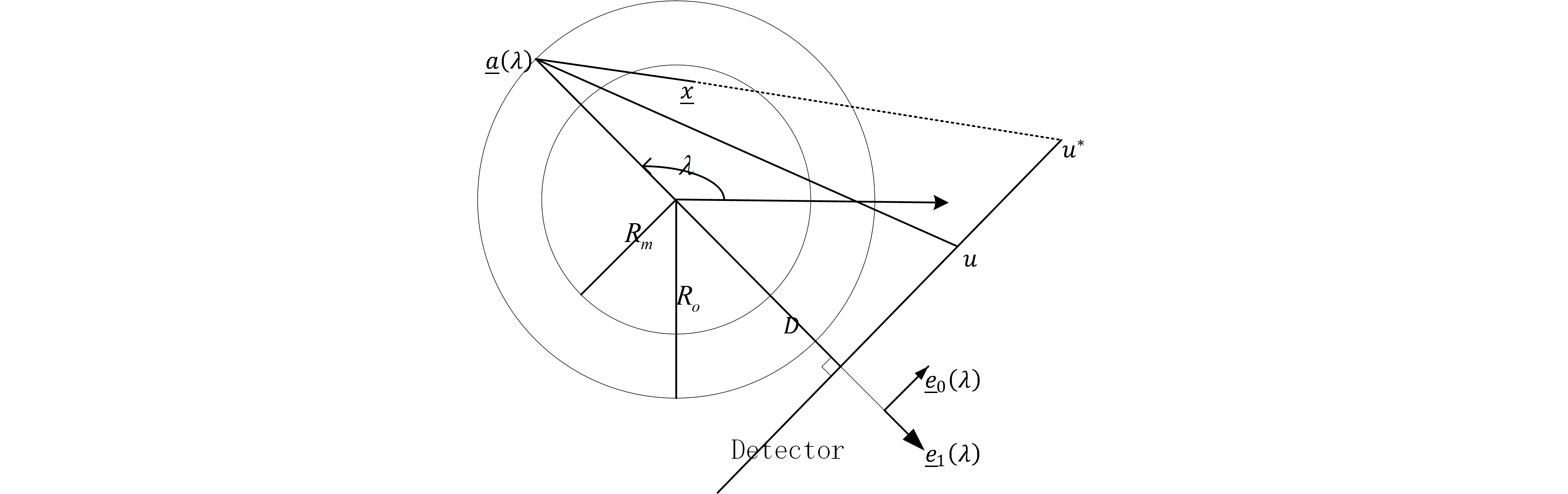}\\			
	\caption{Geometry of data acquisition by using an equi-space line detector.}
	\label{fig3}
\end{figure}

Let $\underline{a}(\lambda)=\left(R_{o} \cos \lambda, R_{o} \sin \lambda\right)$ be the position of the X-ray source, $\underline{\theta}(\lambda,u)=\frac{u\underline{e}_0(\lambda)+D\underline{e}_1(\lambda)}{\sqrt{u^2+D^2}}$ be the diverging direction, $g_l(\lambda,u)=g(\lambda,\underline{\theta}(\lambda,u))$ be the measured projection data using the straight-line detector, where $u$ is the sampling coordinate parallel to the detector, $D$ is the distance between the X-ray source and the detector, $\underline{e}_0(\lambda)=(-\sin(\lambda),\cos(\lambda))$, $\underline{e}_1(\lambda)=(-\cos(\lambda),-\sin(\lambda))$ (See Fig. \ref{fig3}). By the chain rule, equation (\ref{e16}) can be implemented as
\begin{equation}\label{e27}
g_1(\lambda,u):=g^{\prime}(\lambda, \underline{\theta}(\lambda,u))=\frac{\partial g_l(\lambda,u)}{\partial \lambda}+\frac{u^2+D^2}{D}\frac{\partial g_l(\lambda,u)}{\partial u}.
\end{equation}
Applying the changes of variables $u=D\tan(\gamma)$ and $\bar u=D\tan(\gamma^{\prime})$, equations (\ref{e25}) and (\ref{e26}) can be respectively rewritten as 
\begin{equation}\label{e28}
\begin{aligned}
g_2(\lambda,u):=&g^{F}(\lambda, \underline{\theta}(\lambda,u))\\
=&-\int_{-u_m}^{u_m} \mathrm{d} \bar u \frac{D}{\sqrt{\bar u^2+D^2}}h_{H}(u-\bar u) g_1(\lambda, \bar u)
\end{aligned}
\end{equation}
and
\begin{equation}
f(\underline{x})=-\frac{1}{2 \pi} \int_{\lambda_0}^{\lambda_P} \mathrm{d} \lambda \frac{\varpi(\underline{x},\lambda)}{R_o+\underline{x} \cdot \underline{e}_{1}(\lambda)} g_2\left(\lambda,u^{*} \right),
\end{equation}
where $u_{m}=D \tan \gamma_{m}$, $\gamma_m=\arcsin(R_m/R_o)$ and $u^{*}(\underline{x},\lambda)=D\frac{\underline{x} \cdot \underline{e}_{0}(\lambda)}{R_o+\underline{x} \cdot \underline{e}_{1}(\lambda)}$ is the detector location  of the measured X-ray that diverges from $\underline{a}(\lambda)$ and passes through $\underline{x}$.

\subsection{Circle cone-beam algorithm}
The fan-beam algorithm (\ref{e20}) can be extended for circle cone-beam CT reconstruction heuristically. In the following, we give the detailed circle cone-beam reconstruction algorithms for 
flat-plane  and curve-plane detectors, respectively.

\begin{figure}[h]
	\includegraphics[scale=0.5]{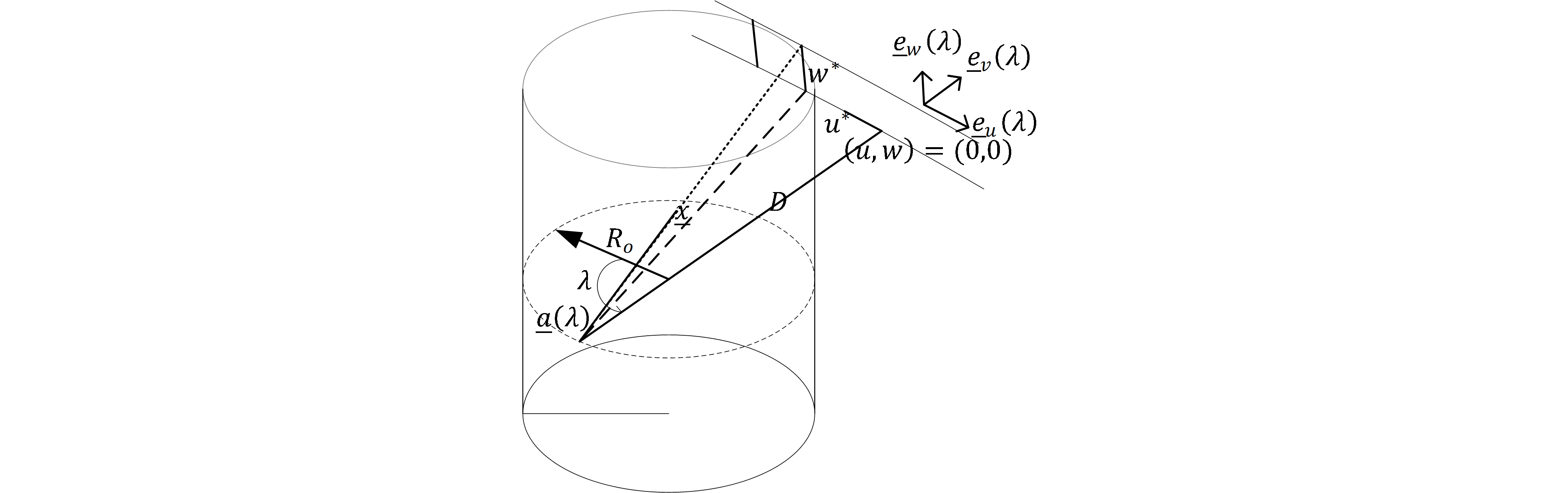}\\			
	\caption{Geometry of data acquisition by using a flat-plane detector.}
	\label{fig4}
\end{figure}

\subsubsection{Implementation for the flat-plane detector}
Let
$\underline{a}(\lambda)=(R_0\cos(\lambda),R_0\cos(\lambda),0)$ be the position of the X-ray source, $(u,v,w)$ be the local detector coordinates with unit vectors
\begin{equation}\label{e30}
\begin{aligned}
\underline{e}_{u}(\lambda)&=(-\sin(\lambda),\cos(\lambda),0),\\ \underline{e}_{v}(\lambda)&=(-\cos(\lambda),-\sin(\lambda),0),\\
\underline{e}_{w}(\lambda)&=(0,0,1),
\end{aligned}
\end{equation}
and $g_f(\lambda,u,w)=g(\lambda,\underline{\theta}(\lambda,u,w))$ be the measured projection data by using a flat-plane detector with diverging direction
\begin{equation}\label{key}
\underline{\theta}(\lambda,u,w)=\frac{u\underline{e}_{u}(\lambda)+D\underline{e}_{v}(\lambda)+w\underline{e}_{w}(\lambda)}{\sqrt{u^2+D^2+w^2}},
\end{equation}
 where $D$ is the distance between the detector plane and the X-ray source (See Fig. \ref{fig4}). To make  formula (\ref{e20}) effective for circle cone-beam CT reconstruction, we need to modify $\underline{\theta}^{\bot}$ in equation (\ref{e19}) and the weighting function $\varpi(\underline x,\lambda)$ in equation (\ref{e21}). 
In this paper, we heuristically set
\begin{equation}\label{key}
\underline{\theta}^{\bot}(\lambda,u,w)=\underline\theta(\lambda,u,w)\times (\frac{D\underline{e}_{v}(\lambda)+w\underline{e}_{w}(\lambda)}{\sqrt{D^2+w^2}}\times \underline{e}_{u}(\lambda))
\end{equation}
and 
\begin{equation}\label{key}
\varpi_{3d}(\underline x,\lambda) =\varpi(\text{Proj}(\underline x),\lambda),
\end{equation}
where $\text{Proj}(\underline x)$ represents projecting $\underline x$ on the circle plane formed by the trajectory of the X-ray source. Then, as derived in  Appendix A, the circle cone-beam reconstruction algorithm for the flat-plane detector  can be implemented as follows.
\begin{itemize}
	\item step1: derivative at constant direction
	\begin{equation}\label{e34}
	\begin{aligned}
	&g_1(\lambda,u,w):=g^{\prime}(\lambda,\underline\theta(\lambda,u,w))\\
	&=\frac{g_f(\lambda,u,w)}{\partial \lambda}+\frac{u^2+D^2}{D}\frac{g_f(\lambda,u,w)}{\partial u}+\frac{uw}{D}\frac{g_f(\lambda,u,w)}{\partial w}.
	\end{aligned}
	\end{equation}
	\item step2: convolution with Hilbert filter:
	\begin{equation}\label{e35}
	\begin{aligned}
	&g_2(\lambda,u,w):=g^{F}(\lambda, \underline{\theta}(\lambda,u,w))\\
	&=\int_{-u_m}^{u_m} \mathrm{d} \bar u \frac{D}{\sqrt{\bar u^2+D^2+w^2}}h_{H}(u-\bar u) g_1(\lambda, \bar u,w),
	\end{aligned}
	\end{equation}
	where $u_{m}=D \tan \gamma_{m}$, $\gamma_m=\arcsin(R_m/R_o)$.
	\item step3: backprojection
	\begin{equation}\label{key}
	\begin{aligned}
	f(\underline{x})=\frac{1}{2 \pi} \int_{\lambda_0}^{\lambda_P} \mathrm{d} \lambda \frac{\varpi_{3d}(\underline{x},\lambda)}{v^{*}(\underline x, \lambda)} g_2\left(\lambda,u^{*},w^{*} \right),
	\end{aligned}
	\end{equation}
	where $(u^{*}(\underline x, \lambda),w^{*}(\underline x, \lambda))$ is the position in the detector of the measured X-ray  that  diverges from $\underline a(\lambda)$  and passes through $\underline x$, which can be calculated by 
	\begin{equation}\label{key}
	\begin{aligned}
	v^{*}(\underline x, \lambda)=&R_o+\underline x \cdot \underline{e}_{v}(\lambda),\\
	u^{*}(\underline x, \lambda)=&\frac{D}{v^{*}(\underline x, \lambda)}(\underline x \cdot \underline{e}_{u}(\lambda)),\\
	w^{*}(\underline x, \lambda)=&\frac{D}{v^{*}(\underline x, \lambda)}(\underline x \cdot \underline{e}_{w}(\lambda)).
	\end{aligned}
	\end{equation}
\end{itemize}

\subsubsection{Implementation for the curve-plane detector}
\begin{figure}[h]
	\includegraphics[scale=0.5]{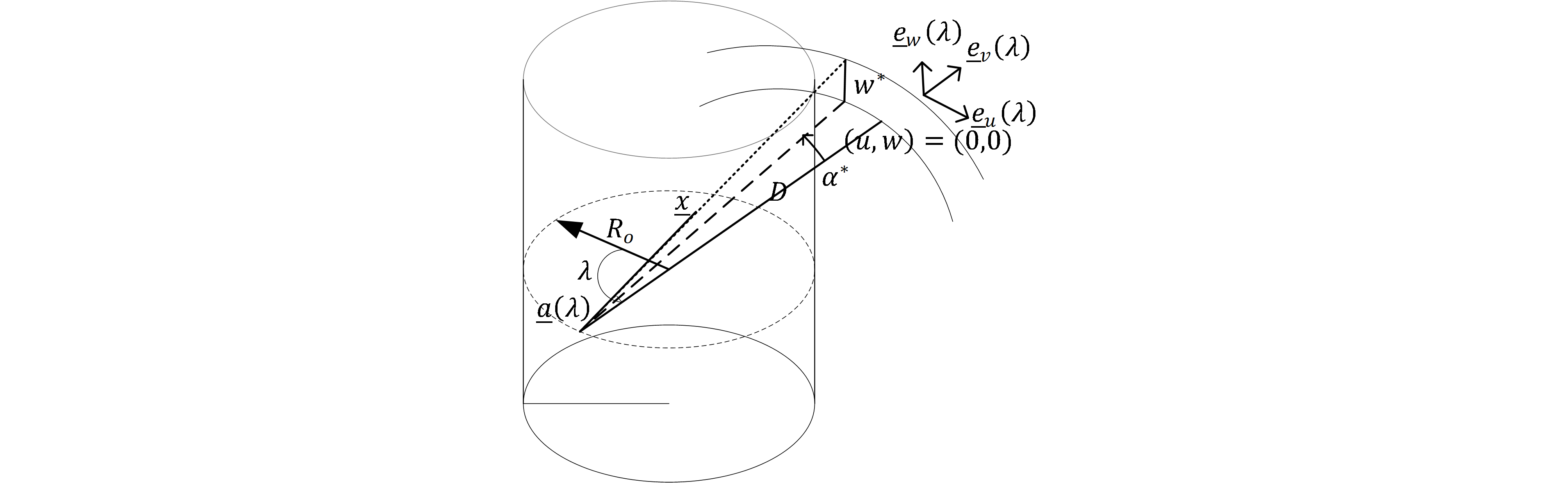}\\			
	\caption{Geometry of data acquisition by using a curve-plane detector.}
	\label{fig5}
\end{figure}
The curved-plane detector  array consists of $N_{row}\times N_{cols}$ detectors. The detector columns  are
perpendicular to the trajectory circle of the X-ray source, while the detector rows form circle arcs parallel to each other and to
the trajectory circle, where the center of the arc on the trajectory circle plane coincides with the X-ray source.
Let
$\underline{a}(\lambda)=(R_0\cos(\lambda),R_0\cos(\lambda),0)$ be the position of the X-ray source, $(\alpha,v,w)$ be the local detector coordinates with unit vectors defined in equation (\ref{e30})
and $g_c(\lambda,\alpha,w)=g(\lambda,\underline{\theta}(\lambda,\alpha,w))$ be the measured projection data by using a curve-plane detector with diverging direction
\begin{equation}\label{key}
\underline{\theta}(\lambda,\alpha,w)=\frac{D\sin(\alpha)\underline{e}_{u}(\lambda)+D\cos(\alpha)\underline{e}_{v}(\lambda)+w\underline{e}_{w}(\lambda)}{\sqrt{D^2+w^2}},
\end{equation}
where $D$ is the distance between the detector plane and the X-ray source (See Fig. \ref{fig5}). The curve-plane detector coordinates $(\alpha,v_c,w_c)$ can be be converted to the flat detector coordinates $(u,v_f,w_f)$ via 
\begin{equation}\label{e39}
u=D\tan(\alpha),~v_f=v_c, ~w_f=\frac{w_c}{\cos(\alpha)}.
\end{equation}
Applying the changes of variable in equation (\ref{e39}), we can obtain the circle cone-beam reconstruction algorithm for the curve-plane detector as follows.
 \begin{itemize}
 	\item step1: derivative at constant direction
 	\begin{equation}\label{e40}
 	\begin{aligned}
 	g_1(\lambda,\alpha,w):&=g^{\prime}(\lambda,\underline\theta(\lambda,\alpha,w))\\
 	&=\frac{g_c(\lambda,\alpha,w)}{\partial \lambda}+\frac{g_c(\lambda,\alpha,w)}{\partial \alpha}.
 	\end{aligned}
 	\end{equation}
 	\item step2: convolution with Hilbert filter:
 	\begin{equation}\label{e41}
 	\begin{aligned}
 	&g_2(\lambda,\alpha,w):=g^{F}(\lambda, \underline{\theta}(\lambda,\alpha,w))\\
 	&=\int_{0}^{2\pi} \mathrm{d} \bar \alpha \frac{D}{\sqrt{\bar D^2+w^2}}h_{H}(\sin(\alpha-\bar \alpha)) g_1(\lambda, \bar \alpha,w).
 	\end{aligned}
 	\end{equation}
 	\item step3: backprojection
 	\begin{equation}\label{key}
 	\begin{aligned}
 	f(\underline{x})=\frac{1}{2 \pi} \int_{\lambda_0}^{\lambda_P} \mathrm{d} \lambda \frac{\varpi_{3d}(\underline{x},\lambda)}{v^{*}(\underline x, \lambda)} \cos(\alpha^*)g_2\left(\lambda,\alpha^{*},w^{*} \right),
 	\end{aligned}
 	\end{equation}
 	where $(\alpha^{*}(\underline x, \lambda),w^{*}(\underline x, \lambda))$ is the position in the detector of the measured X-ray  that  diverges from $\underline a(\lambda)$  and passes through $\underline x$, which can be calculated by 
 	\begin{equation}\label{key}
 	\begin{aligned}
 	v^{*}(\underline x, \lambda)=&R_o+\underline x \cdot \underline{e}_{v}(\lambda),\\
 	\alpha^{*}(\underline x, \lambda)=&\arctan{(\frac{\underline x \cdot \underline{e}_{u}(\lambda)}{v^{*}(\underline x, \lambda)})},\\
 	w^{*}(\underline x, \lambda)=&\frac{D\cos(\alpha^*)}{v^{*}(\underline x, \lambda)}(\underline x \cdot \underline{e}_{w}(\lambda)).
 	\end{aligned}
 	\end{equation}
 \end{itemize}

\begin{figure*}[h]
	\centering{
		\subfloat[Original]{
			\label{figfan1_a}
			\includegraphics[width=0.49\columnwidth]{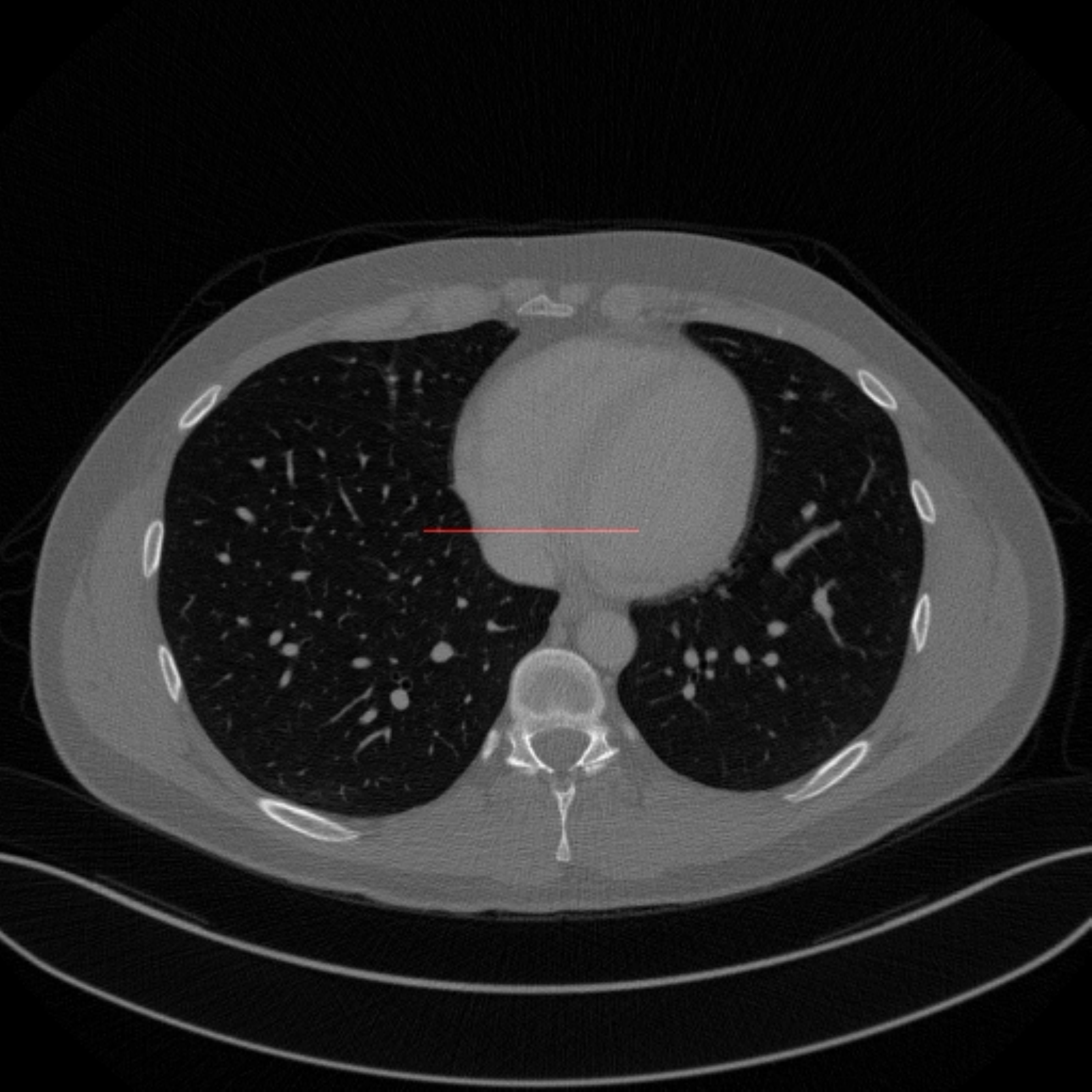}}		
		\subfloat[CFA]{ 
			\label{figfan1_b}
			\includegraphics[width=0.49\columnwidth]{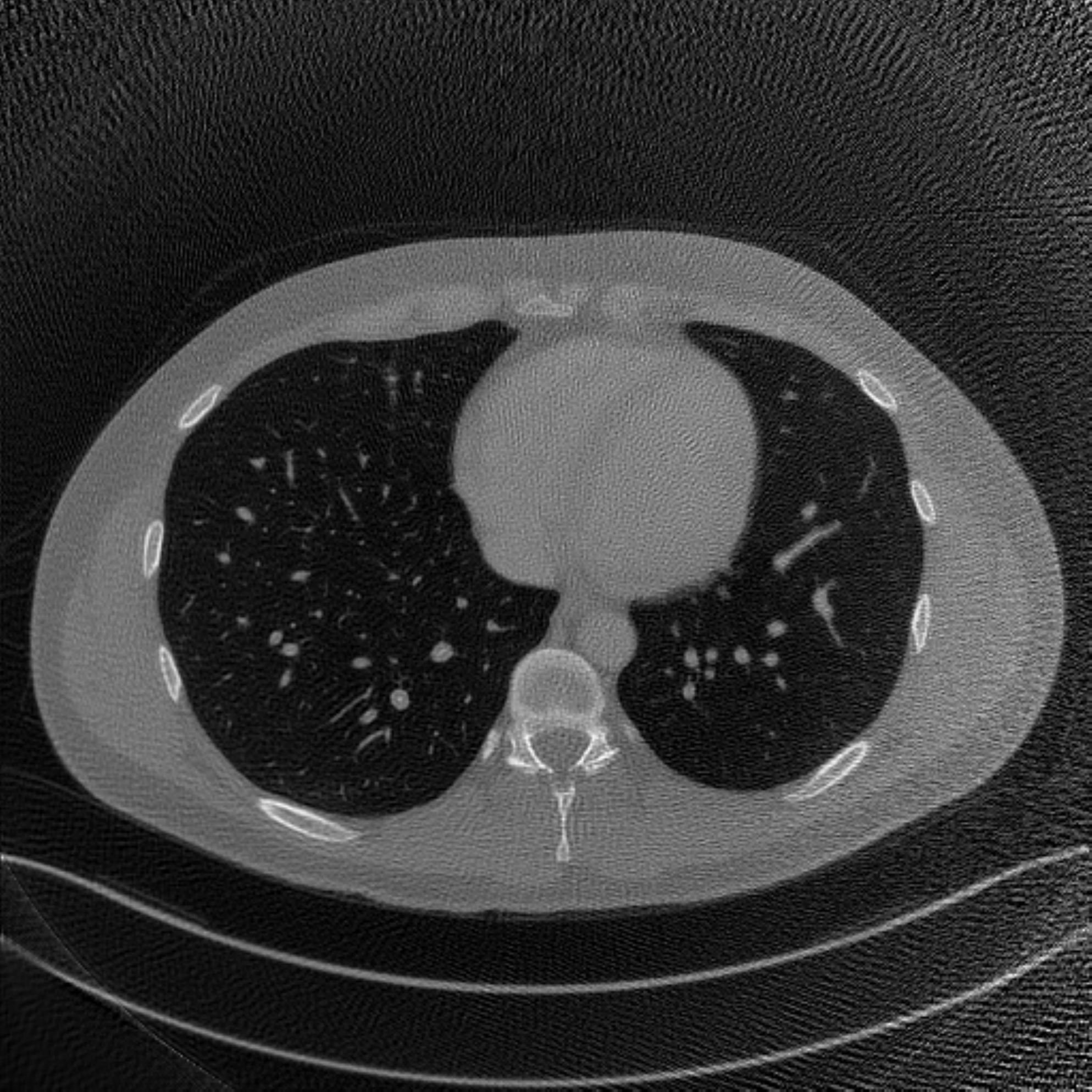}}		
		\subfloat[ACE]{ 
			\label{figfan1_c}
			\includegraphics[width=0.49\columnwidth]{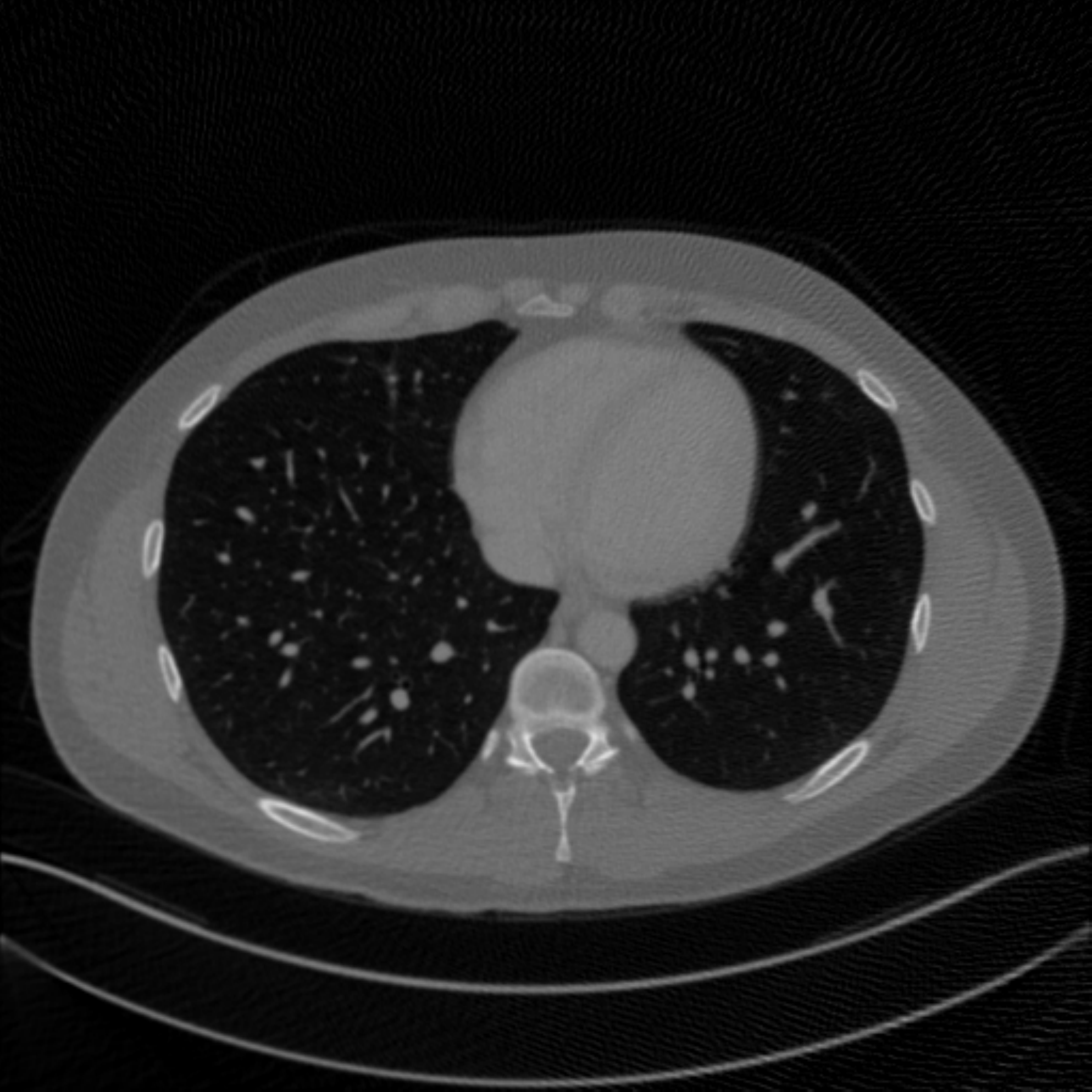}}
		\subfloat[Ours]{ 
			\label{figfan1_d}
			\includegraphics[width=0.49\columnwidth]{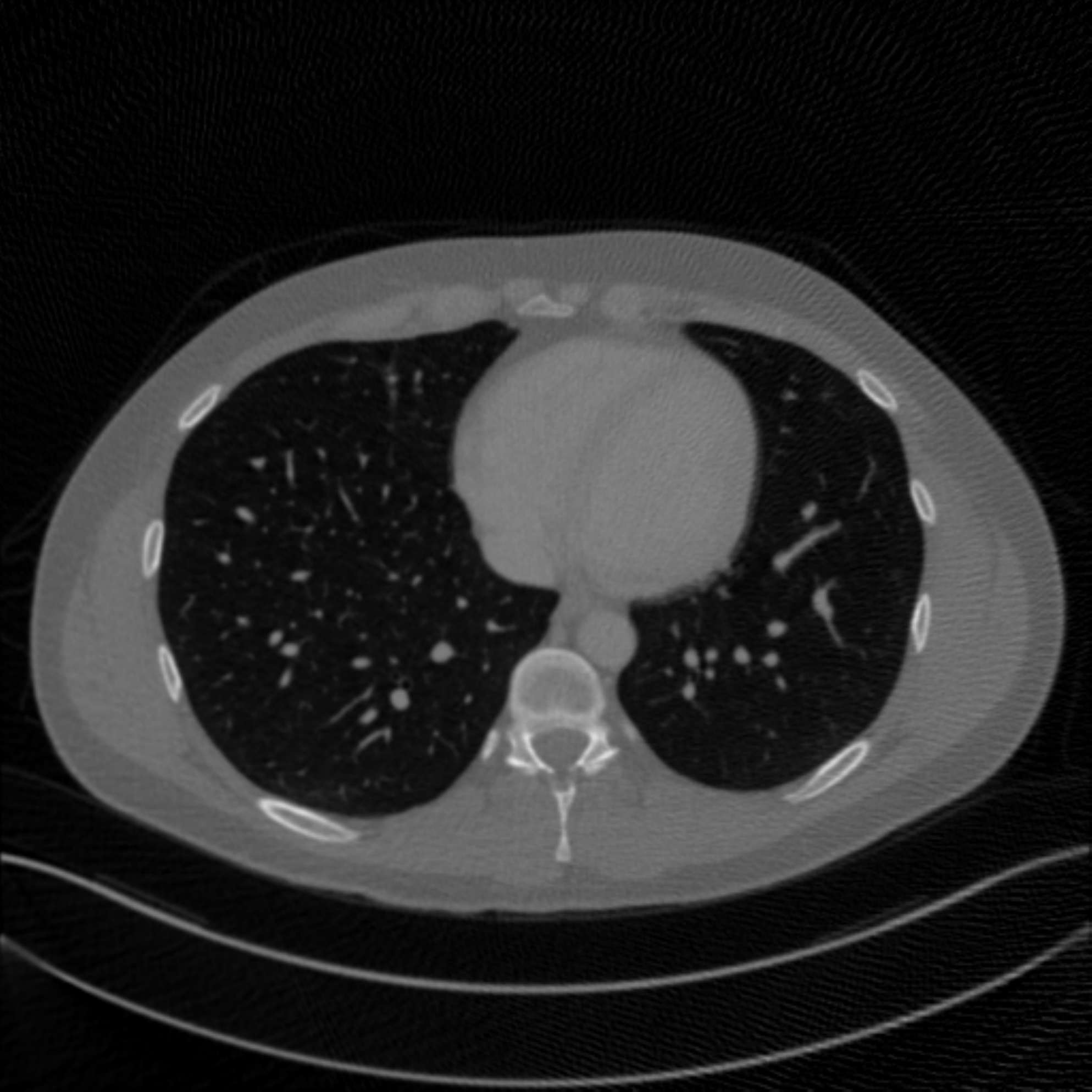}}
	}
	\caption{CT images reconstructed from short-scan fan-beam projection data.}
	\label{figfan1}
\end{figure*}

\begin{figure}[h]
	\centering{
		\includegraphics[width=1.9\columnwidth]{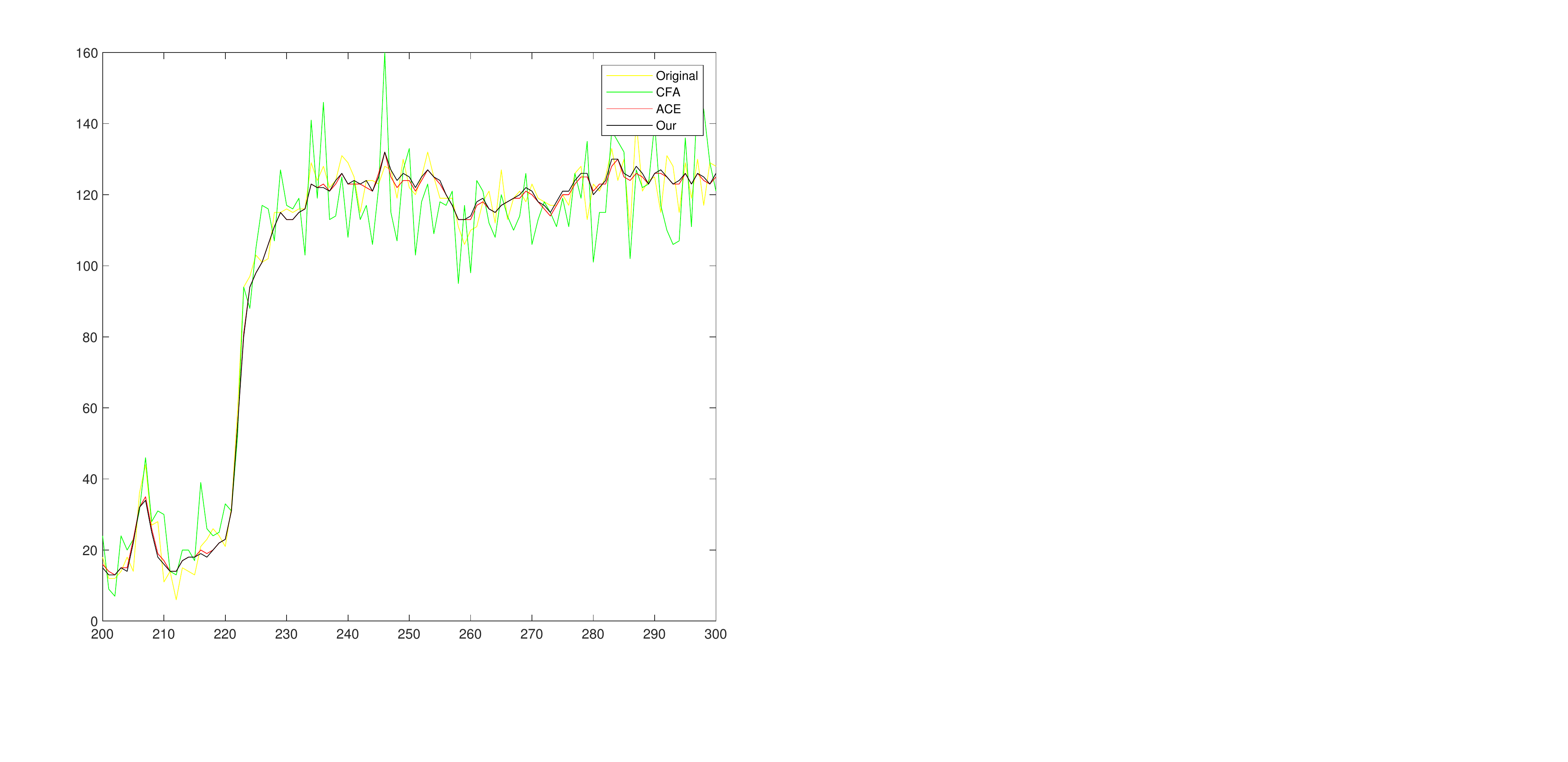}
	}
	\caption{1D intensity profile passing through the red  line
		in Fig. \ref{figfan1_a}.}
	\label{plot}
\end{figure}

\begin{figure*}[h]
	\centering{
		\subfloat[Original]{
			\label{figfan2_a}
			\includegraphics[width=0.49\columnwidth]{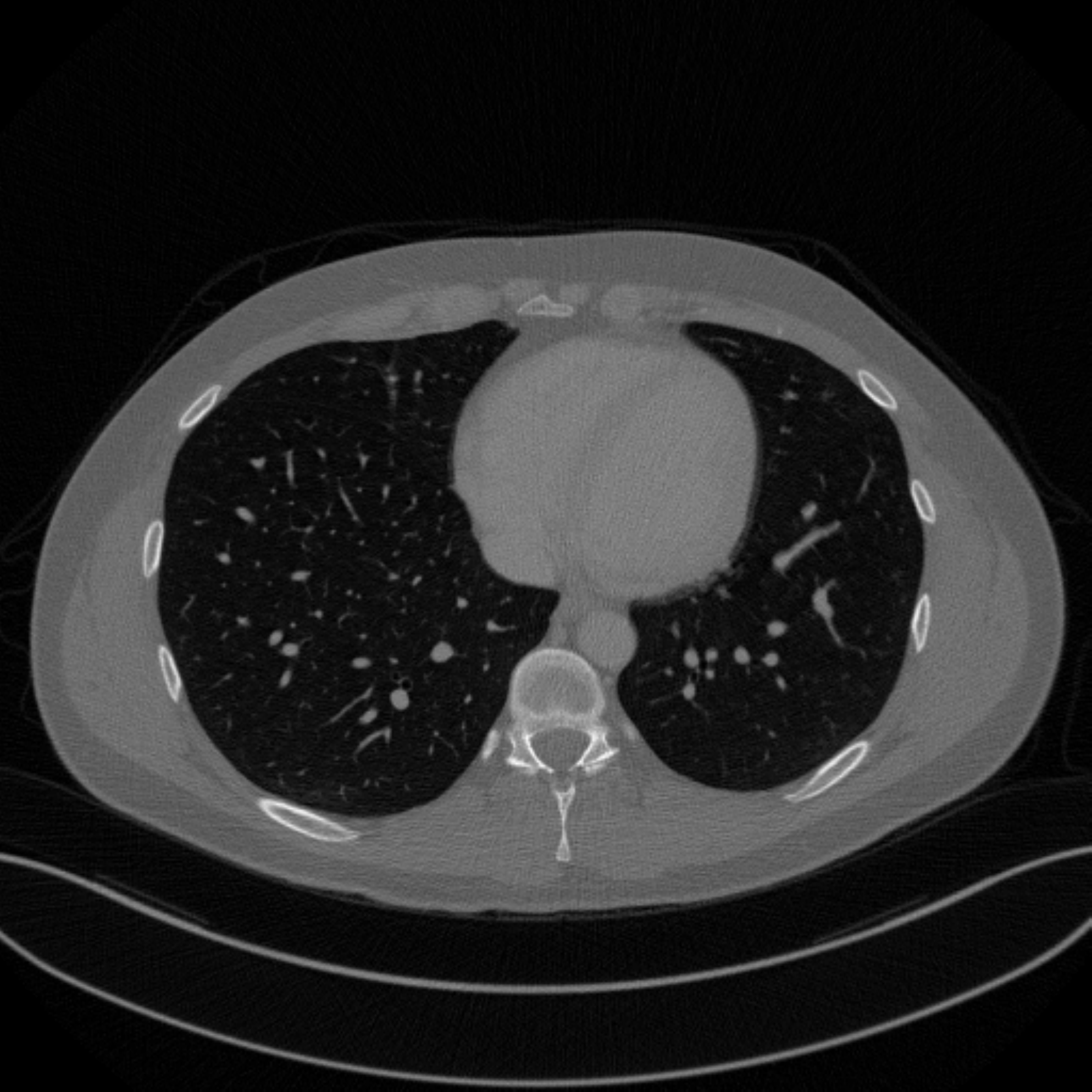}}		
		\subfloat[CFA]{ 
			\label{figfan2_b}
			\includegraphics[width=0.49\columnwidth]{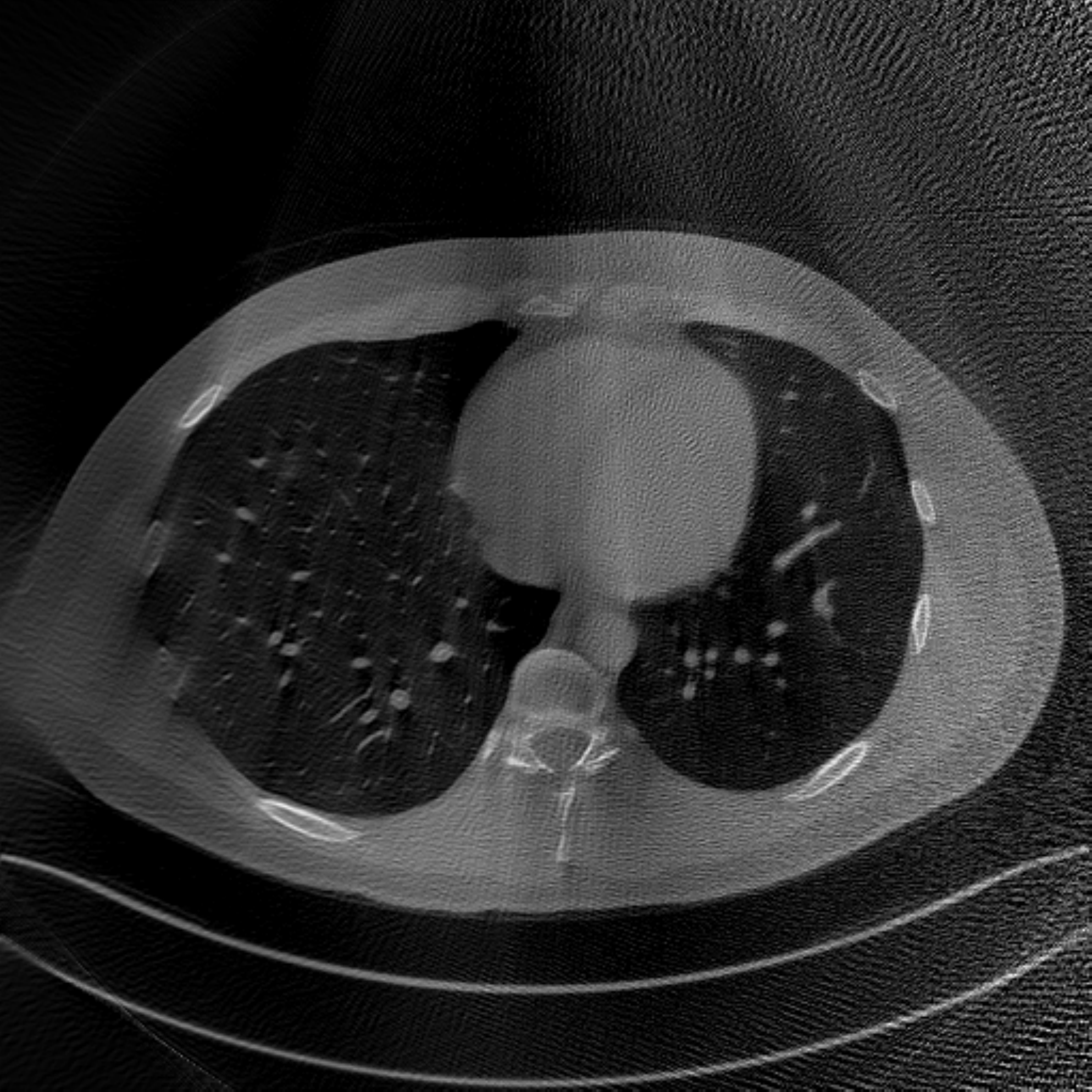}}		
		\subfloat[ACE]{ 
			\label{figfan2_c}
			\includegraphics[width=0.49\columnwidth]{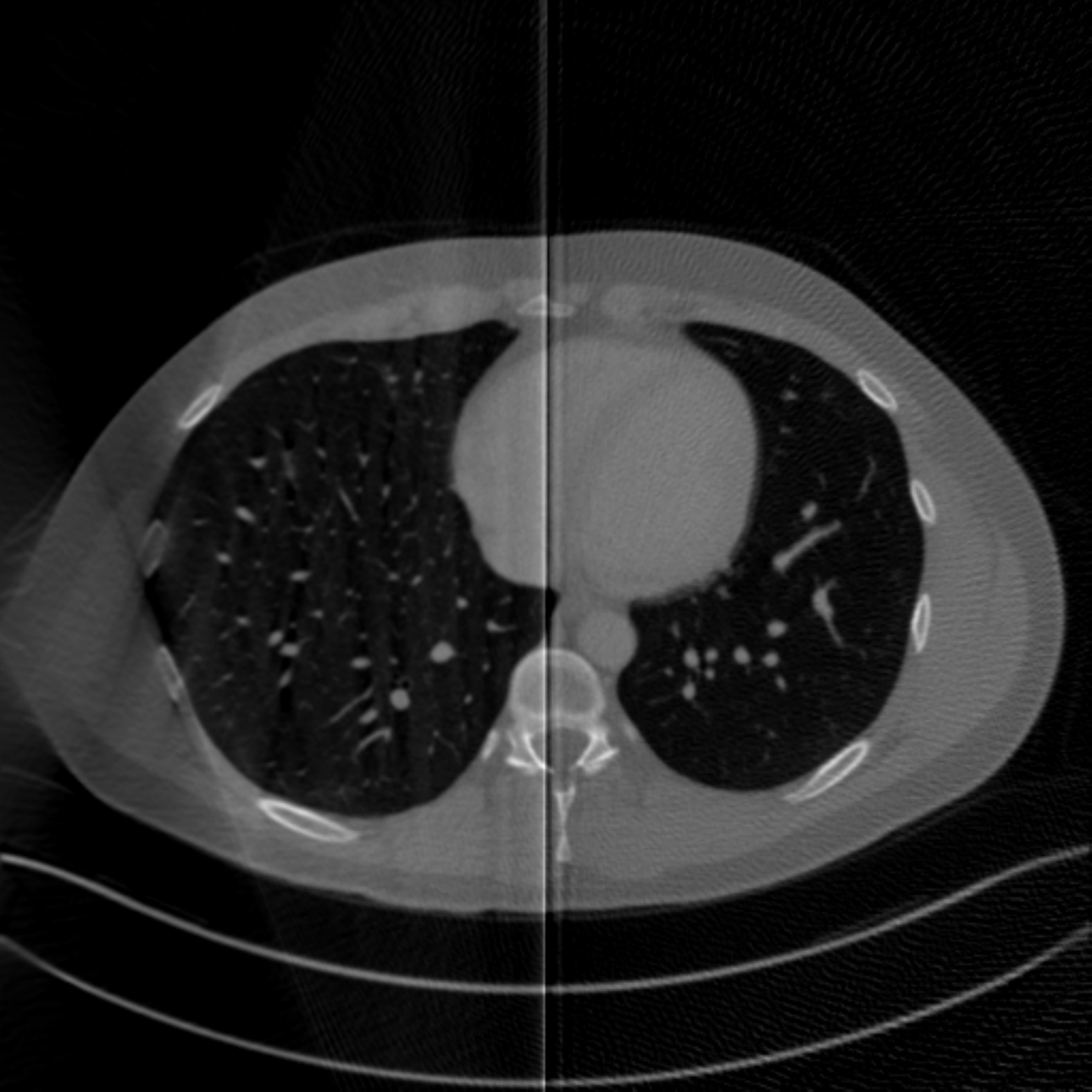}}
		\subfloat[Ours]{ 
			\label{figfan2_d}
			\includegraphics[width=0.49\columnwidth]{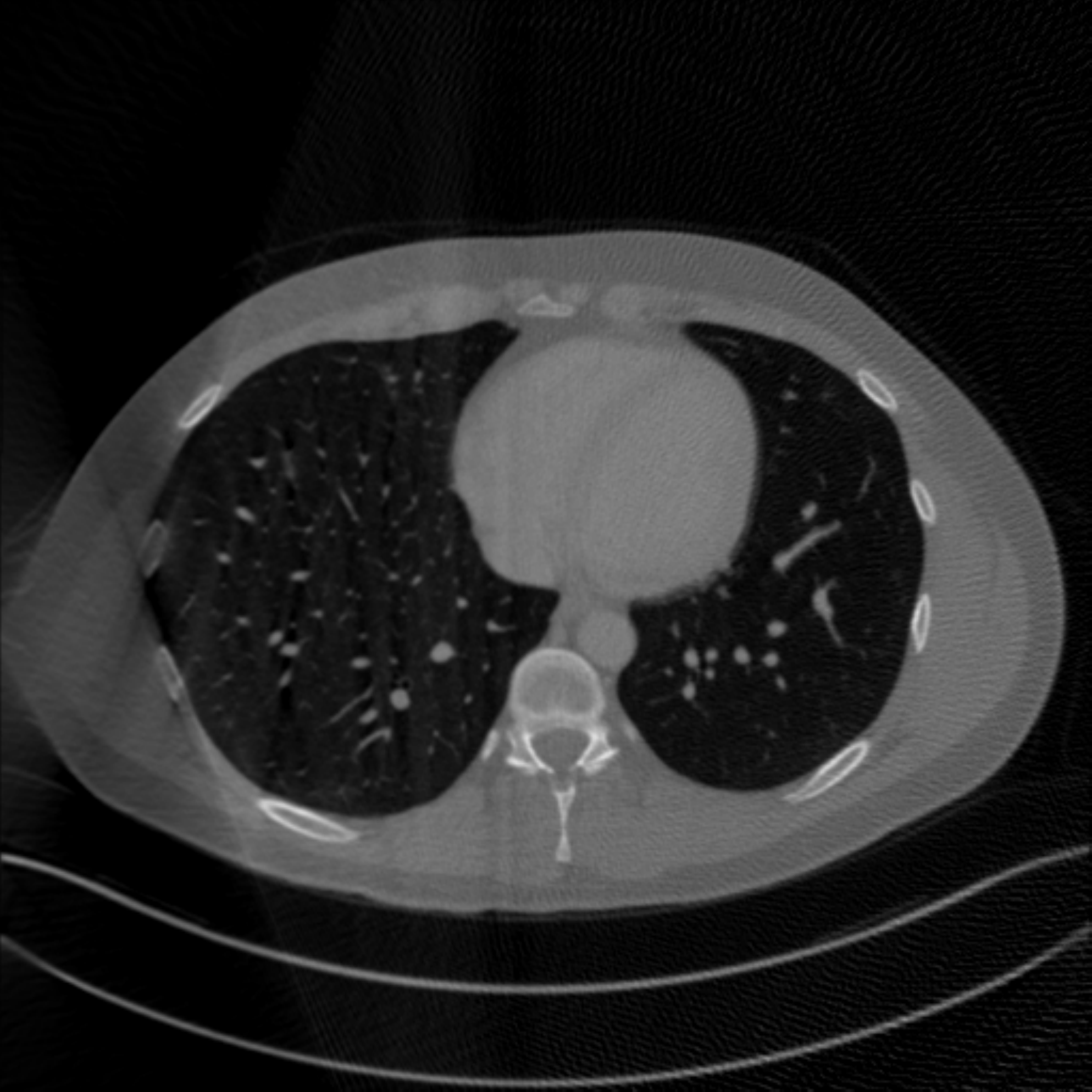}}
	}
	\caption{CT images reconstructed  from super-short-scan fan-beam projection data.}
	\label{figfan2}
\end{figure*}

\subsection{Additional discrete schemes}
In order to implement  our CT reconstruction algorithms, we need to give the  discrete definitions of  
the derivatives $\frac{\partial g_c(\lambda,\gamma)}{\partial \lambda}$ and  $\frac{\partial g_c(\lambda,\gamma)}{\partial \gamma}$ in equation (\ref{e24}), $\frac{\partial g_l(\lambda,u)}{\partial \lambda}$ and  $\frac{\partial g_l(\lambda,u)}{\partial u}$ in equation (\ref{e27}),
$\frac{g_f(\lambda,u,w)}{\partial \lambda}$, $\frac{g_f(\lambda,u,w)}{\partial u}$ and $\frac{g_f(\lambda,u,w)}{\partial w}$ in equation (\ref{e34}), and  $\frac{g_c(\lambda,\alpha,w)}{\partial \lambda}$ and $\frac{g_c(\lambda,\alpha,w)}{\partial \alpha}$ in equation (\ref{e40}),
the Hilbert filters $h_{H}(u)$ in equations (\ref{e28}) and (\ref{e35}), $h_{H}(\sin( \gamma))$ in equation (\ref{e25}) and $h_{H}(\sin(\alpha))$ in equation (\ref{e41}),
 and the weighting functions $\varpi_1(\underline{x},\lambda)$ in equation (\ref{e22})
and $\varpi_2(\underline{x},\lambda)$ in equation (\ref{e23}).

Let 
\begin{equation}\label{key}
\begin{aligned}
\gamma_i=&i\nabla \gamma, ~~i=-N,-N+1,...,0,...,N-1,N,\\
\alpha_i=&i\nabla \alpha, ~~i=-N,-N+1,...,0,...,N-1,N,\\
u_j=&j\nabla u, ~~j=-M,-M+1,...,0,...,M-1,M,\\
w_k=&k\nabla w, ~~k=-L,-L+1,...,0,...,L-1,L
\end{aligned}
\end{equation}
be the discrete coordinates of the detectors and
\begin{equation}\label{key}
\lambda_s=s\nabla \lambda,~~s=0,1,...,P
\end{equation}
be the discrete sampling angles. To avoid the space shift in the reconstructed images, we use the centered difference schemes to calculate the derivatives:
 \begin{equation}\label{key}
 \begin{aligned}
 \frac{\partial g_c(\lambda_s,\gamma)}{\partial \lambda}=&\frac{g_c(\lambda_{s+1},\gamma)-g_c(\lambda_{s-1},\gamma)}{2\nabla \lambda},\\
 \frac{\partial g_c(\lambda,\gamma_i)}{\partial \gamma}=&\frac{g_c(\lambda,\gamma_{i+1})-g_c(\lambda,\gamma_{i-1})}{2\nabla \gamma},\\
 \frac{\partial g_l(\lambda_s,u)}{\partial \lambda}=&\frac{g_l(\lambda_{s+1},u)-g_l(\lambda_{s-1},u)}{2\nabla \lambda},\\
 \frac{\partial g_l(\lambda,u_j)}{\partial \lambda}=&\frac{g_l(\lambda,u_{j+1})-g_l(\lambda,u_{j-1})}{2\nabla u},\\ 
 \frac{g_f(\lambda_s,u,w)}{\partial \lambda}=&\frac{g_f(\lambda_{s+1},u,w)-g_f(\lambda_{s-1},u,w)}{2\nabla \lambda},\\
 \frac{g_f(\lambda,u_j,w)}{\partial u}=&\frac{g_f(\lambda,u_{j+1},w)-g_f(\lambda,u_{j-1},w)}{2\nabla u},\\
  \frac{g_f(\lambda,u,w_k)}{\partial u}=&\frac{g_f(\lambda,u,w_{k+1})-g_f(\lambda,u,w_{k-1})}{2\nabla w},\\
  \frac{g_c(\lambda_s,\alpha,w)}{\partial \lambda}=&\frac{g_c(\lambda_{s+1},\alpha,w)-g_c(\lambda_{s-1},\alpha,w)}{2\nabla \lambda},\\
   \frac{g_c(\lambda,\alpha_i,w)}{\partial \alpha}=&\frac{g_c(\lambda,\alpha_{i+1},w)-g_c(\lambda,\alpha_{i-1},w)}{2\nabla \alpha}.\\
 \end{aligned}
 \end{equation}

 Let $b$ denote a cut-off frequency for the Hilbert filter $h_{H}(u)$. Then,  we may write \cite{Wunderlich_thekatsevich}:
 \begin{equation}
 \begin{aligned}
 k_{H}(u) & \approx-\int_{-b}^{b} i \operatorname{sgn}(\sigma) e^{i 2 \pi \sigma u} d \sigma \\
 &=\int_{-b}^{0} i e^{i 2 \pi \sigma u} d \sigma-\int_{0}^{b} i e^{i 2 \pi \sigma u} d \sigma \\
 &=\left[\frac{1}{2 \pi u} e^{i 2 \pi \sigma t}\right]_{\sigma=-b}^{0}-\left[\frac{1}{2 \pi u} e^{i 2 \pi \sigma t}\right]_{\sigma=0}^{b} \\
 &=\frac{1}{2 \pi u}(1-e^{-i 2 \pi b u}-e^{i 2 \pi b u}+1)\\
 &=\frac{1}{2 \pi u}(2-2 \cos (2 \pi b u)) \\
 &=\frac{1}{\pi u}(1-\cos (2 \pi b u)).
 \end{aligned}
 \end{equation}
For $h_{H}(\sin( \gamma))$ and $h_{H}(\sin( \gamma))$, we have
\begin{equation}\label{key}
h_{H}(\sin( \gamma))=\frac{\gamma}{\sin(\gamma)}h_{H}(\gamma)=\frac{1-\cos(2\pi b \gamma)}{\pi\sin(\gamma)}
\end{equation}
and so
\begin{equation}\label{key}
h_{H}(\sin( \alpha))=\frac{1-\cos(2\pi b \alpha)}{\pi\sin(\alpha)}.
\end{equation}
In this paper,  we set $b=1/(2\nabla u), 1/(2\nabla \gamma)$ and  $1/(2\nabla \alpha)$ for $h_{H}(u)$, $h_{H}(\sin( \gamma))$ and $h_{H}(\sin( \alpha))$, respectively. Therefore, the discrete definitions of the Hilbert filters are:
\begin{equation}\label{key}
\begin{aligned}
k_{H}(u_j)=&\frac{1-\cos (\pi u_j /\nabla u)}{\pi u_j},\\
h_{H}(\sin( \gamma_i))=&\frac{1-\cos(\pi \gamma_i/\nabla \gamma)}{\pi\sin(\gamma_i)},\\
h_{H}(\sin( \alpha_i))=&\frac{1-\cos(\pi \alpha_i/\nabla \alpha)}{\pi\sin(\alpha_i)}.\\
\end{aligned}
\end{equation}

For the weighting functions $\varpi_1(\underline{x},\lambda)$ and $\varpi_2(\underline{x},\lambda)$, since the endpoints $\underline{a}(\lambda(\underline{x},\lambda_0))$ and $\underline{a}(\lambda(\underline{x},\lambda_P))$ may not coincide with the  sampling points  $a(\lambda_s)$, we need to interpolate them at the endpoints of the sampling angles $\lambda(\underline{x},\lambda_0)$ and $\lambda(\underline{x},\lambda_P)$. Let
\begin{equation}\label{key}
\begin{aligned}
s(\underline{x},\lambda_0)=&(\lambda(\underline{x},\lambda_0)-\lambda_0)/\nabla \lambda,\\ 
\lfloor s(\underline{x},\lambda_0)\rfloor=&\text{floor}(s(\underline{x},\lambda_0)),\\
\lceil s(\underline{x},\lambda_0)\rceil=&\text{floor}(s(\underline{x},\lambda_0))+1,\\
s(\underline{x},\lambda_P)=&(\lambda(\underline{x},\lambda_P)-\lambda_0)/\nabla \lambda, \\ 
\lfloor s(\underline{x},\lambda_P)\rfloor=&\text{floor}(s(\underline{x},\lambda_P)),\\
\lceil s(\underline{x},\lambda_P)\rceil=&\text{floor}(s(\underline{x},\lambda_P))+1.\\
\end{aligned}
\end{equation}
Then, we set
\begin{equation}\label{key}
\begin{aligned}
\varpi_1(\underline{x},\lambda_s)=&\left\{\begin{array}{l}
1,~~~~~~~~~~~~~~~~\text{if}~ s\in [0,\lfloor s(\underline{x},\lambda_0)\rfloor],\\
s-\lfloor s(\underline{x},\lambda_0)\rfloor,~\text{if}~ s\in (\lfloor s(\underline{x},\lambda_0)\rfloor,\lceil s(\underline{x},\lambda_0)\rceil),\\
0,~~~~~~~~~~~~~~~~\text{else},
\end{array}\right.\\
\varpi_2(\underline{x},\lambda_s)=&\left\{\begin{array}{l}
1,~~~~~~~~~~~~~~~~~\text{if}~ s\in [\lceil s(\underline{x},\lambda_P)\rceil,P],\\
s-\lfloor s(\underline{x},\lambda_P)\rfloor,~\text{if}~ s\in (\lfloor s(\underline{x},\lambda_P)\rfloor,\lceil s(\underline{x},\lambda_P)\rceil),\\
0,~~~~~~~~~~~~~~~~~\text{else},
\end{array}\right.\\
\end{aligned}
\end{equation}
 where $\lambda(\underline{x},\lambda_0)$ and $\lambda(\underline{x},\lambda_P)$ can be calculated via Appendix B.
 
 \begin{figure*}[htbp]
 	\centering{
 		\subfloat{\includegraphics[width=0.49\columnwidth]{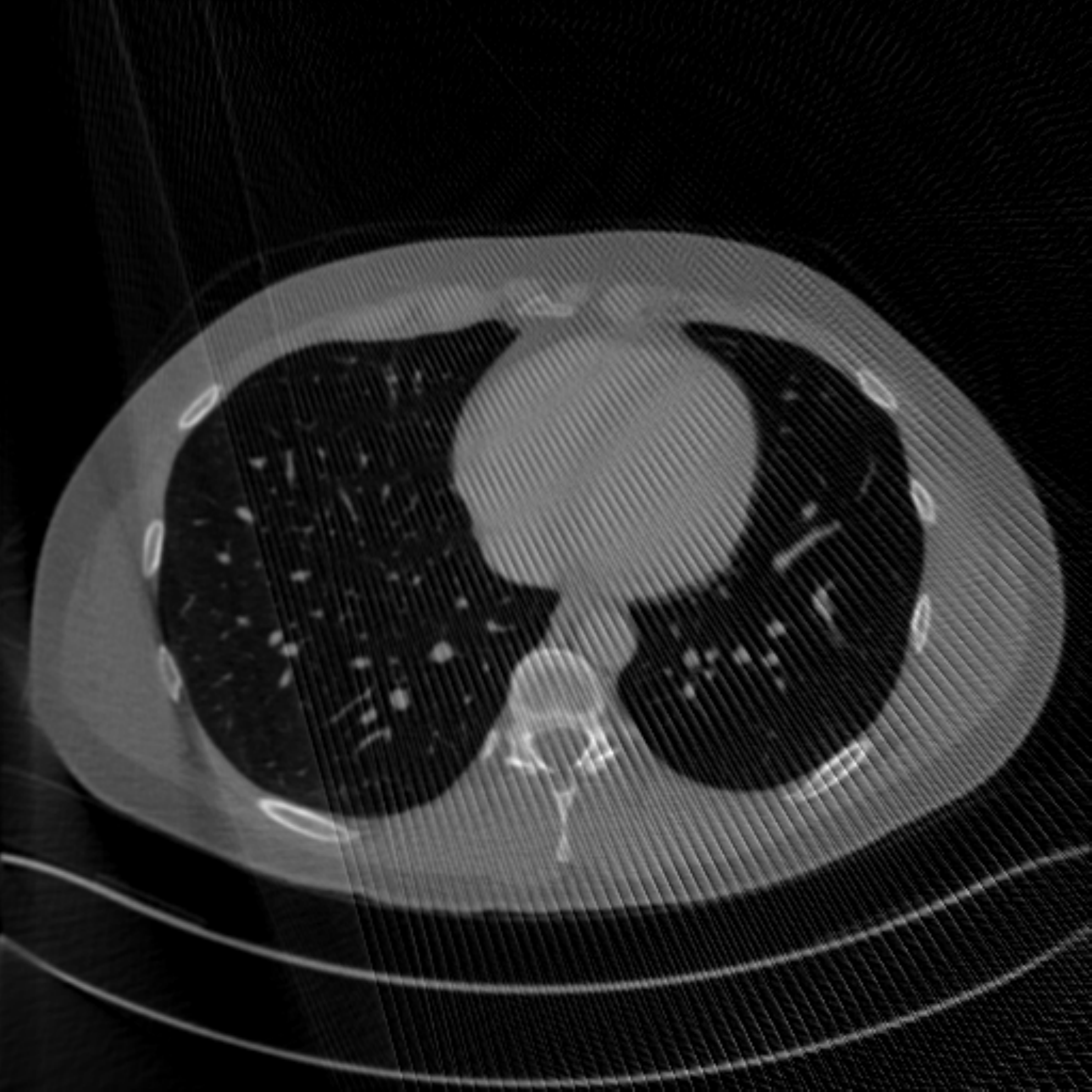}}
 		\subfloat{\includegraphics[width=0.49\columnwidth]{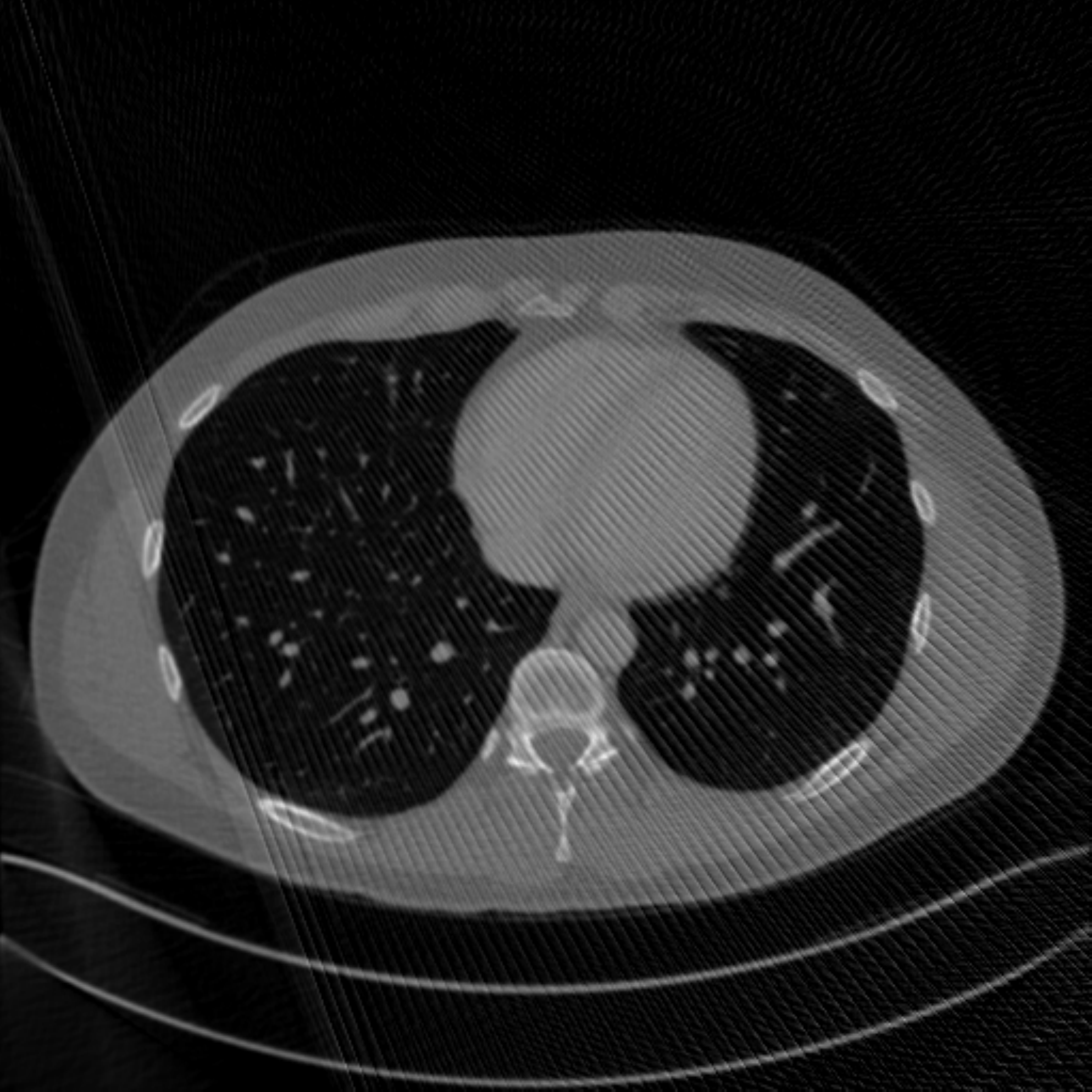}}
 		\subfloat{\includegraphics[width=0.49\columnwidth]{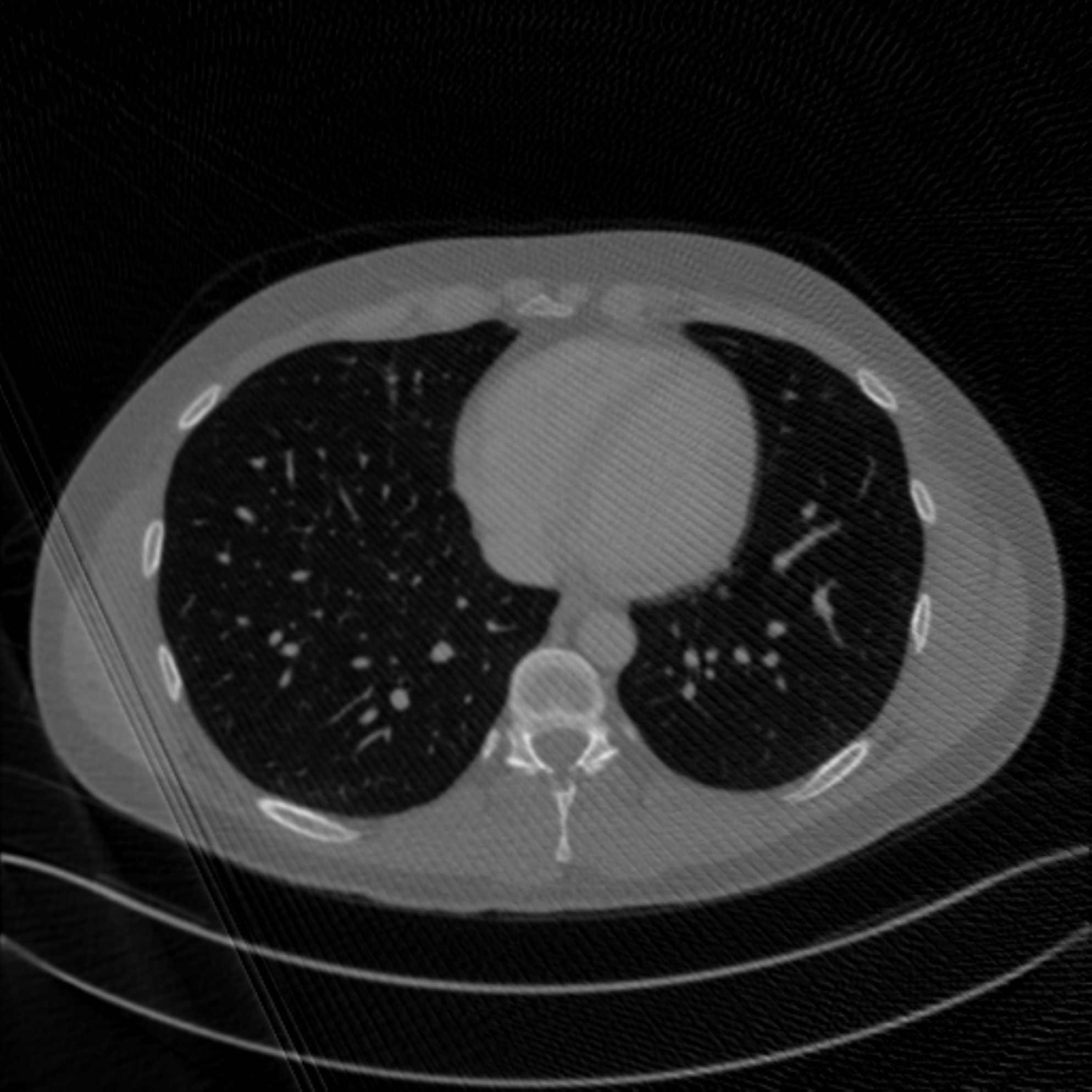}}
 		\subfloat{\includegraphics[width=0.49\columnwidth]{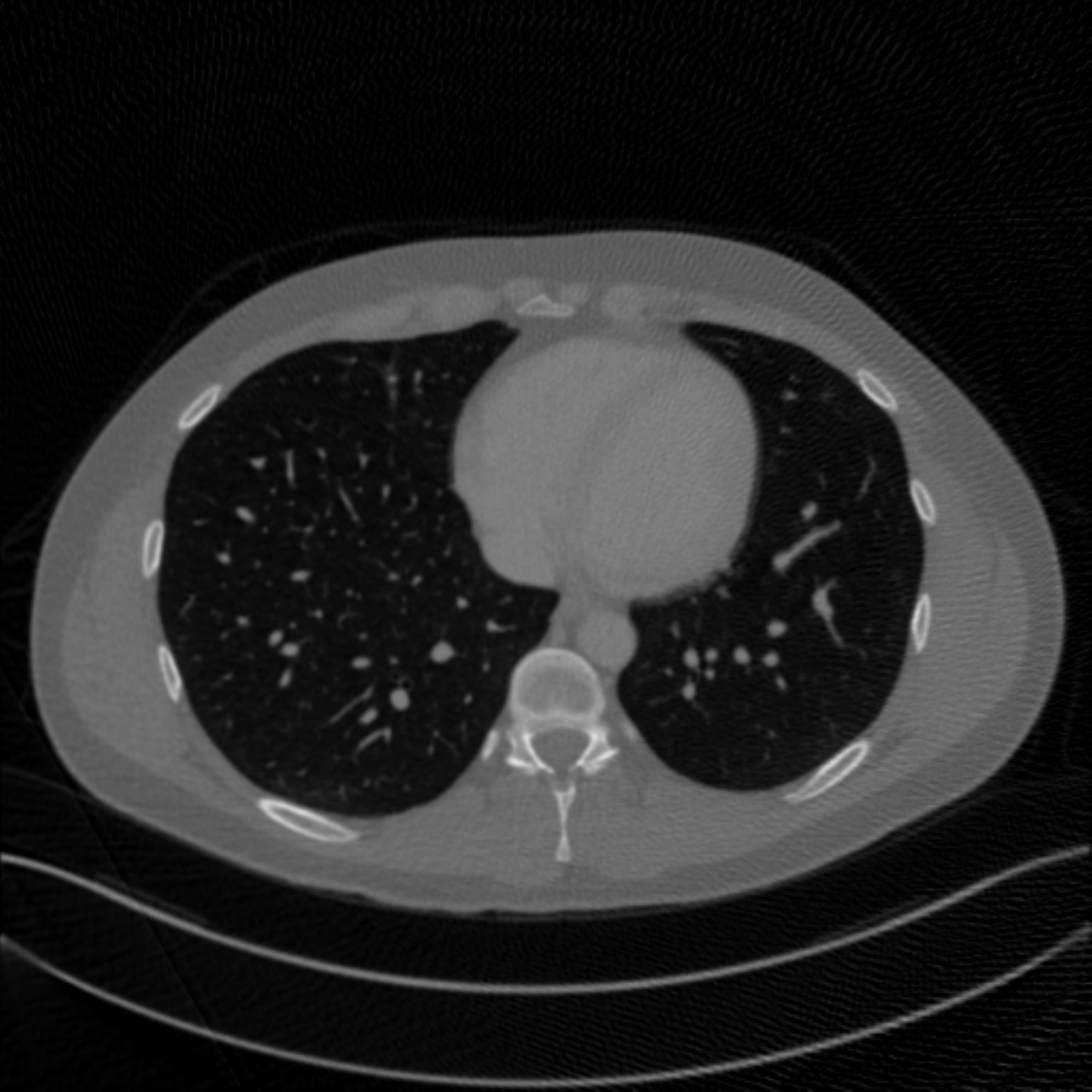}}
 		
 		\vspace{-3mm}
 		\setcounter{subfigure}{0}
 		\subfloat[{$\beta$=[0:210]}]{\includegraphics[width=0.49\columnwidth]{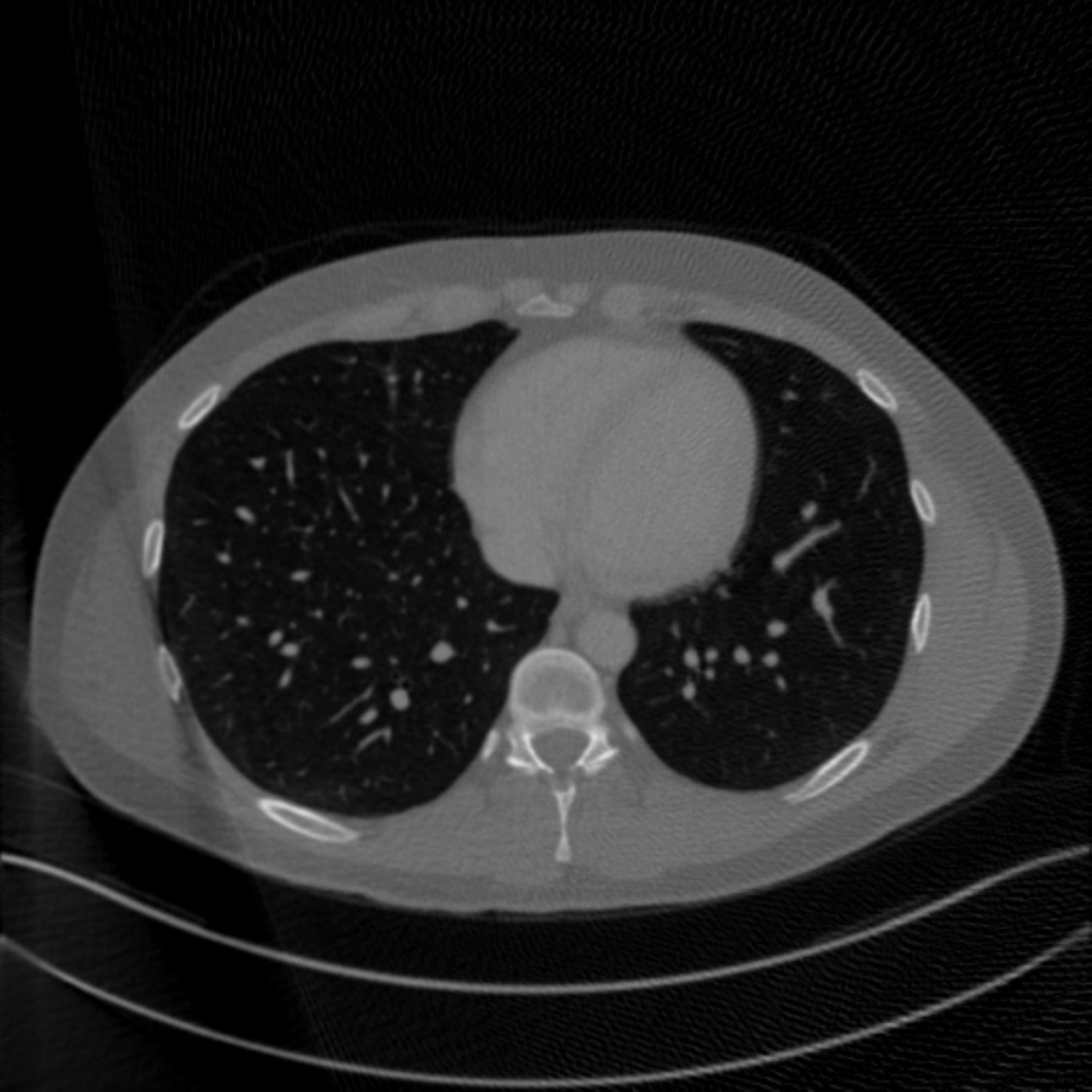}}
 		\subfloat[{$\beta$=[0:220]}]{\includegraphics[width=0.49\columnwidth]{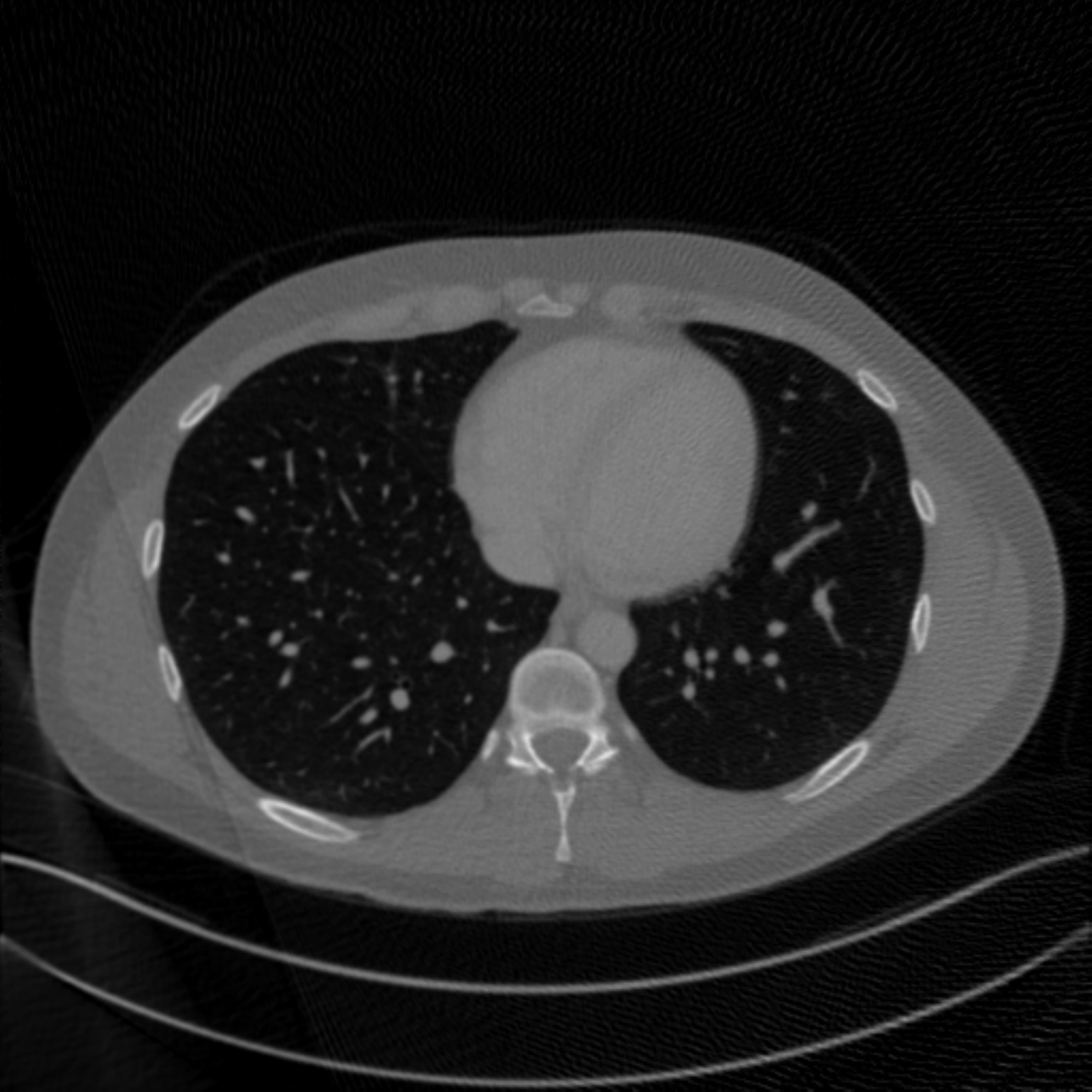}}
 		\subfloat[{$\beta$=[0:230]}]{\includegraphics[width=0.49\columnwidth]{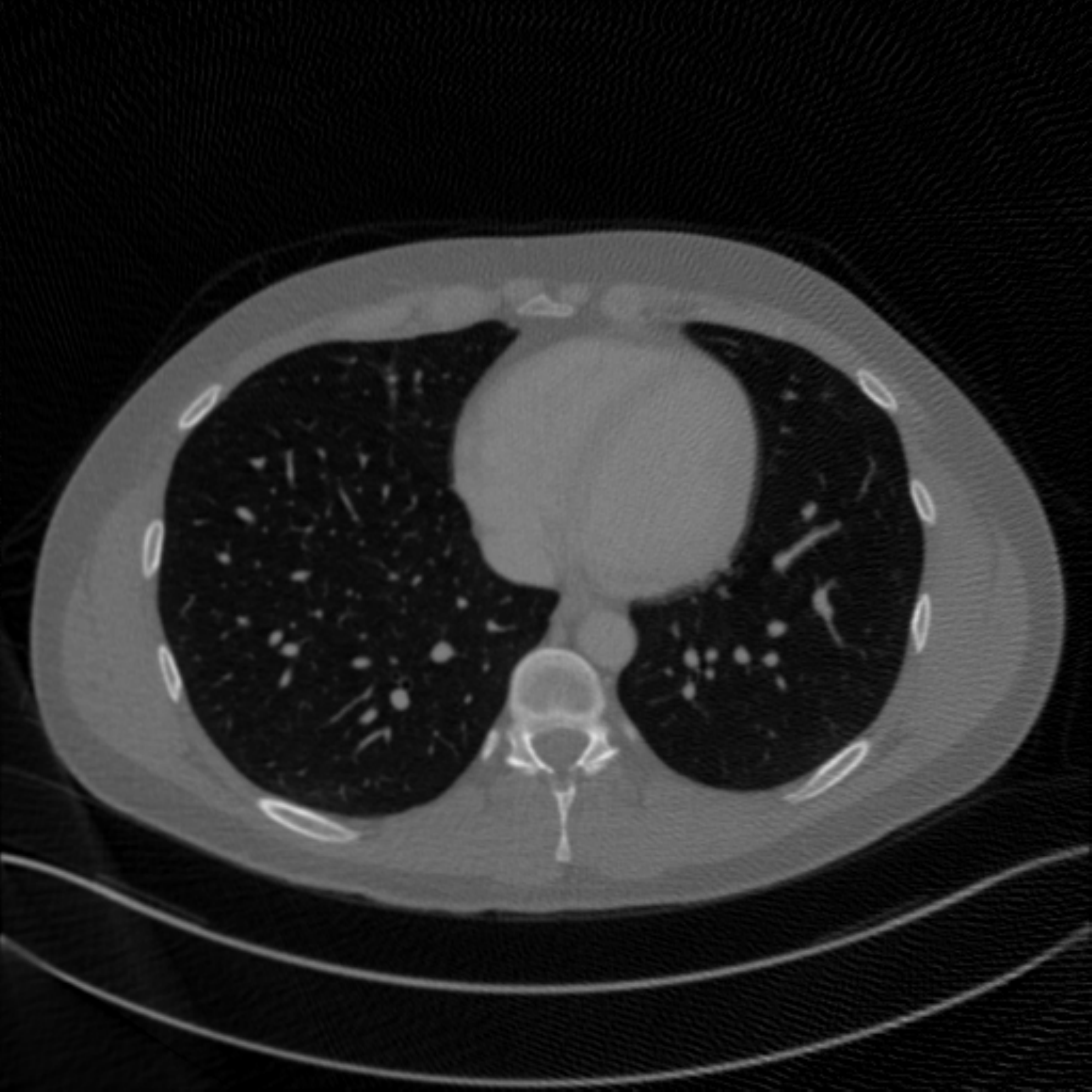}}
 		\subfloat[{$\beta$=[0:240]}]{\includegraphics[width=0.49\columnwidth]{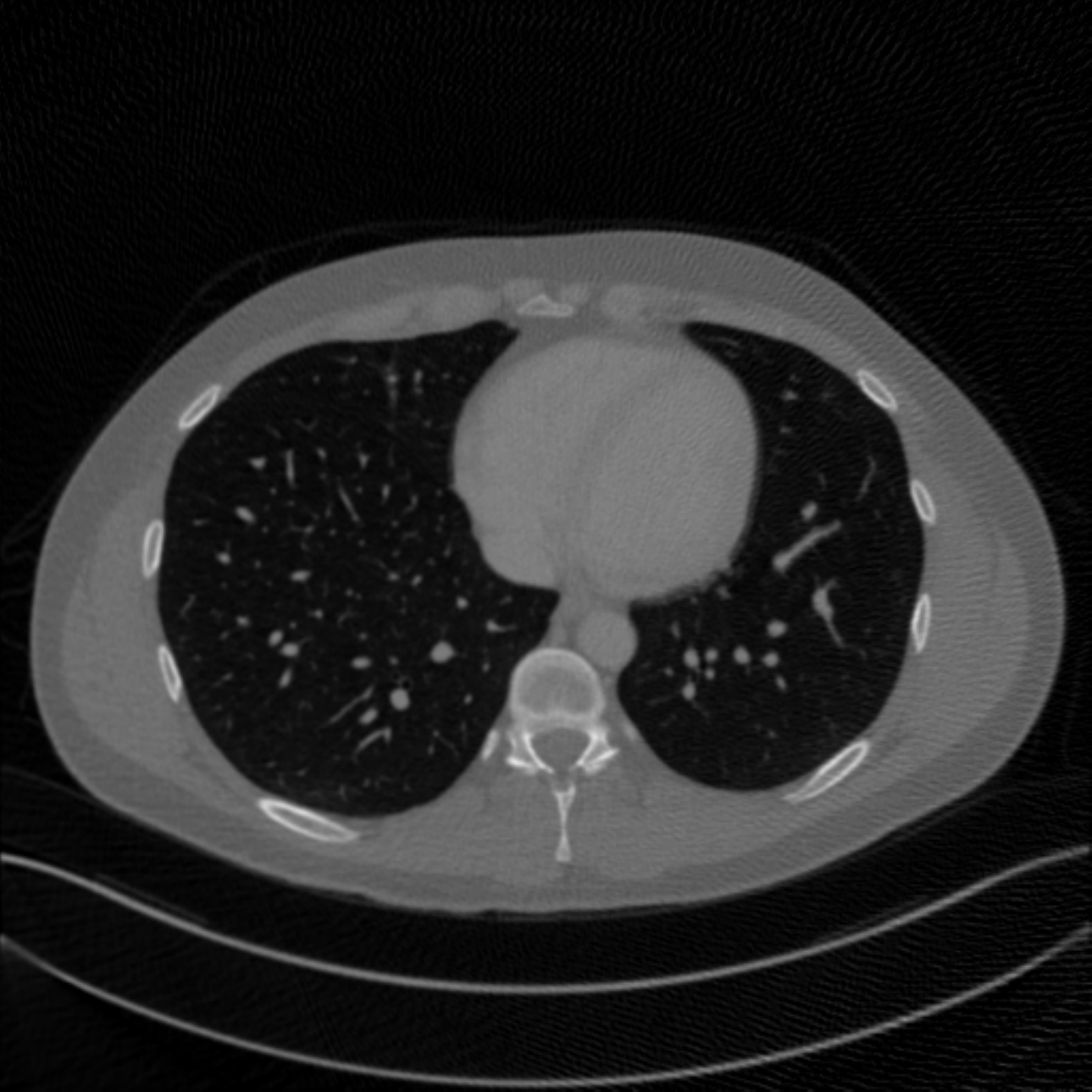}}	
 	}
 	\caption{CT images reconstructed from fan-beam projections with different $\beta$. The images in the first row are  reconstructed by ACE and in the second row  by ours.}
 	\label{figfan3}
 \end{figure*}
 
 \begin{figure*}[h]
 	\centering{
 		\subfloat[var=1]{
 			\includegraphics[width=0.49\columnwidth]{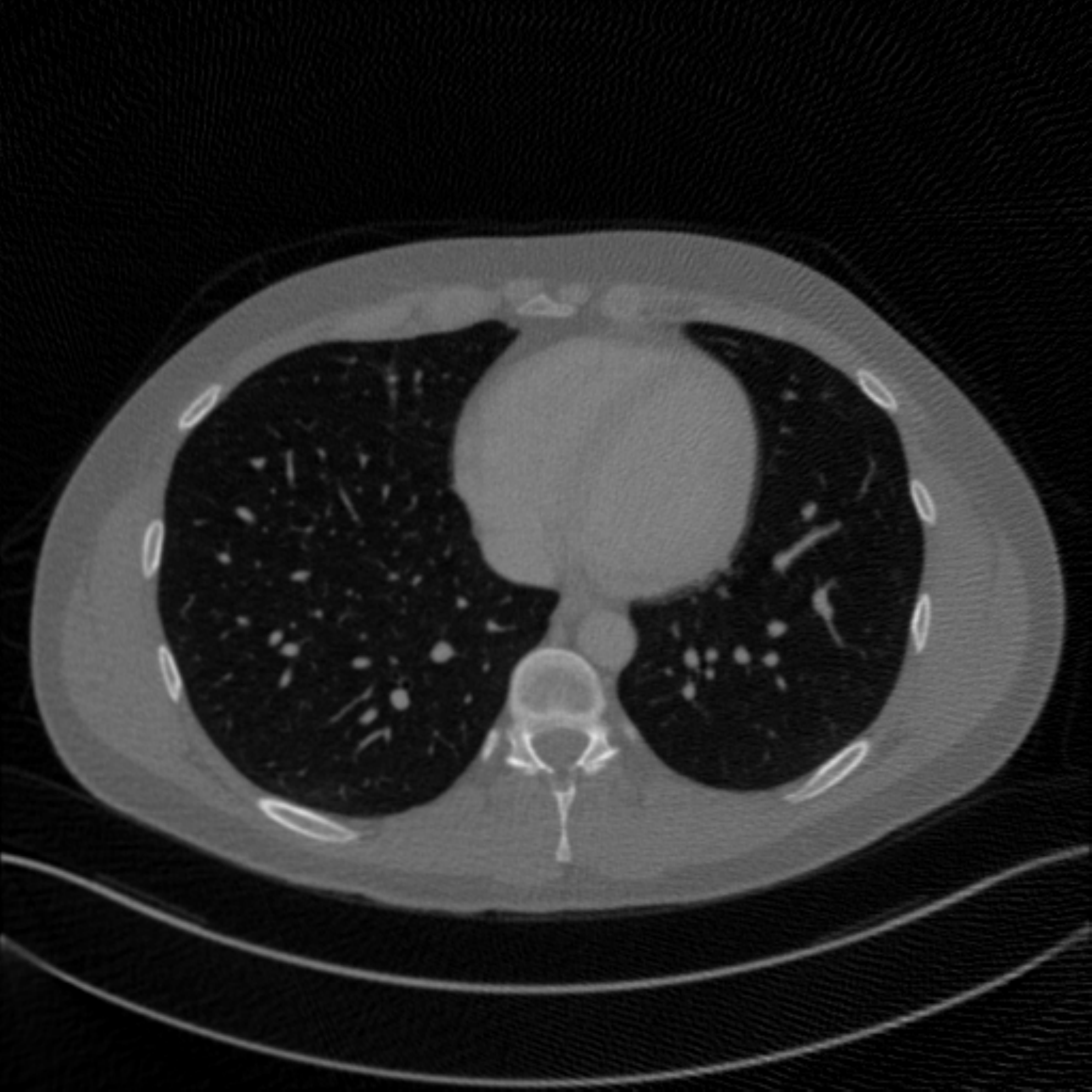}}		
 		\subfloat[var=10]{
 			
 			\includegraphics[width=0.49\columnwidth]{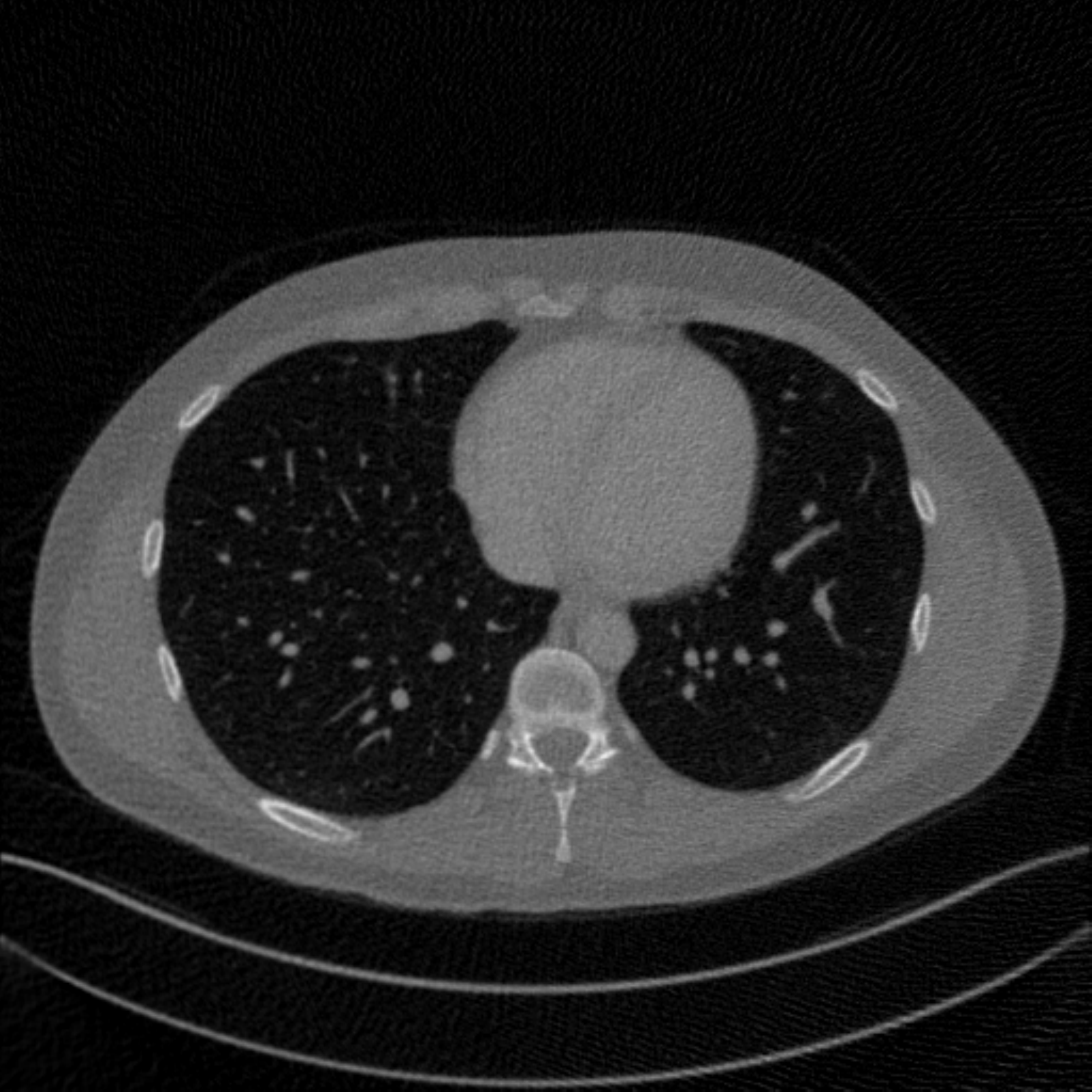}}
 		\subfloat[var=100]{
 			\includegraphics[width=0.49\columnwidth]{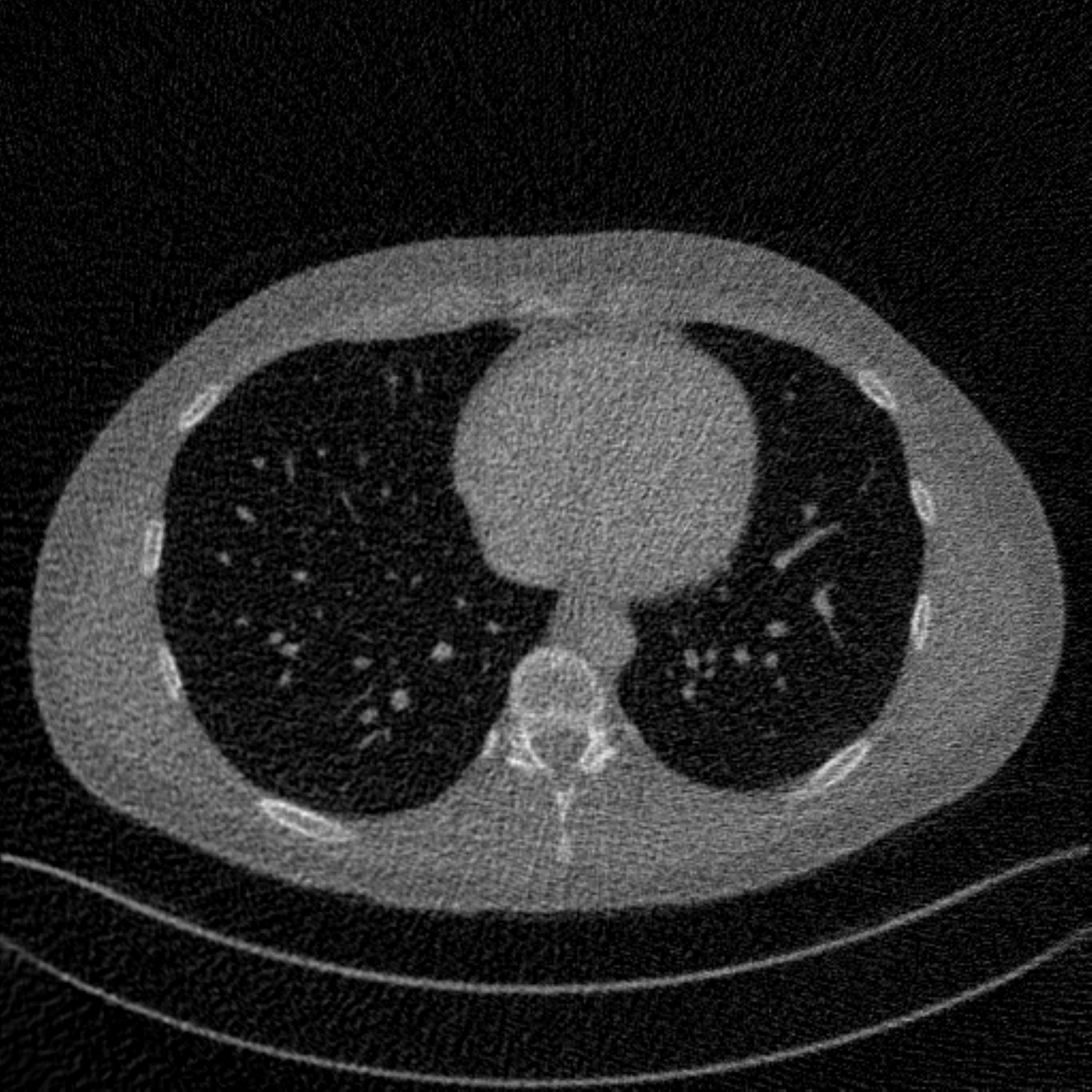}}
 		\subfloat[var=200]{
 			\includegraphics[width=0.49\columnwidth]{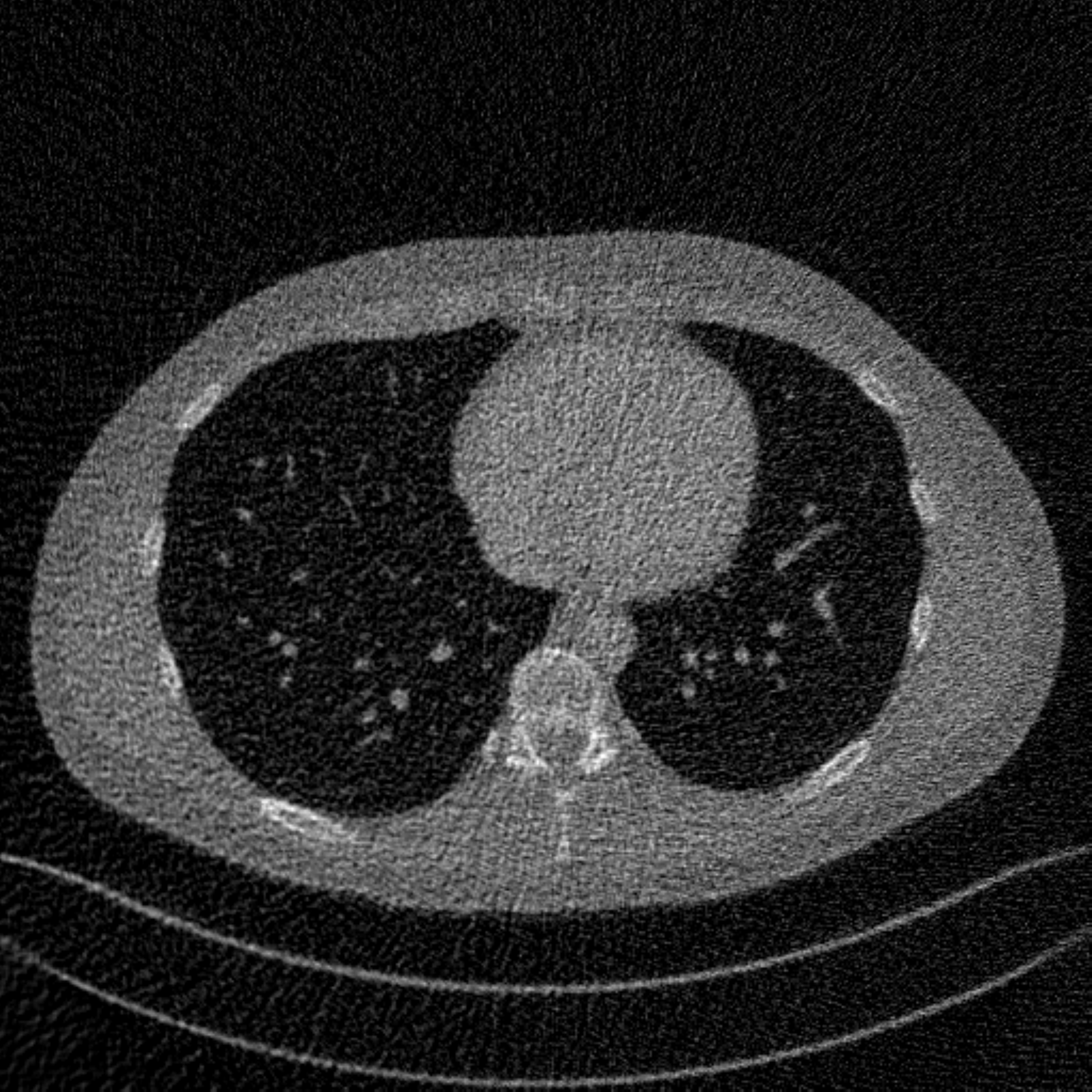}}
 	}
 	\caption{CT images reconstructed  from noisy short-scan fan-beam projection data with different variances.}
 	\label{figfan4}
 \end{figure*}
 
\section{Experimental Results}
In this section, we give some simulation results to verify the effectiveness of our algorithms. For 2D fan-beam CT reconstructions, we do the experiments on  the simulated projection data measured by  
equi-angular curved-line detectors while for  3D circle cone-beam, the simualted projection data measured by  
flat-plane detectors are used to reconstruct CT images. The
codes for implementing our methods can be downloaded from
https://github.com/wangwei-cmd/arc-based-CT-reconstruction. We  also provide the codes that implement the compared methods coded by ourselves in this paper.
\subsection{Fan-beam with curved-line detectors}
In this subsection, we present some   CT images reconstructed from the fan-beam projection data measured by  
equi-angular curved-line detectors under short-scan and super-short-scan   trajectories, and compare the results with those of the conventional fan-beam algorithm with Parker-extended weighting function (CFA) \cite{RN4} and Noo's algorithm (ACE) \cite{ISI:000177297700011}.

To test the performances of our method and the compared algorithms, we  randomly choose 500 full dose CT images (of size $512\times512$) from “the 2016 NIH-AAPM-Mayo Clinic Low Dose CT Grand Challenge”  \cite{data} as the  original images. The parameters for the fan-beam CT with the equi-angular curved-line detector are set as follows: $R_o=500$, $R_m=\sqrt 2\times256$ and so $$\gamma_m=\arctan(R_m/R_o)=35.90^{\circ}.$$ We set $\gamma=[-36:0.1:36]\times\pi/180$ for the sampling positions on   the $\gamma$ coordinate. 
For the sampling positions on the $\lambda$ coordinate, we set $$\lambda=[0:1:180+2\times36]\times\pi/180$$ for short-scan and $$\lambda=[0:1:180]\times\pi/180$$ for super-short scan. The hyper-parameter $d$ in ACE \cite{ISI:000177297700011} is set as $d=6\times\pi/180$.

In Fig. \ref{figfan1}, we show one set of  CT images reconstructed by 
CFA, ACE and ours from the short-scan fan-beam projection data. We can observe that the visual effects of the reconstructed images by ACE and ours are very similar while that by CFA has a lot of artifacts. To objectively estimate the qualities of the reconstructed images by the three methods, the average peak signal to noise ratio (PSNR) and structural similarity (SSIM) of the 500 reconstructed CT images are listed in Table \ref{T1}. We can see that our method has a slightly higher average PSNR and SSIM than those of ACE. The average PSNR and SSIM of CFA are far lower than those of ACE and ours. It's because that the CFA algorithm needs higher sampling rate on the $\lambda$ coordinate. By experiments, we find that when setting $\lambda=[0:0.25:180+2\times36]\times\pi/180$, the PSNR and SSIM of CFA can be raised to the same level of ACE and ours.
 
In Fig. \ref{plot}, we plot the 1D  line intensity profile passing through the red  line in Fig. \ref{figfan1_a}. We can observe that the 
  intensity lines of ACE and ours  are almost coincident and 
 resemble more
 closely to the one of the original compared to that of CFA.  
 
In Fig. \ref{figfan2}, a set of  CT images reconstructed by 
CFA, ACE and ours from the super-short-scan fan-beam projections are presented. We can observe that all the three images reconstructed by CFA, ACE and ours suffer from intensity  inhomogeneity  on the left part of the reconstructed images due to the data incompleteness. Moreover, the  stripe visual effect can be easily observed in Fig. \ref{figfan2_b}. In Fig. \ref{figfan2_c}, a vertical line can be observed. This may be caused by data missing at the endpoints of  the scanning arcs. Compared to CFA and ACE, our method has the best visual effect as can be seen from Fig. \ref{figfan2_d}. In Table \ref{T2}, the average PSNR and SSIM of the  CT images reconstructed by CFA, ACE and ours are, respectively, listed, from which we can see that our method has the highest average PSNR and SSIM.

To better demonstrate  the artifacts of ACE caused by data incompleteness, the CT images corresponding to  $\lambda=[0:1:210]\times\pi/180$,  $\lambda=[0:1:220]\times\pi/180$,  $\lambda=[0:1:230]\times\pi/180$ and  $\lambda=[0:1:240]\times\pi/180$ reconstructed by ACE and ours are shown in Fig. \ref{figfan3}. We can observe that when the range of $\lambda$ is lower than $240^{\circ}$, some stripe visual effects can be observed  in the CT images reconstructed by ACE. As the range of $\lambda$ exceeds $240^{\circ}$, the artifacts of ACE caused by the data missing  almost disappear. For our method, even $\lambda=[0:1:210]\times\pi/180$, there exists no stripe artifact in the reconstructed CT images.
 
\begin{figure*}[htbp]
	\centering{
		\subfloat{\includegraphics[width=0.49\columnwidth]{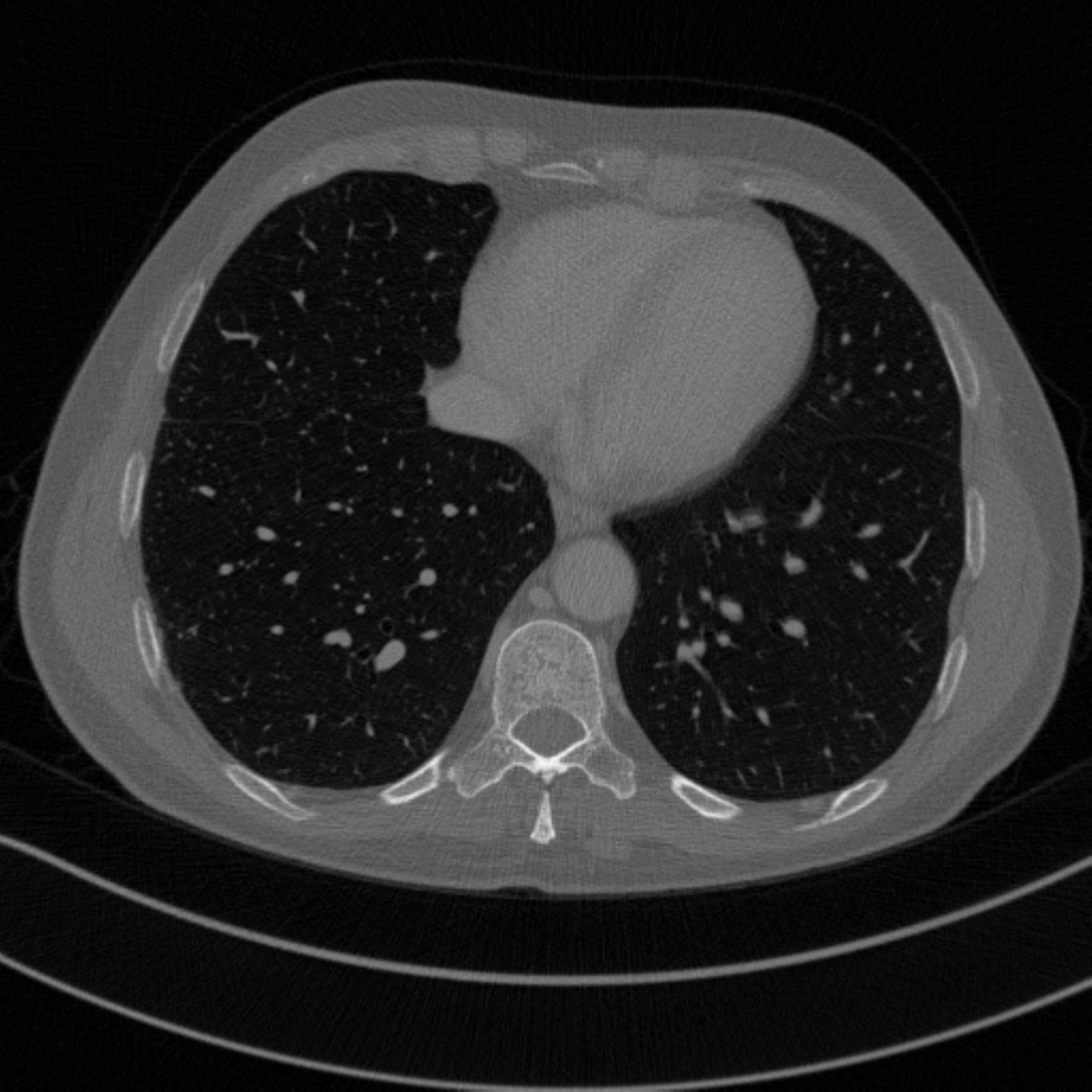}}
		\subfloat{\includegraphics[width=0.49\columnwidth]{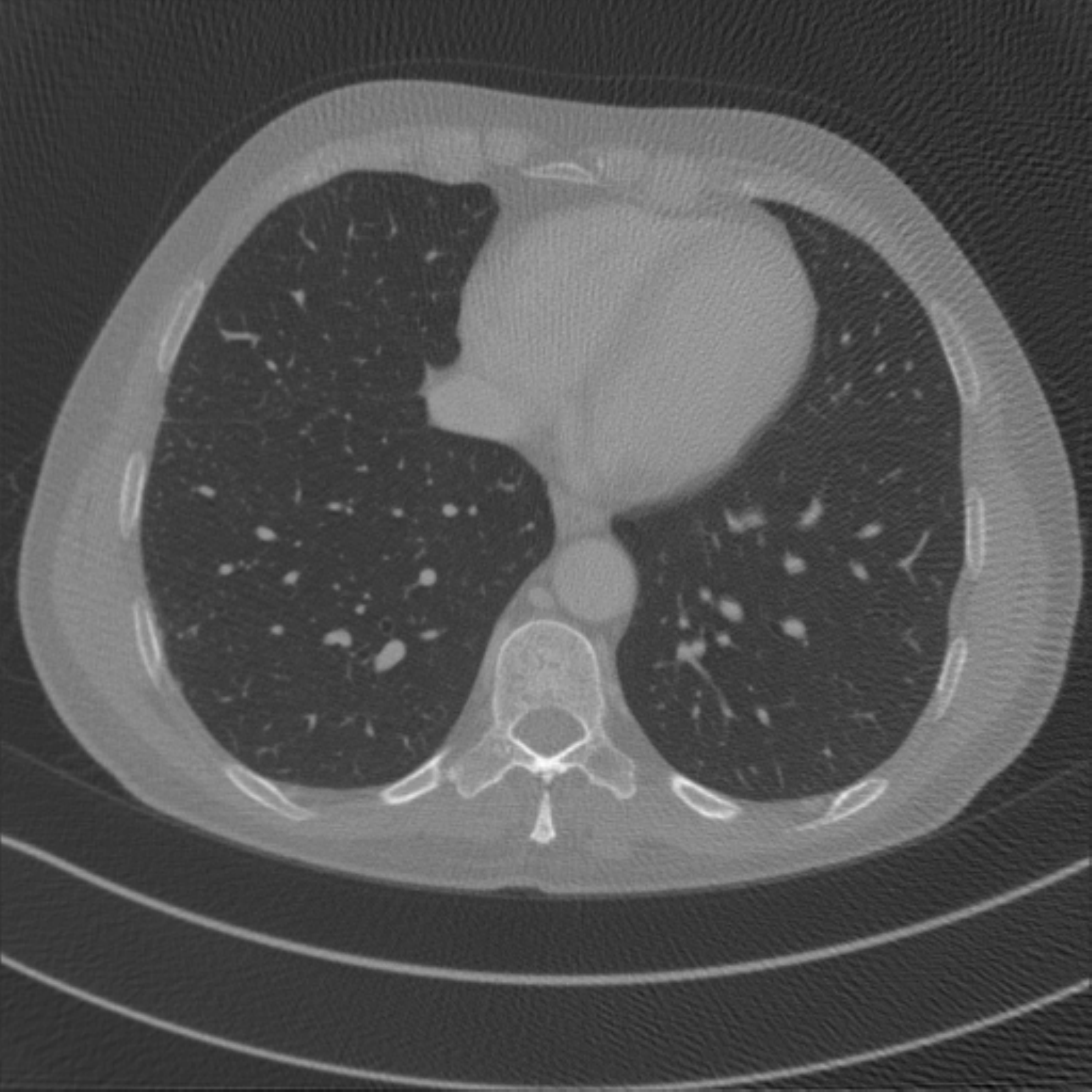}}
		\subfloat{\includegraphics[width=0.49\columnwidth]{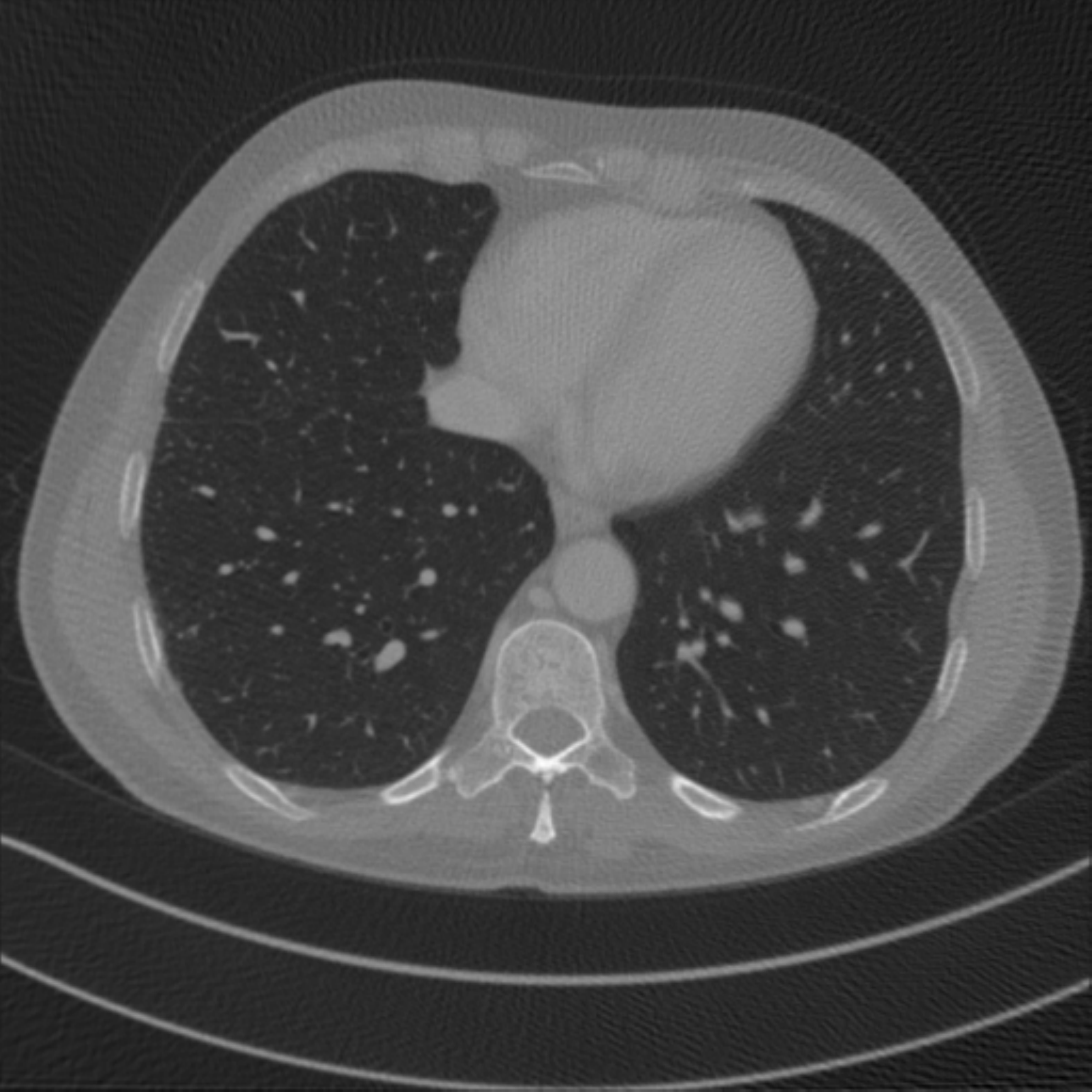}}
		\subfloat{\includegraphics[width=0.49\columnwidth]{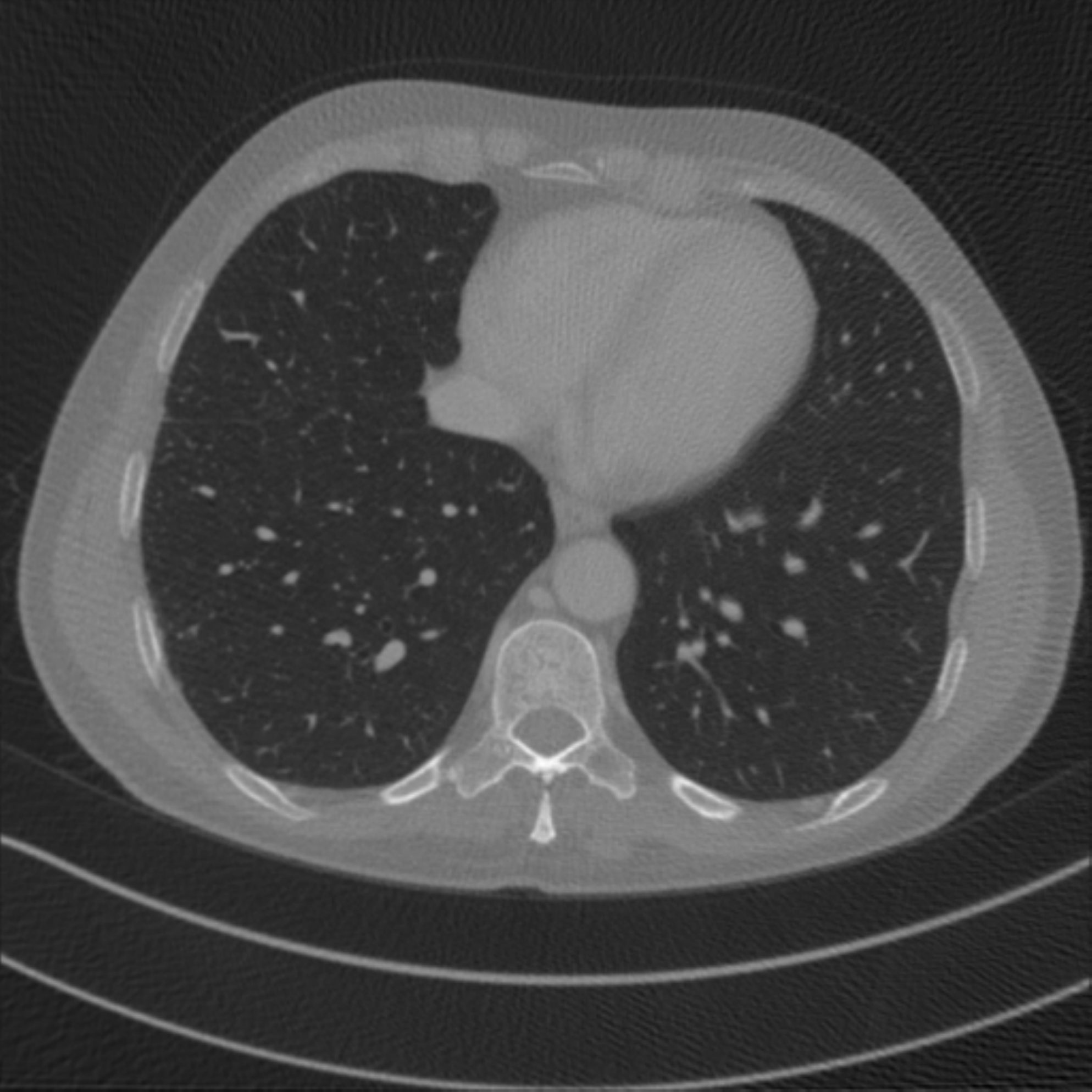}}
		
		\vspace{-3mm}
		\subfloat{\includegraphics[width=0.49\columnwidth]{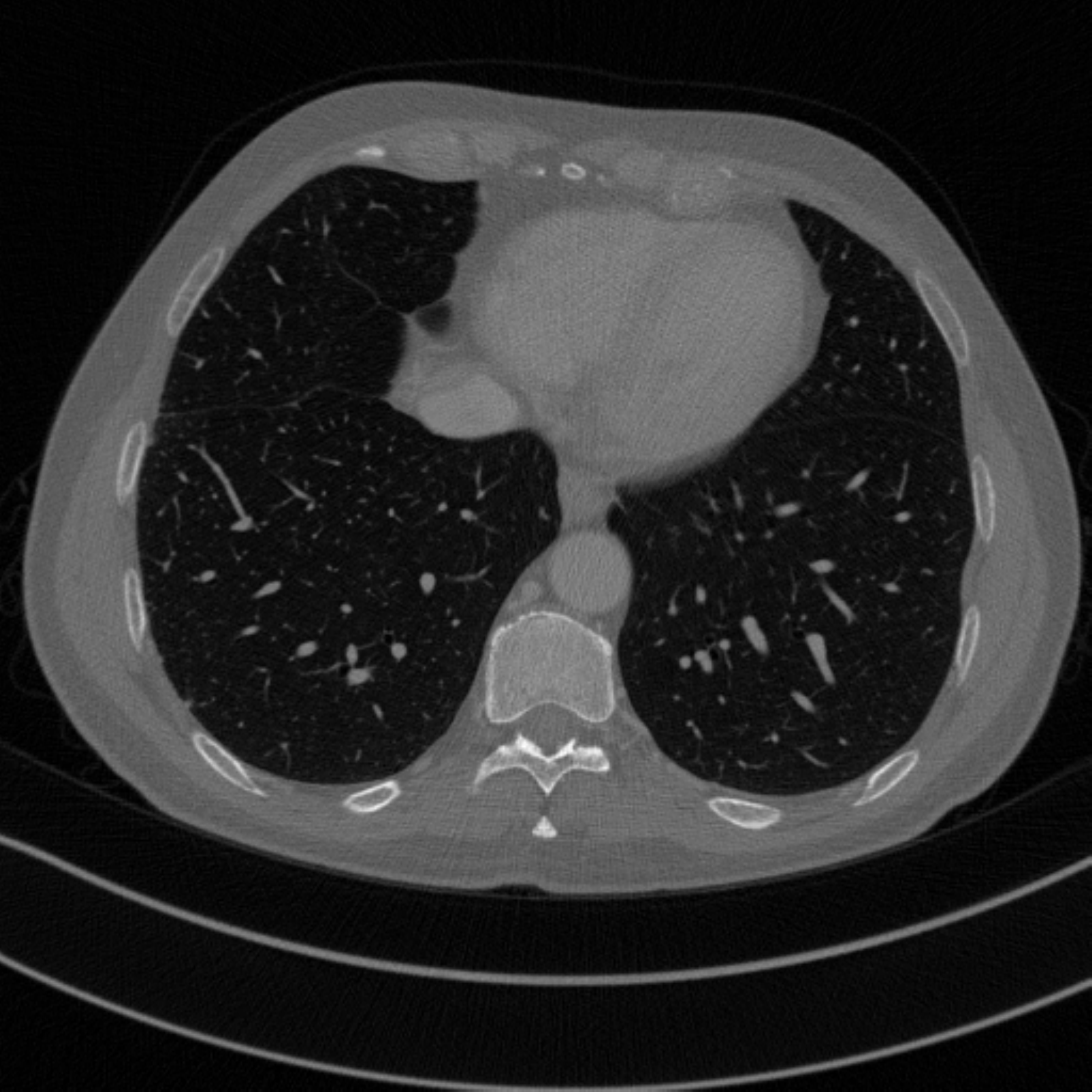}}
		\subfloat{\includegraphics[width=0.49\columnwidth]{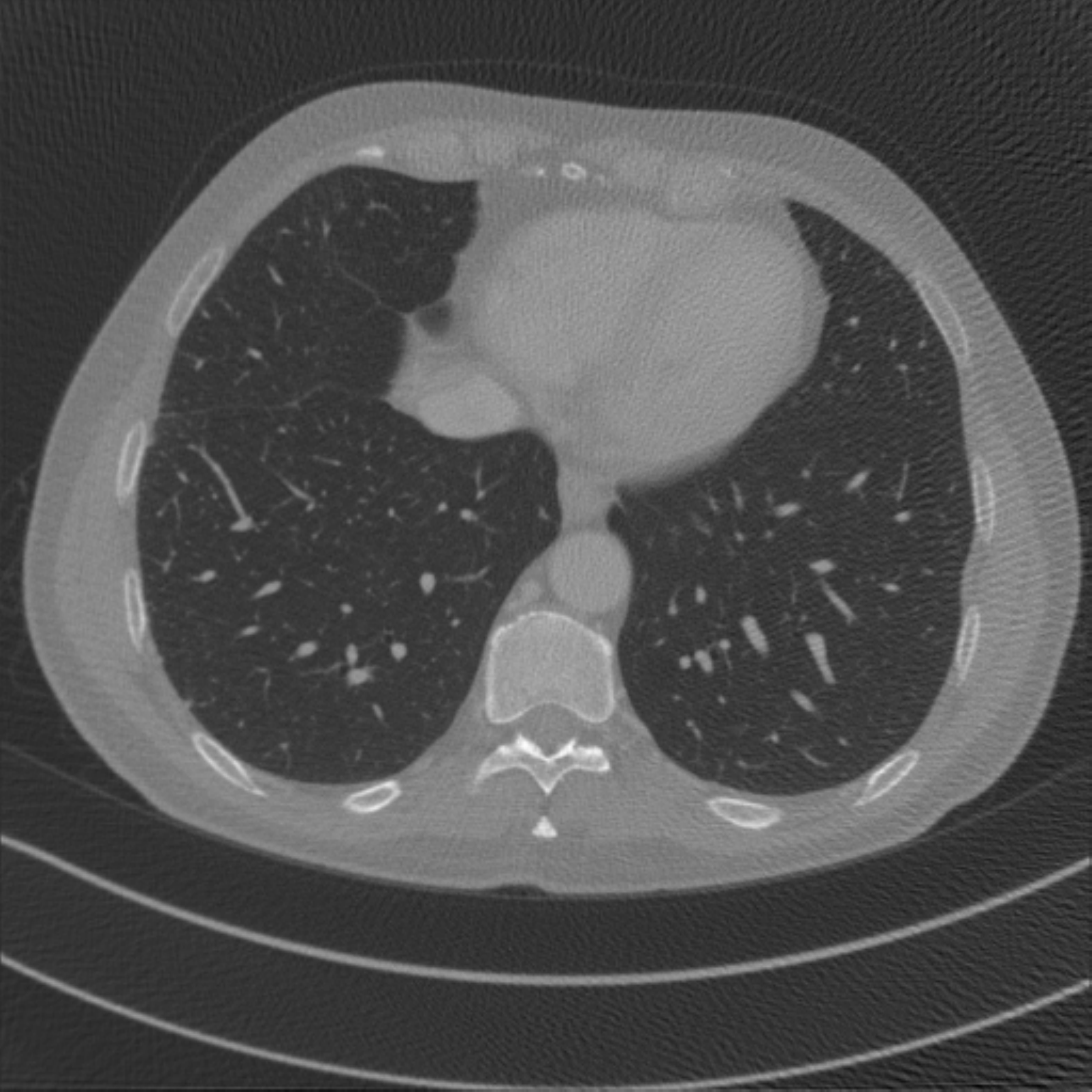}}
		\subfloat{\includegraphics[width=0.49\columnwidth]{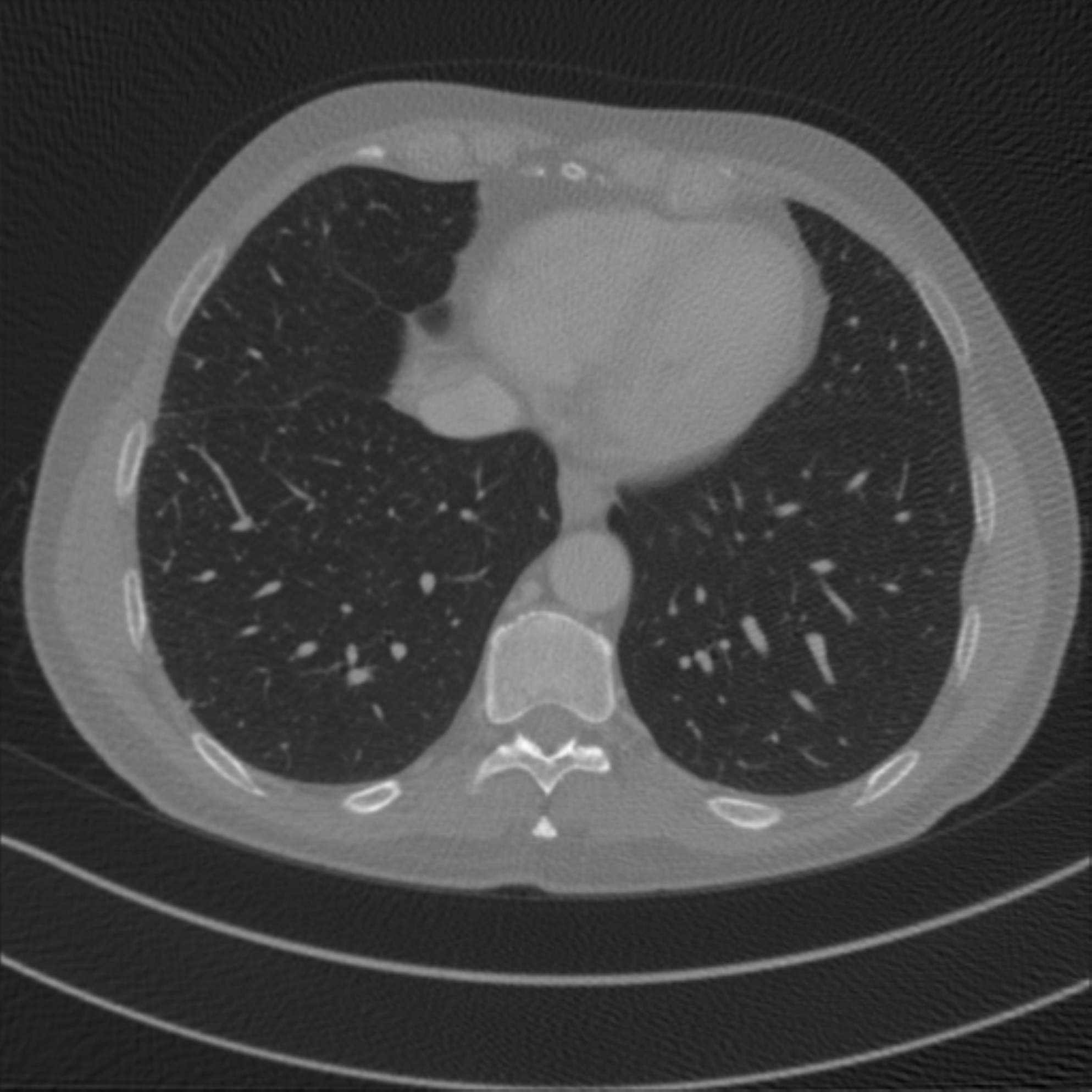}}
		\subfloat{\includegraphics[width=0.49\columnwidth]{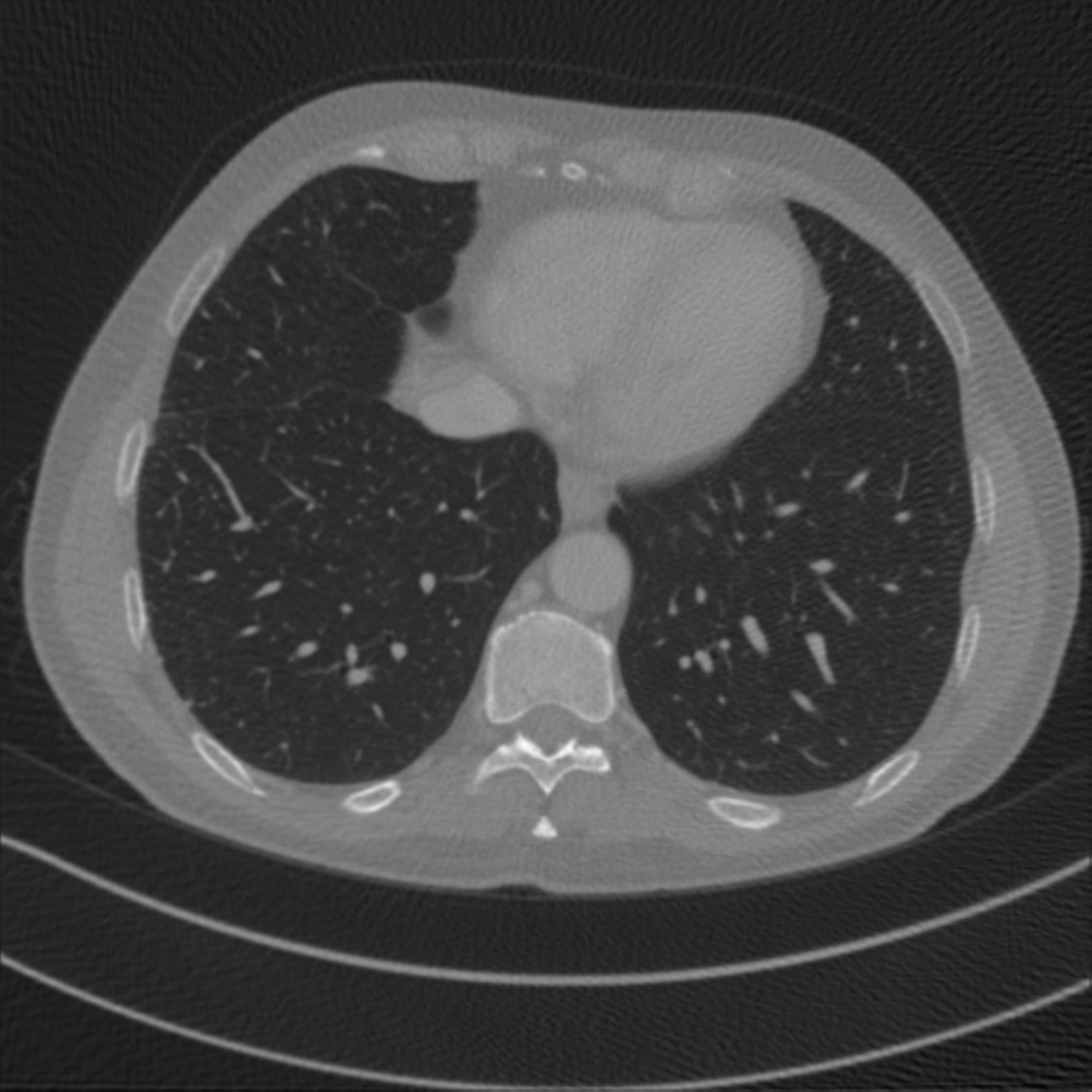}}
		
		\vspace{-3mm}
		\subfloat{\includegraphics[width=0.49\columnwidth]{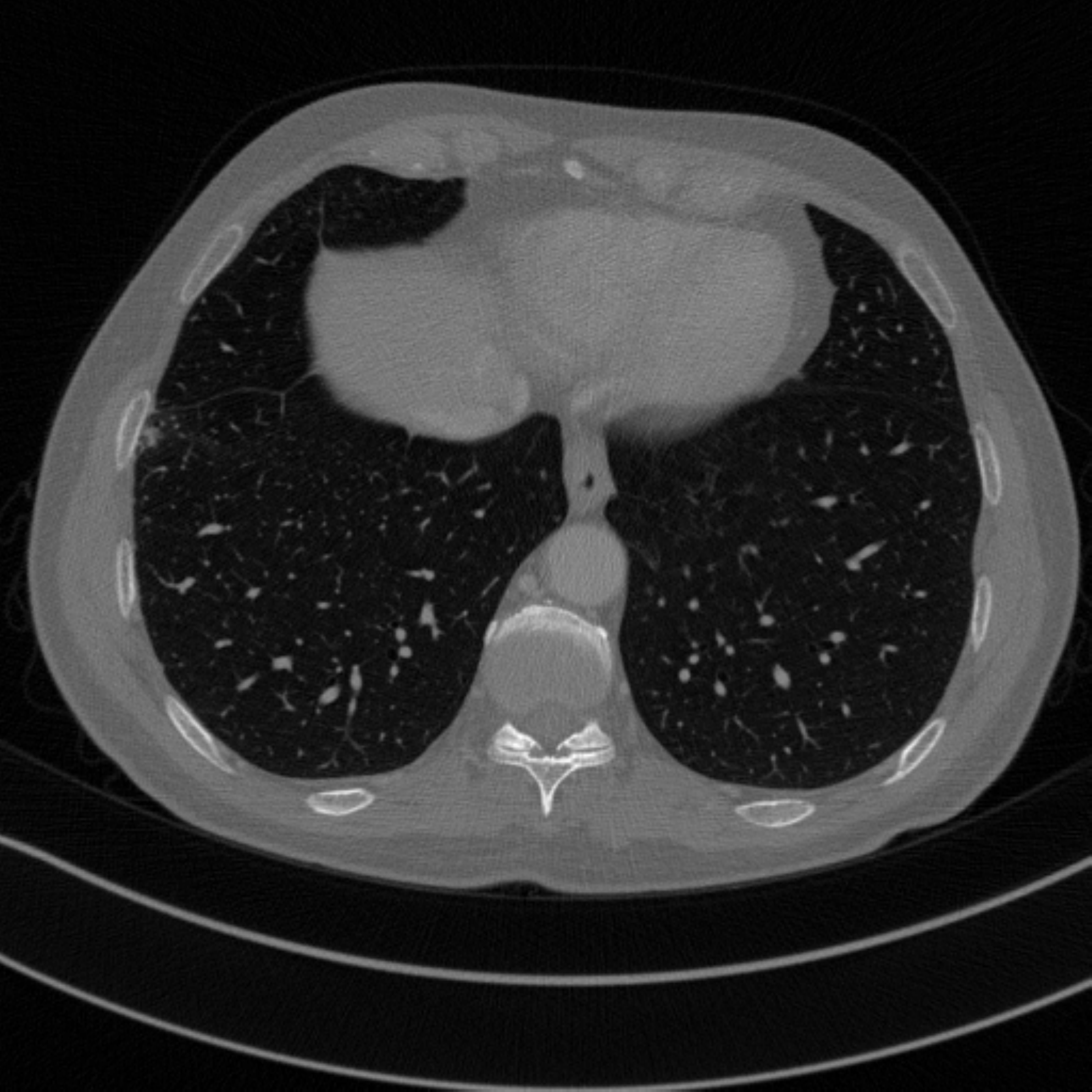}}
		\subfloat{\includegraphics[width=0.49\columnwidth]{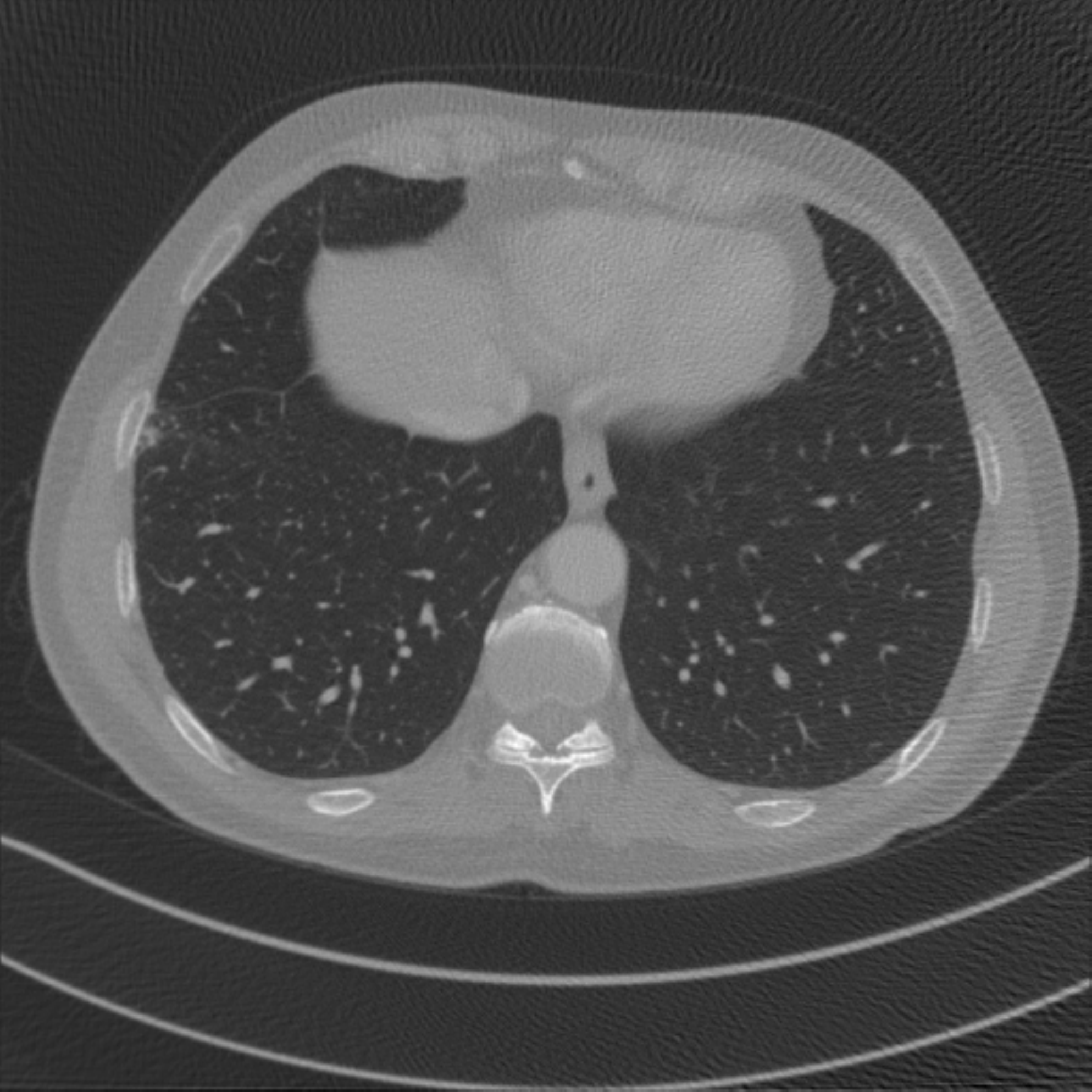}}
		\subfloat{\includegraphics[width=0.49\columnwidth]{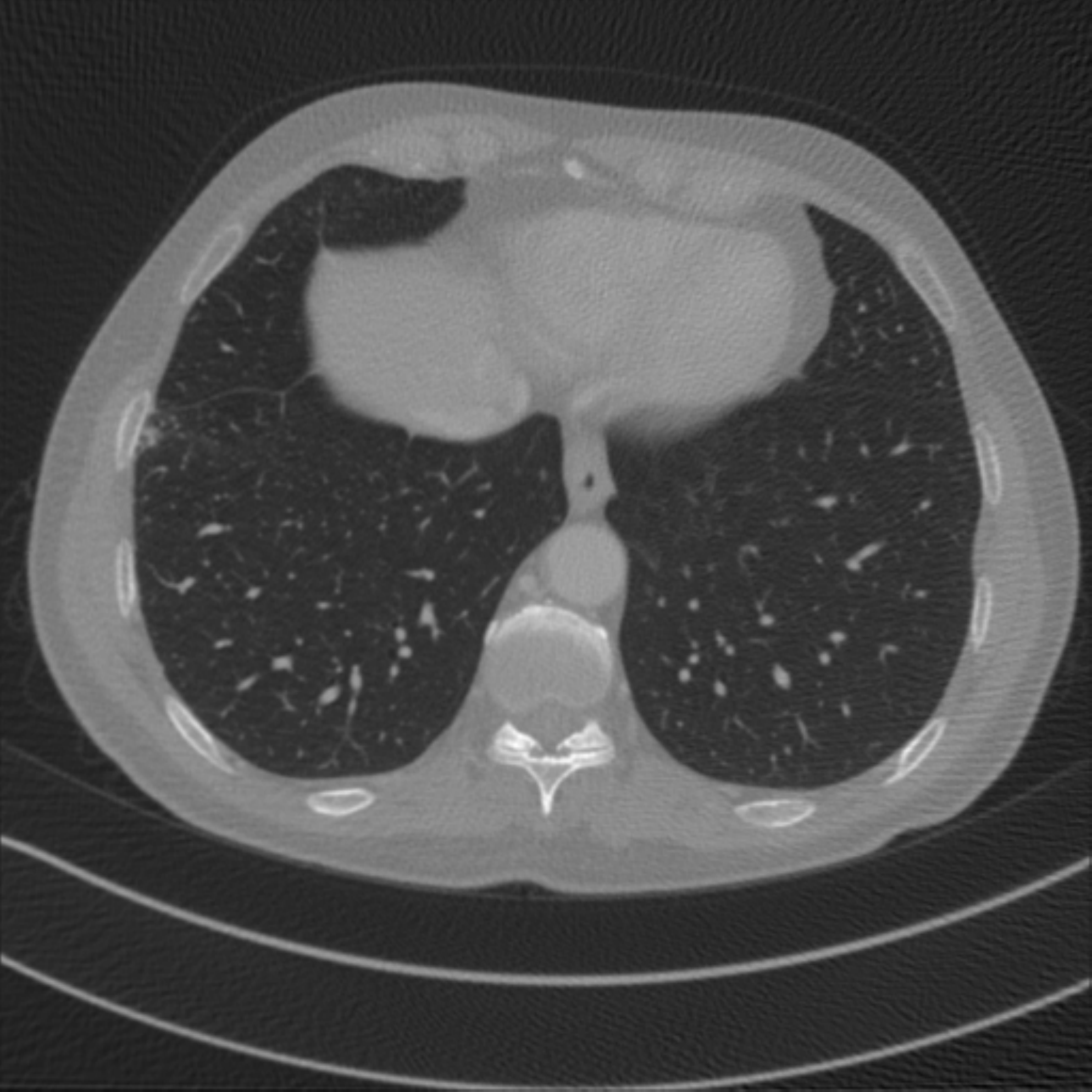}}
		\subfloat{\includegraphics[width=0.49\columnwidth]{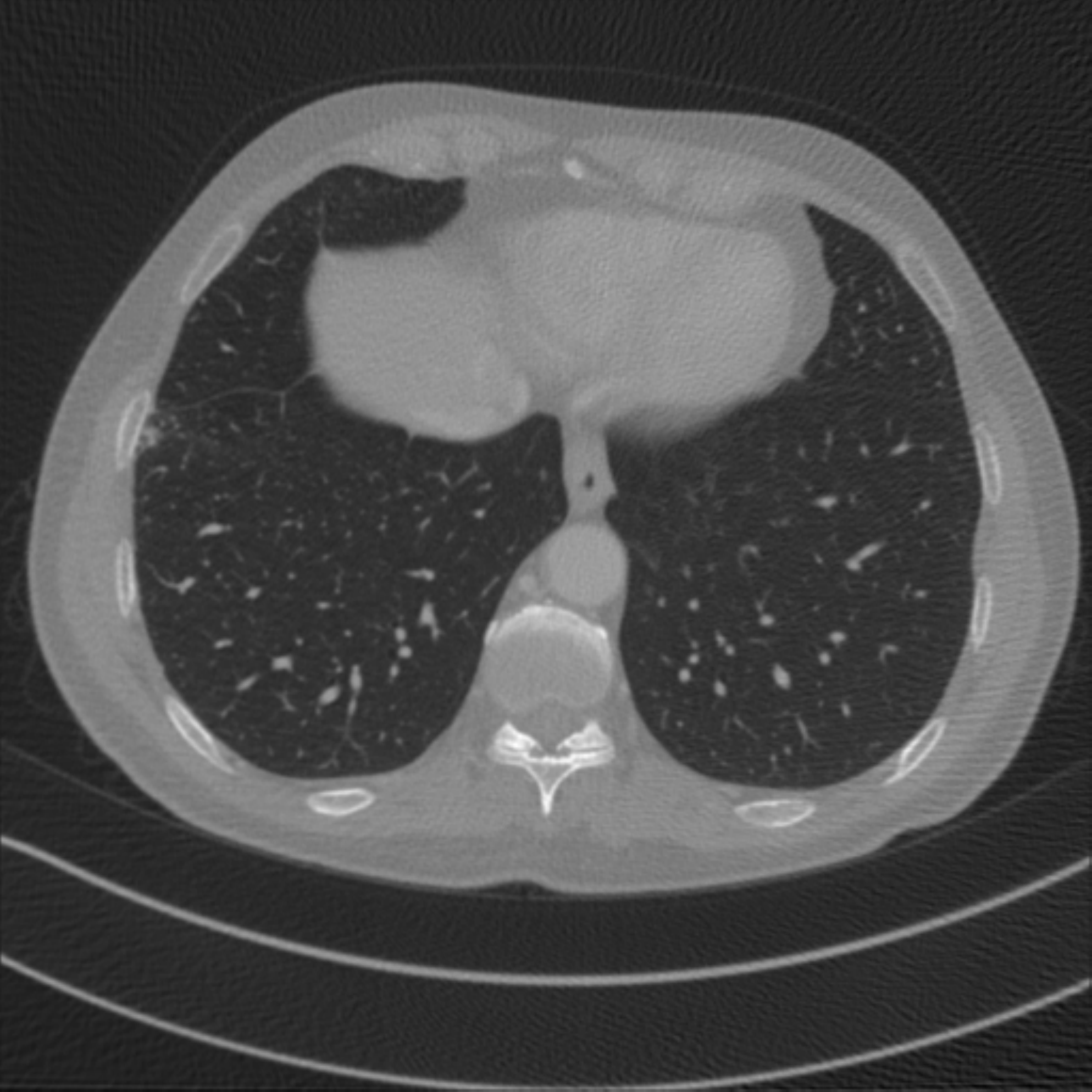}}
		
		\setcounter{subfigure}{0}
		\vspace{-3mm}
		\subfloat[Original]{\includegraphics[width=0.49\columnwidth]{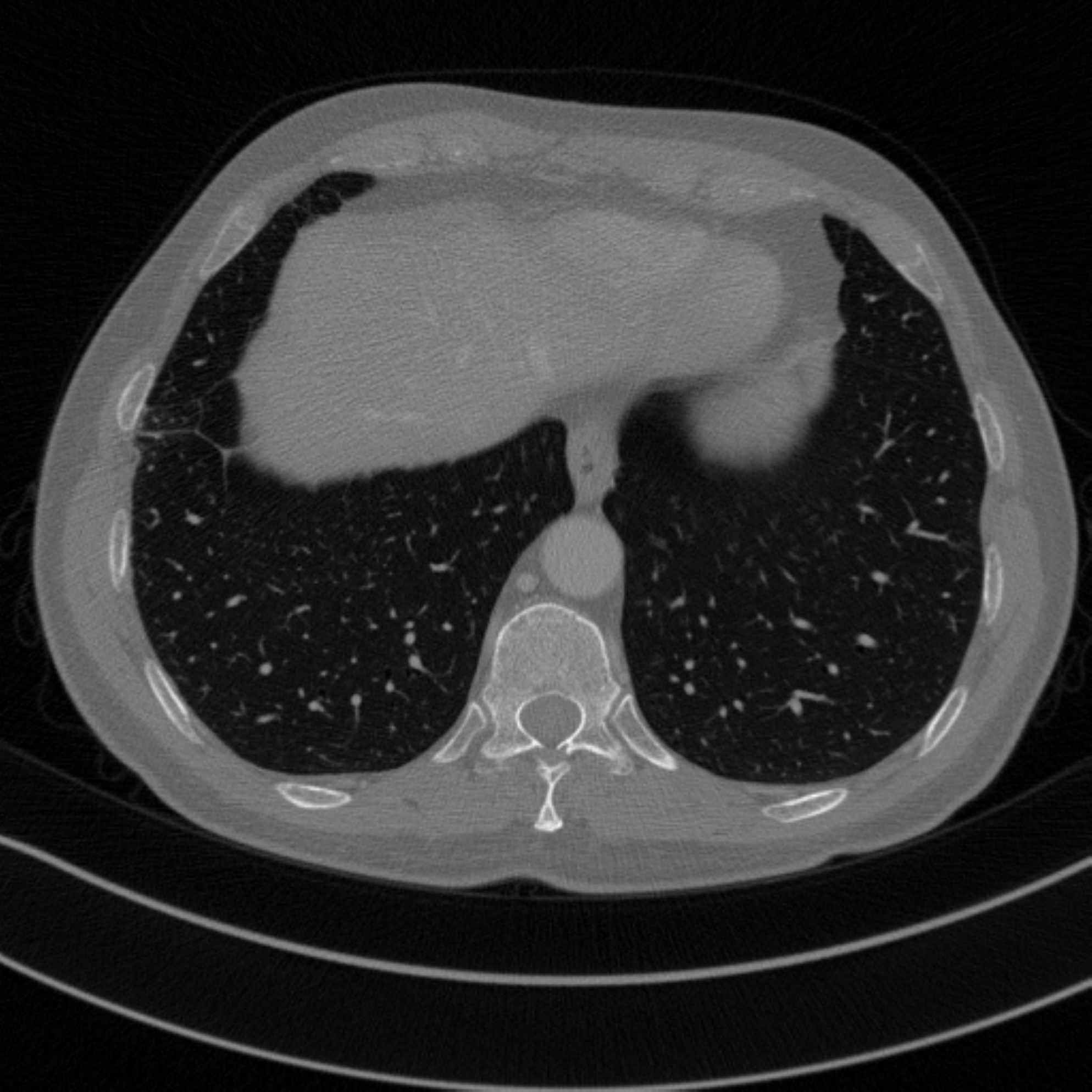}}
		\subfloat[FDK]{\includegraphics[width=0.49\columnwidth]{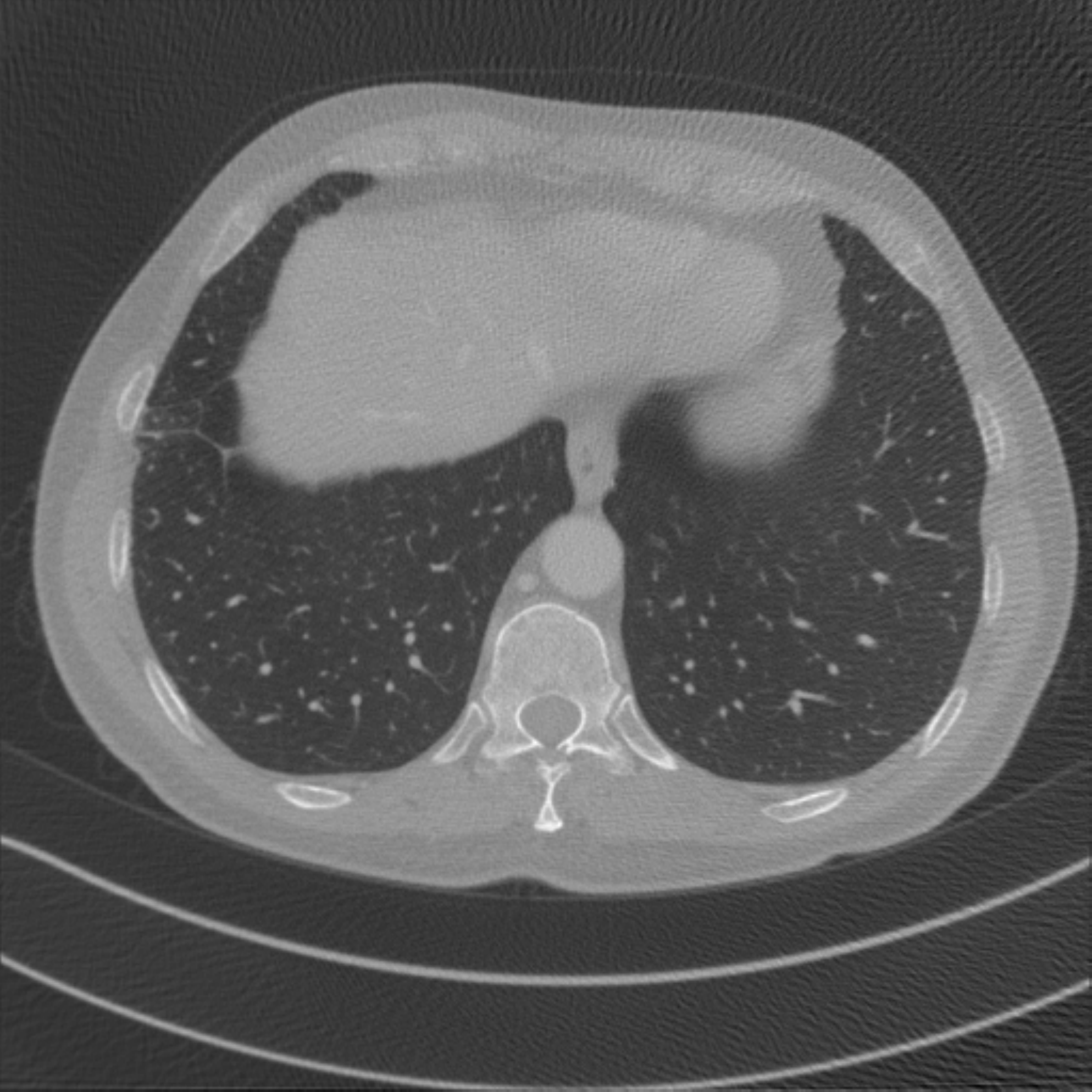}}
		\subfloat[ACE]{\includegraphics[width=0.49\columnwidth]{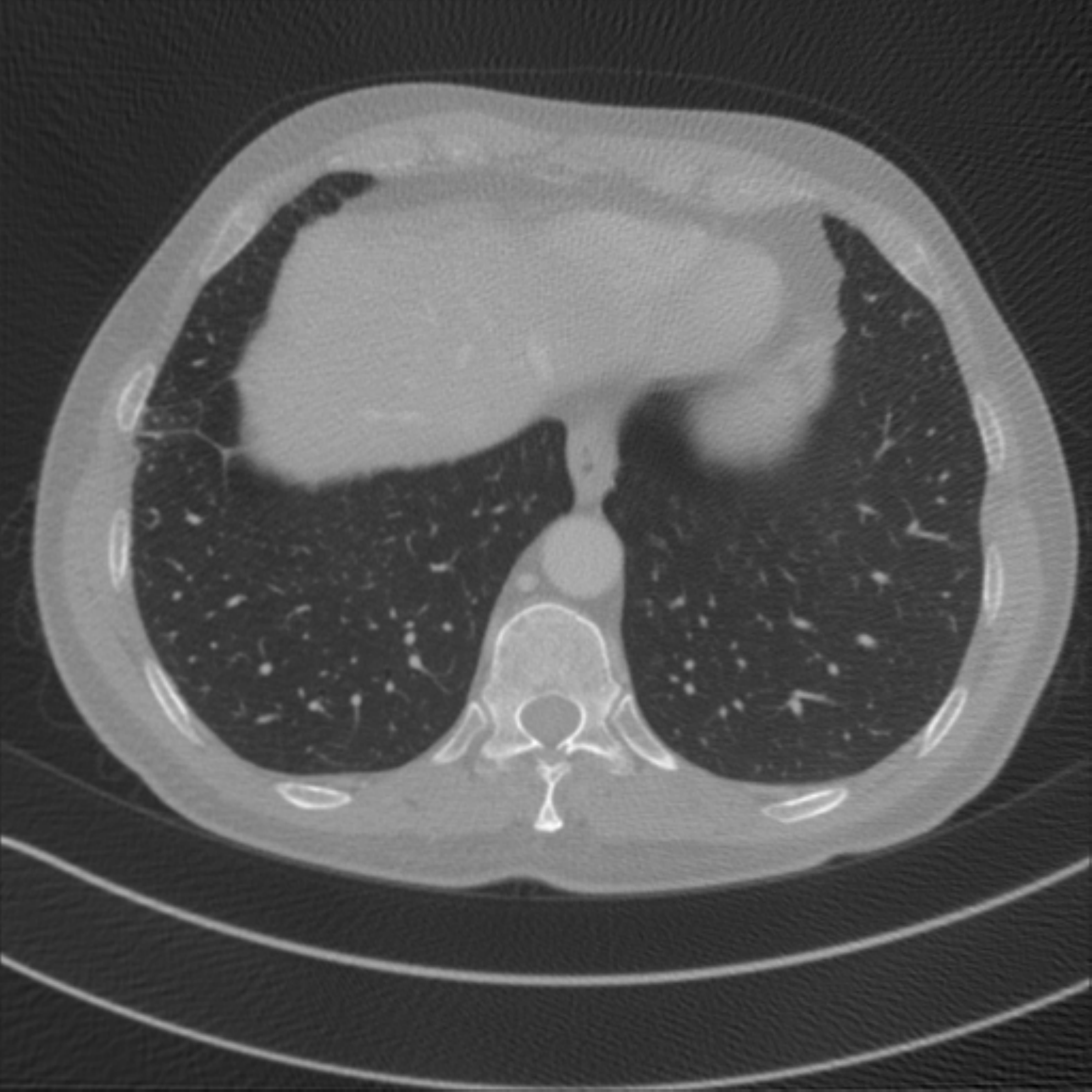}}
		\subfloat[Ours]{\includegraphics[width=0.49\columnwidth]{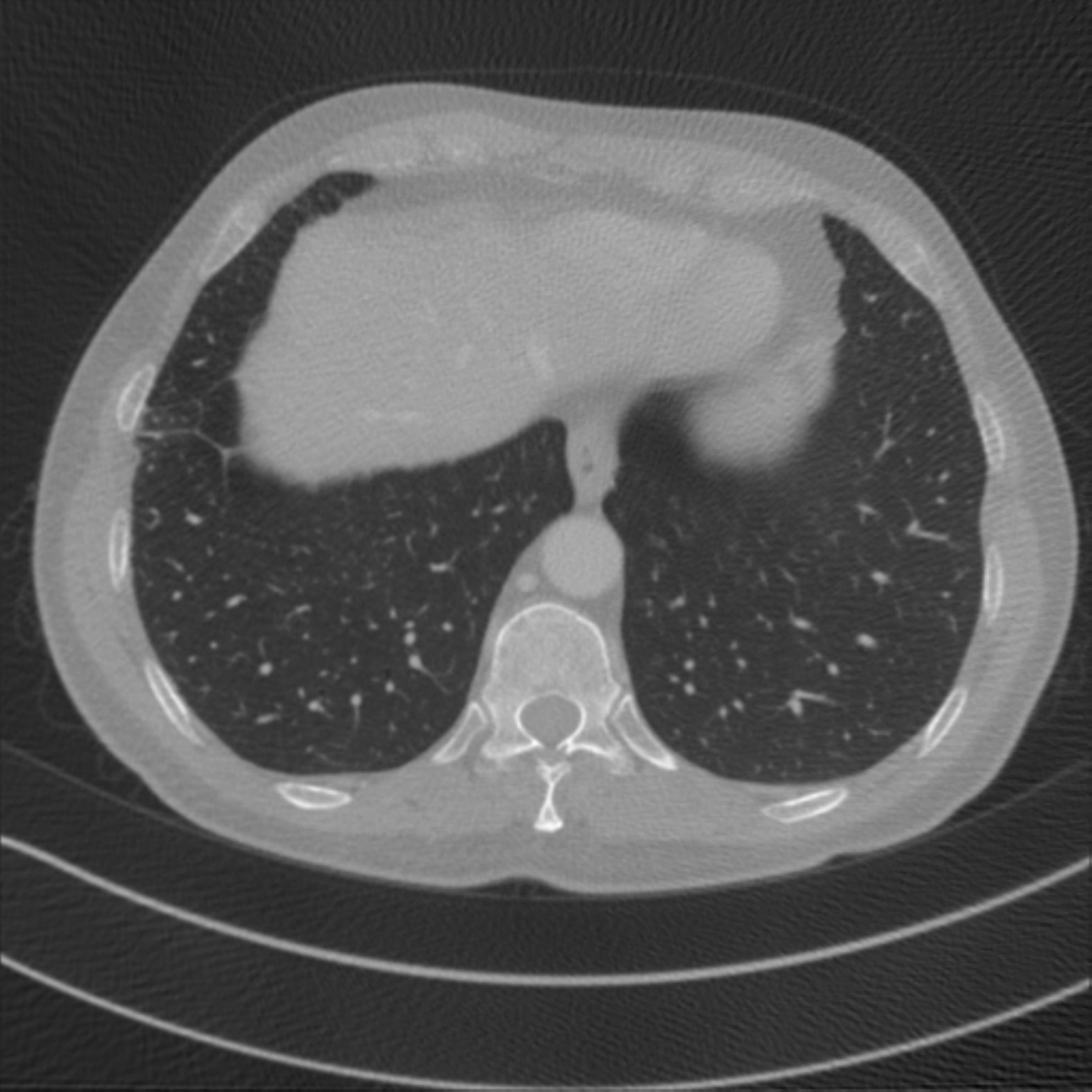}}
	}
	\caption{Sliced CT images reconstructed from short-scan circle cone-beam projection data.}
	\label{figcone1}
\end{figure*}

To test the performances of our method when the projection data is corrupted by noise, we add  Poisson and Gaussian noise   to the short-scan projections via the following formulas \cite{ISI:000311351400019}:
\begin{equation}\label{addnoise}
\begin{aligned}
g&=\exp(-g/M),\\
g&=g+I_0*poission(g)+I_0*Gaussin(m,var/I_0),\\
g&=\log(I_0/g)*M,
\end{aligned}
\end{equation}
where $M$ is the maximal value of the projection data $g$, $m$ and $var$ are the mean and variance of the Gaussian noise, respectively, $I_0$ is the average photon count for the Poisson noise. In this experiment, we set $I_0=10^6$, $m = 0$, and $var=1,10,100,200$, respectively. The reconstructed images from the noisy  projection data are shown in Fig. \ref{figfan4}. We can observe that as the variance $var$ increases, the reconstructed CT image has more noise.
\begin{figure*}[htbp]
	\centering{
		\subfloat{\includegraphics[width=0.49\columnwidth]{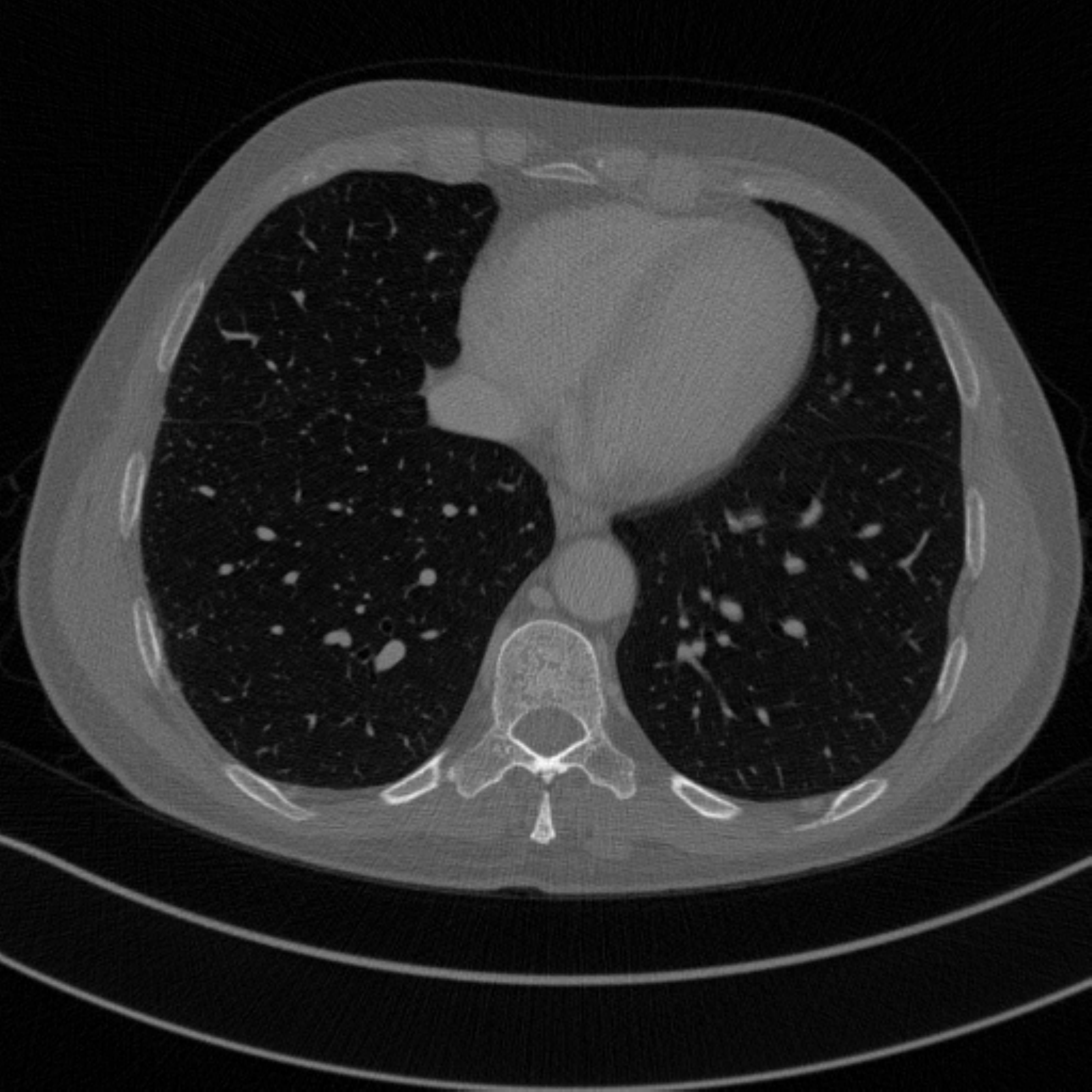}}
		\subfloat{\includegraphics[width=0.49\columnwidth]{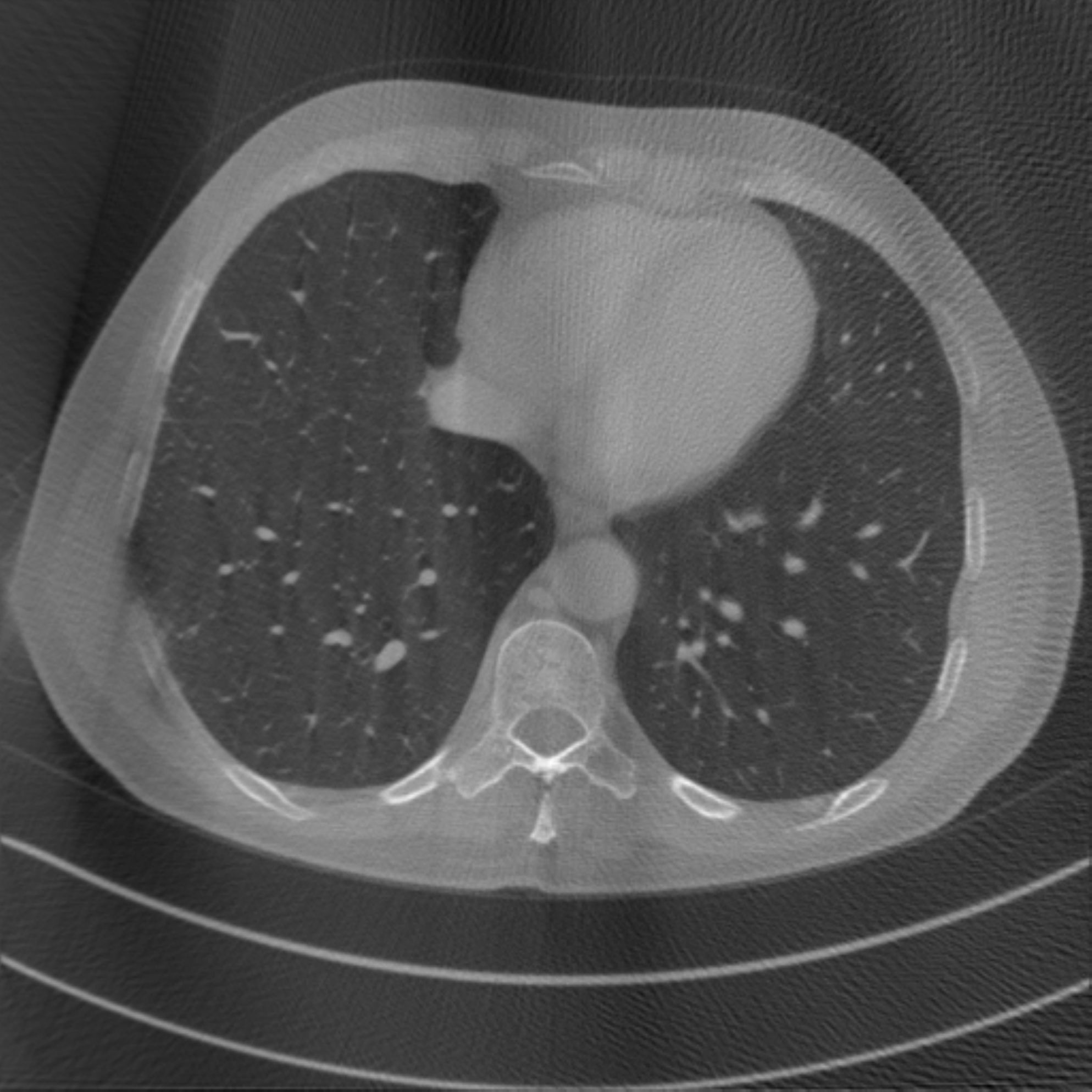}}
		\subfloat{\includegraphics[width=0.49\columnwidth]{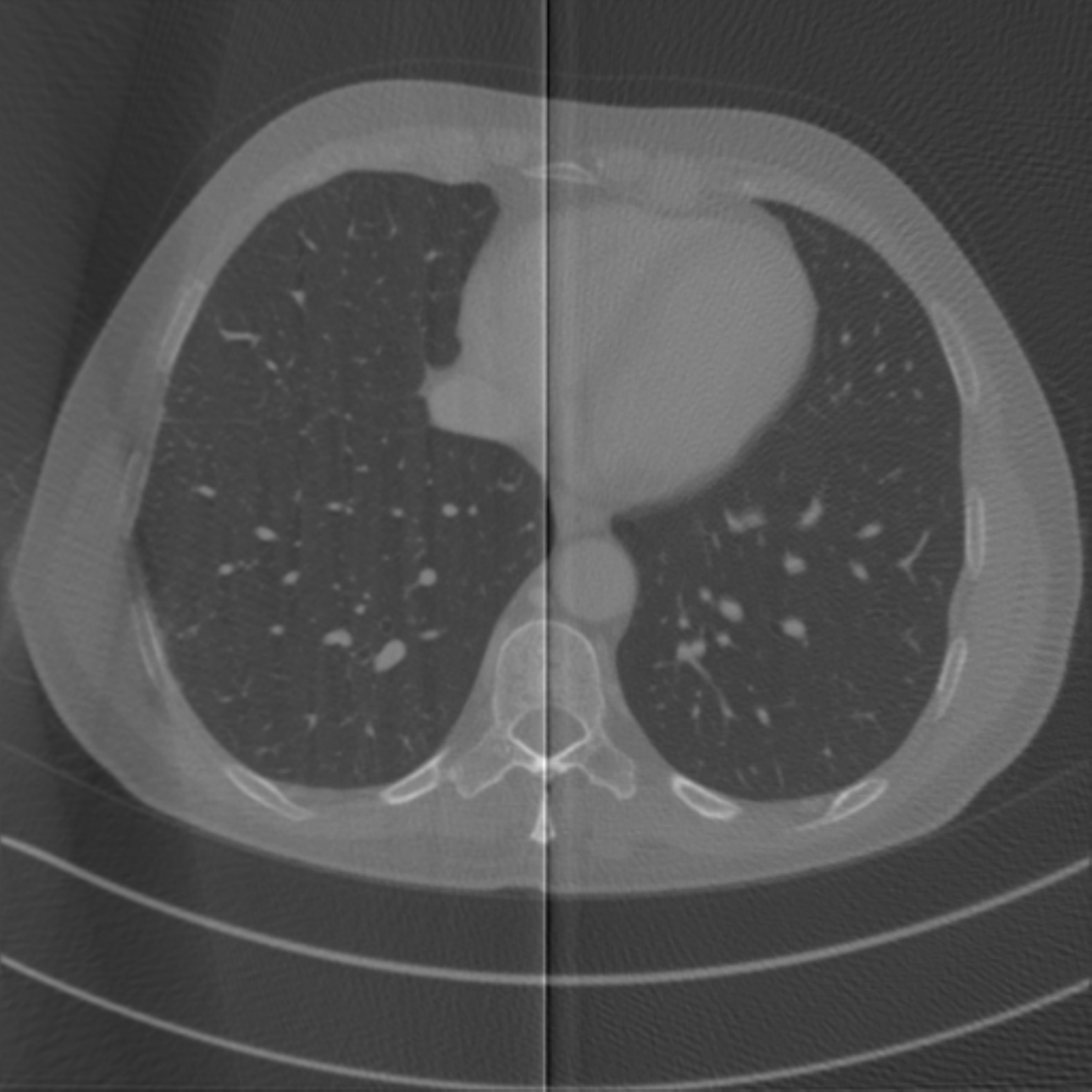}}
		\subfloat{\includegraphics[width=0.49\columnwidth]{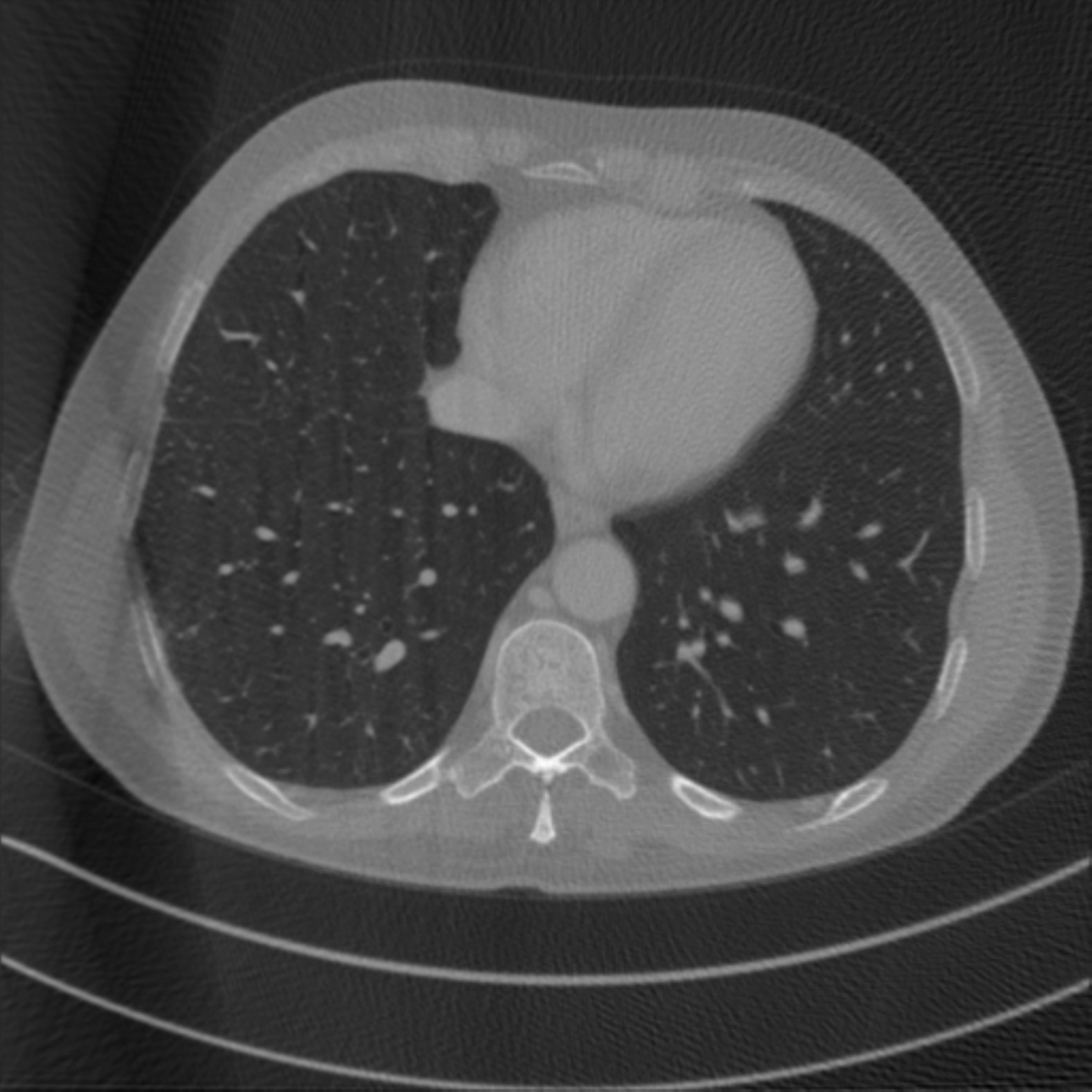}}
		
		\vspace{-3mm}
		\subfloat{\includegraphics[width=0.49\columnwidth]{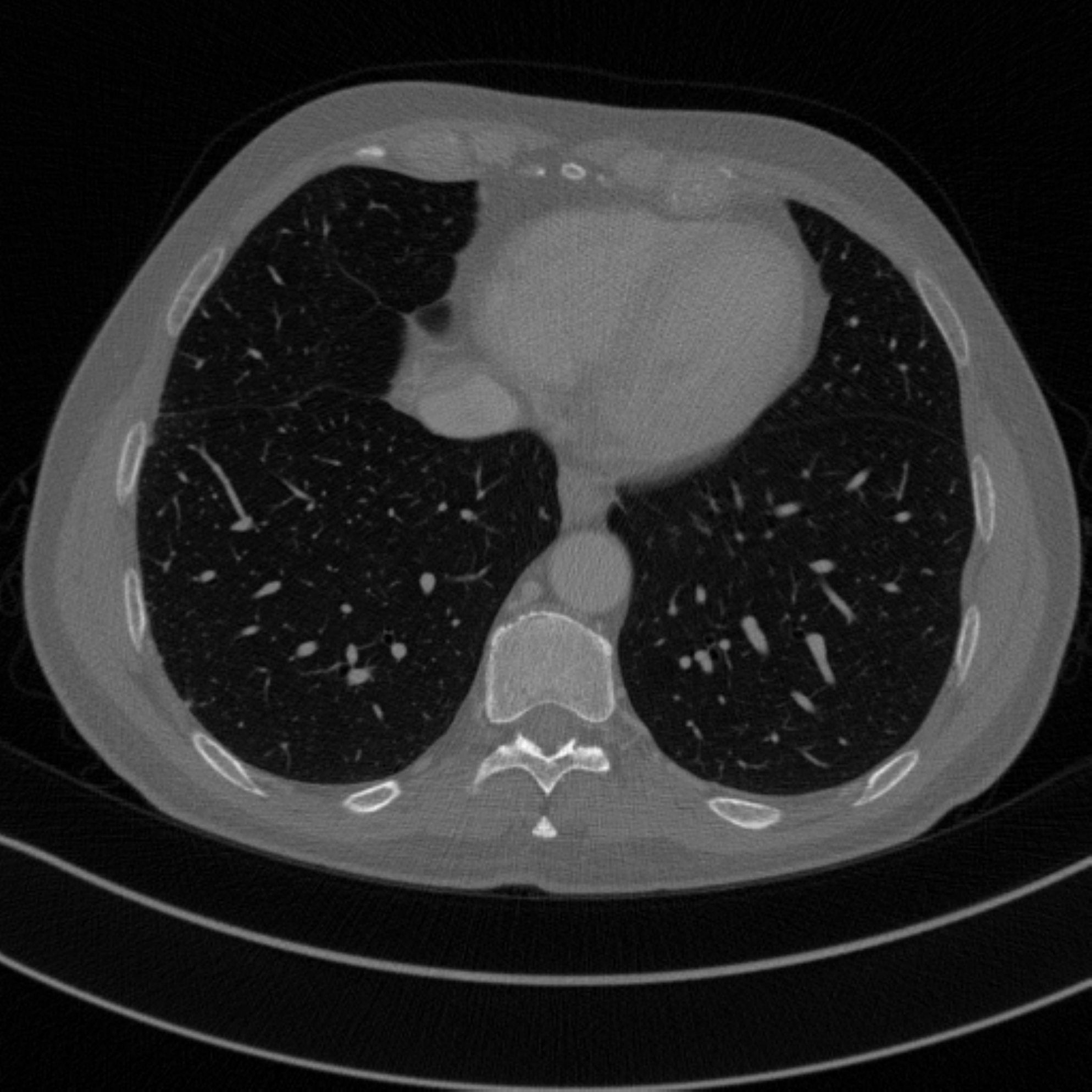}}
		\subfloat{\includegraphics[width=0.49\columnwidth]{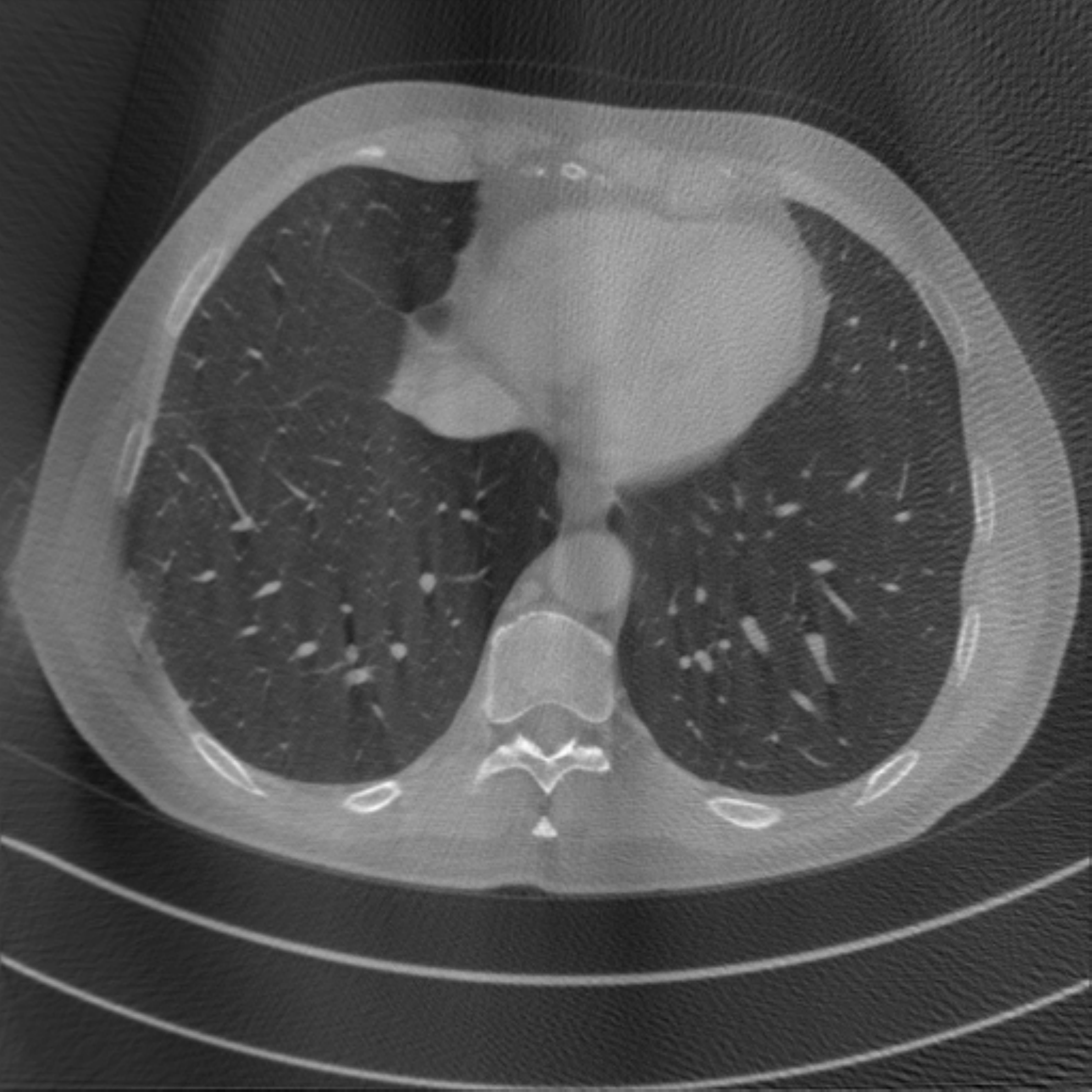}}
		\subfloat{\includegraphics[width=0.49\columnwidth]{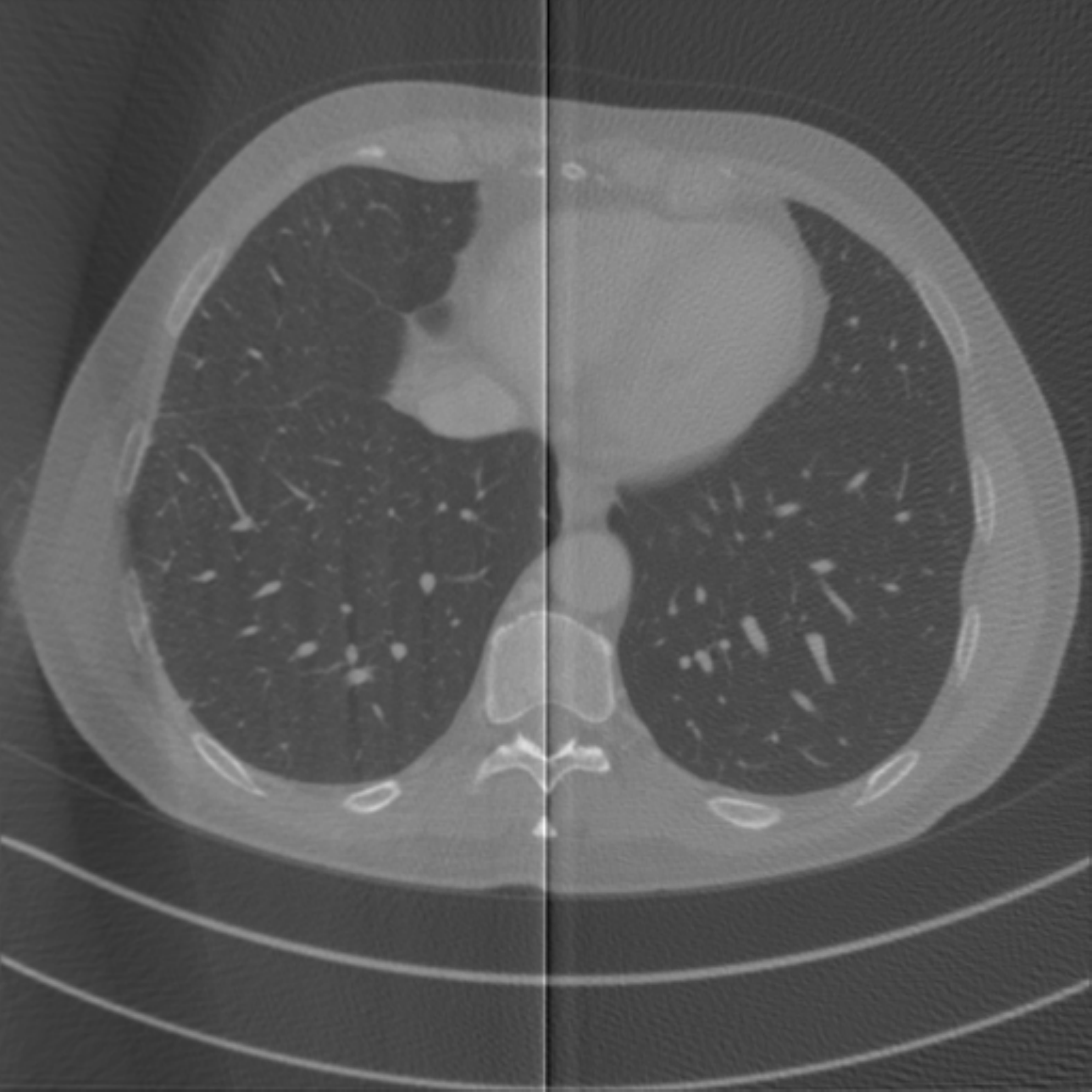}}
		\subfloat{\includegraphics[width=0.49\columnwidth]{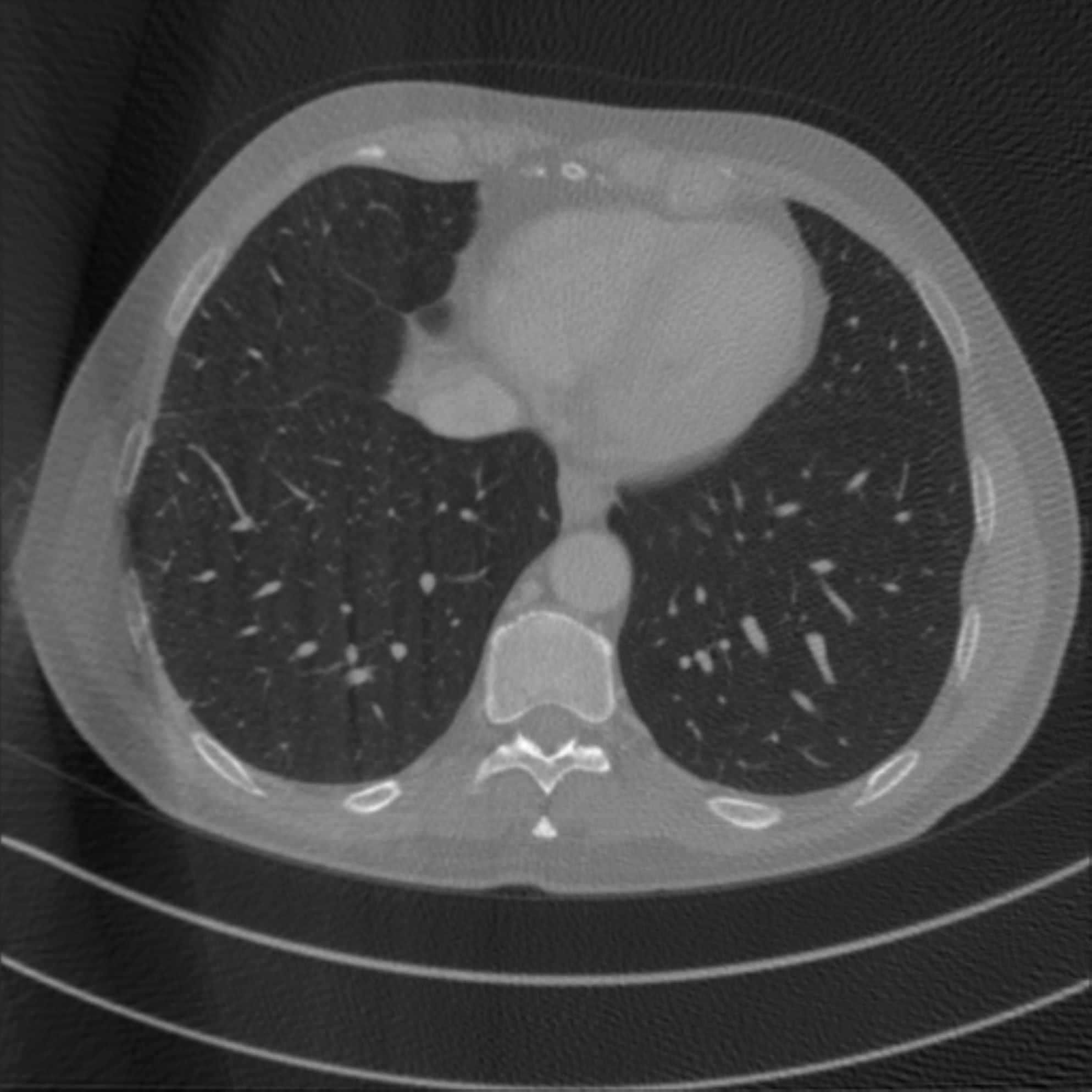}}
		
		\vspace{-3mm}
		\subfloat{\includegraphics[width=0.49\columnwidth]{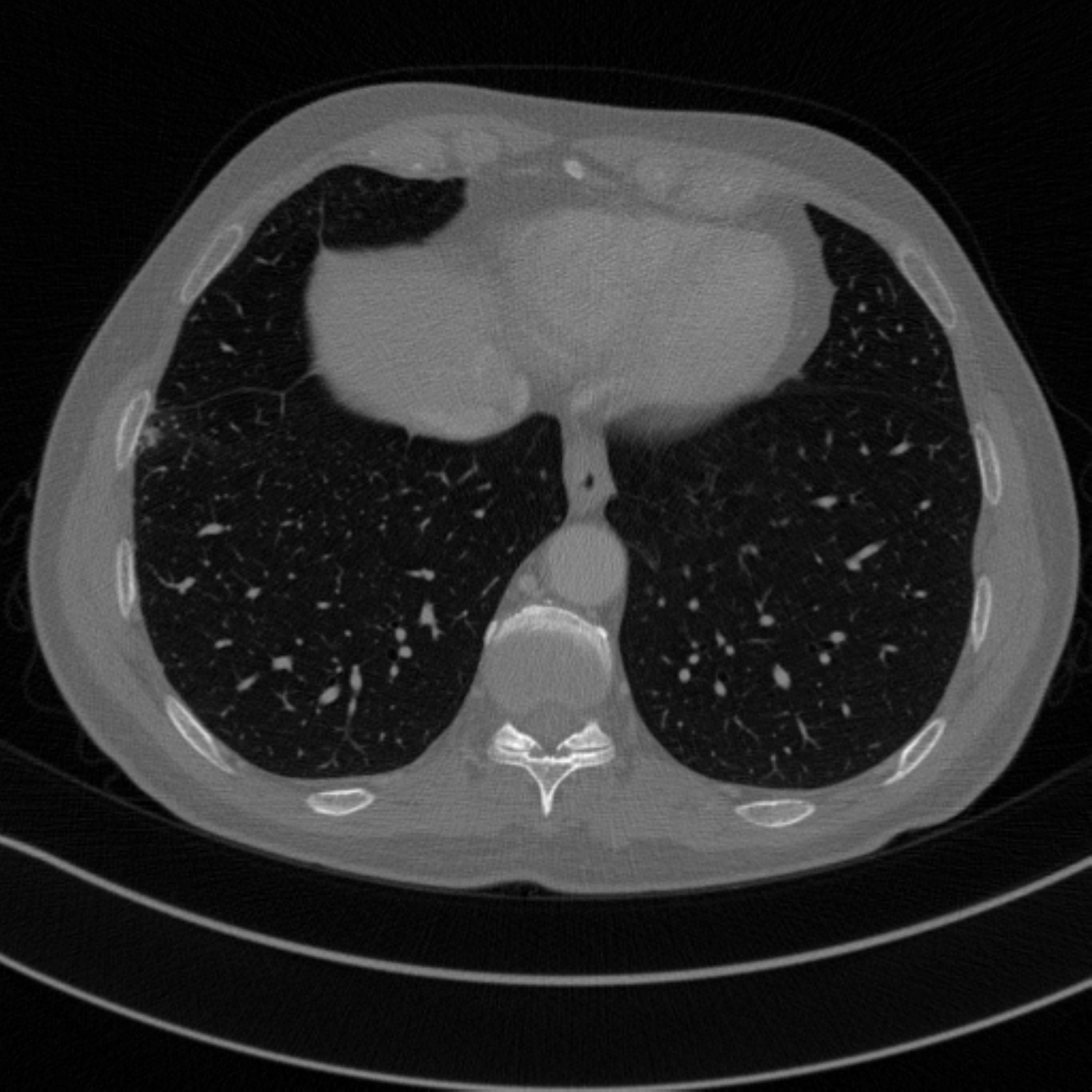}}
		\subfloat{\includegraphics[width=0.49\columnwidth]{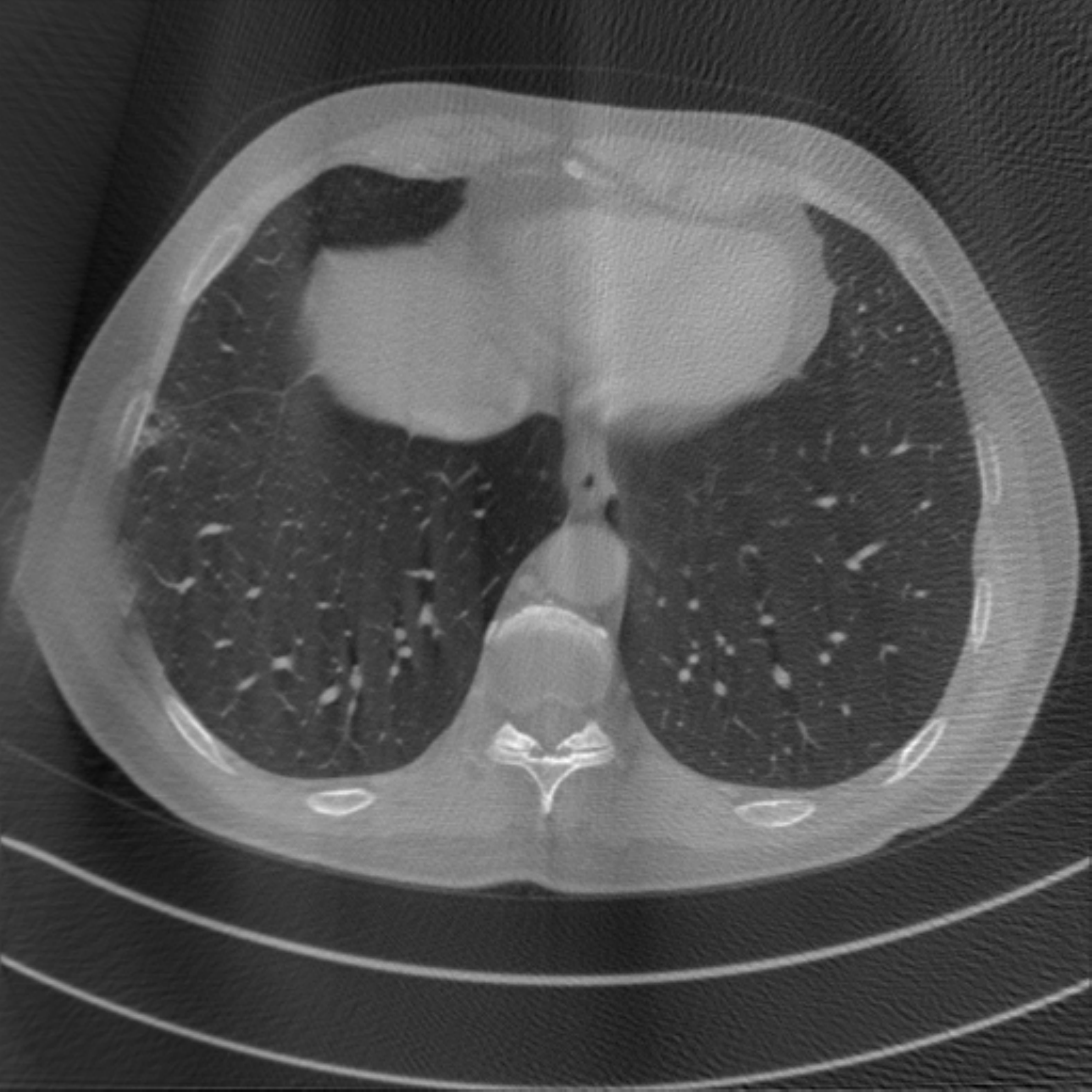}}
		\subfloat{\includegraphics[width=0.49\columnwidth]{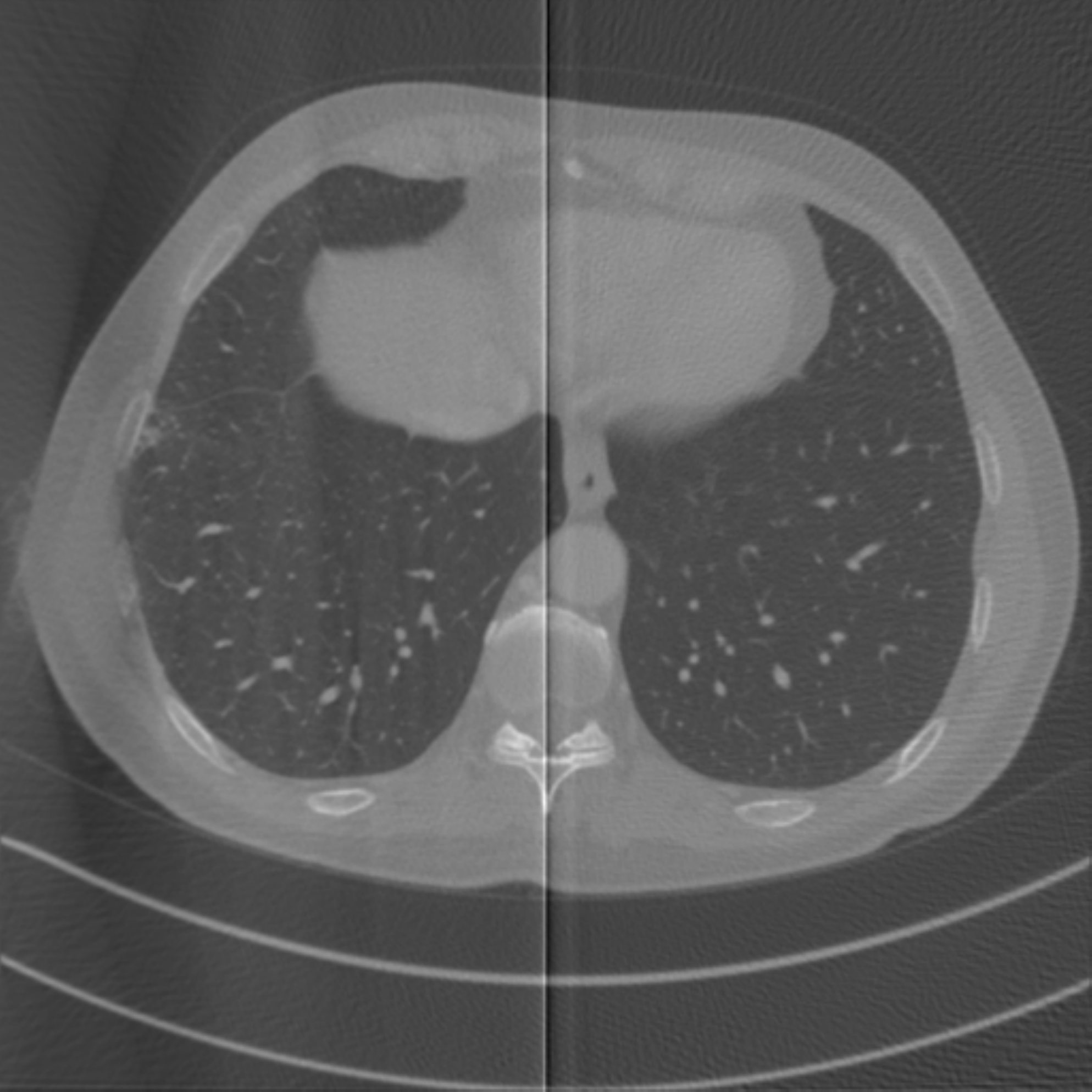}}
		\subfloat{\includegraphics[width=0.49\columnwidth]{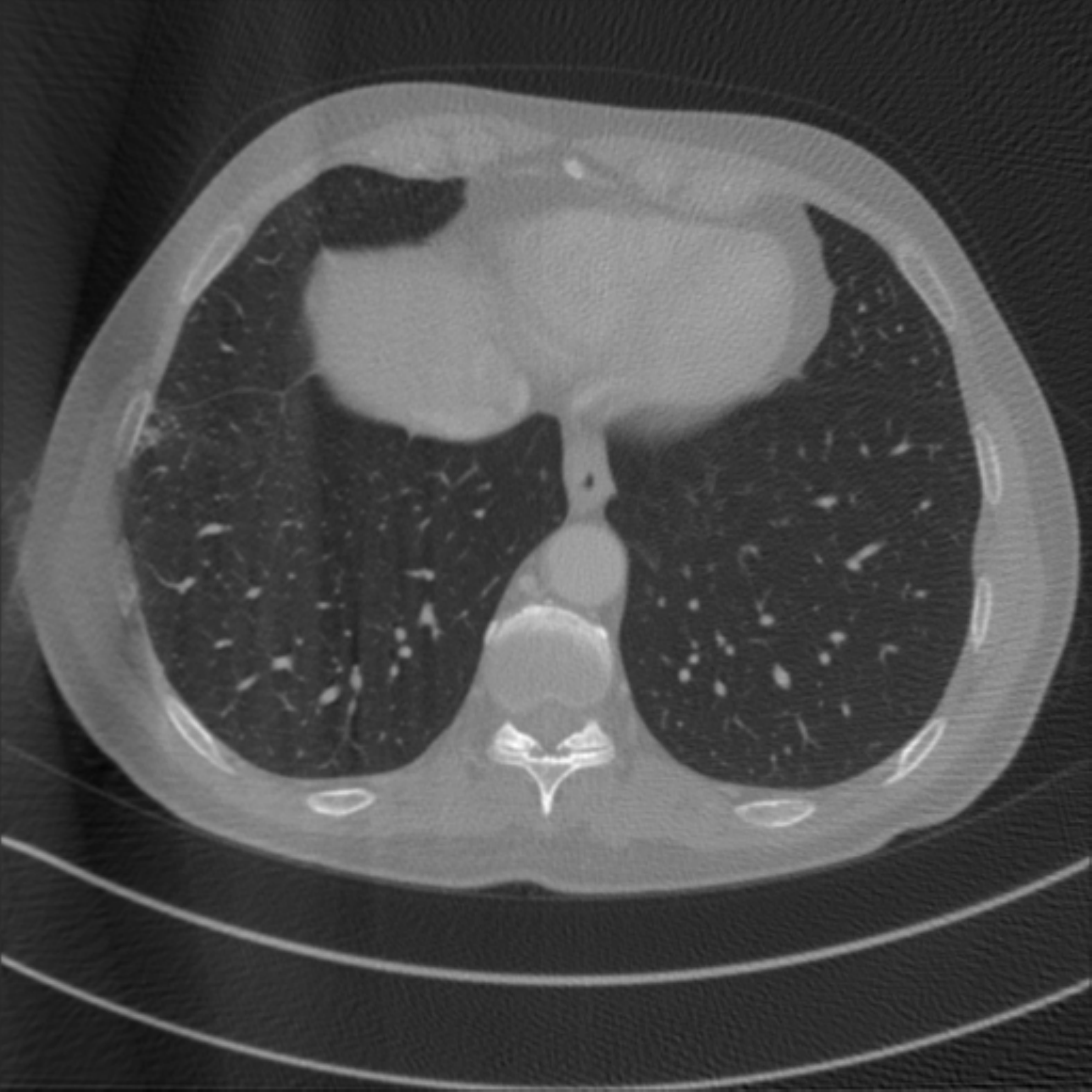}}
		
		\setcounter{subfigure}{0}
		\vspace{-3mm}
		\subfloat[Original]{\includegraphics[width=0.49\columnwidth]{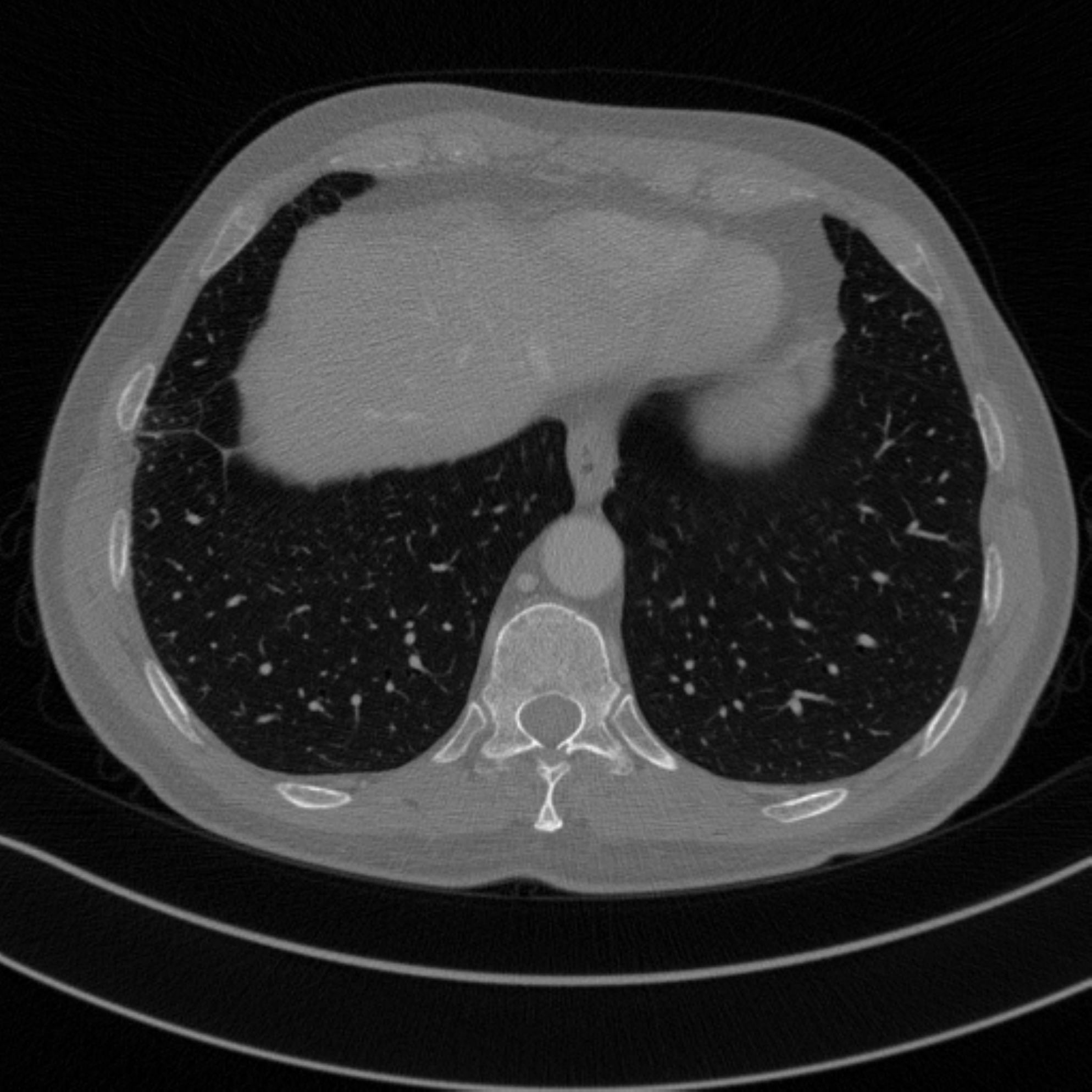}}
		\subfloat[FDK]{\includegraphics[width=0.49\columnwidth]{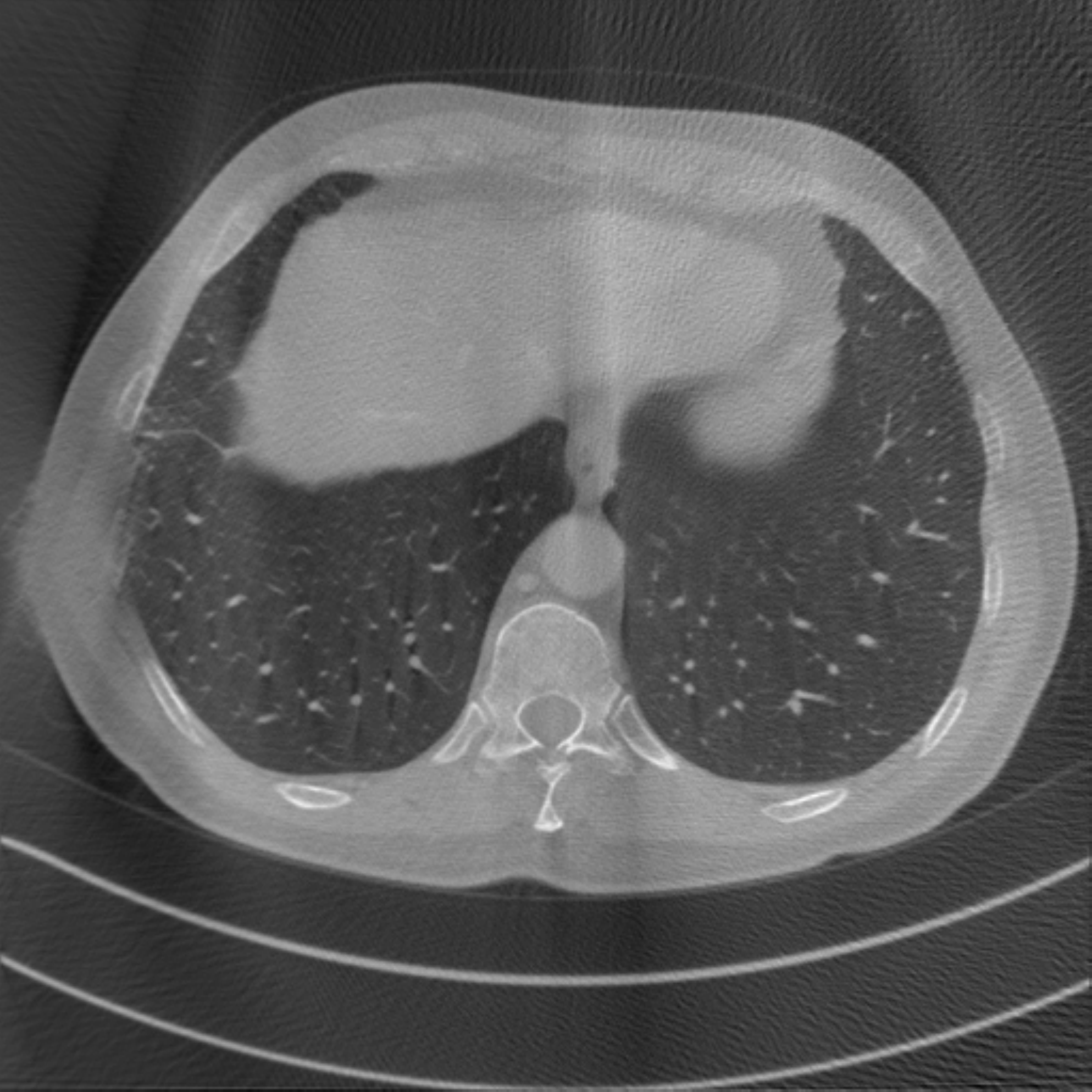}}
		\subfloat[ACE]{\includegraphics[width=0.49\columnwidth]{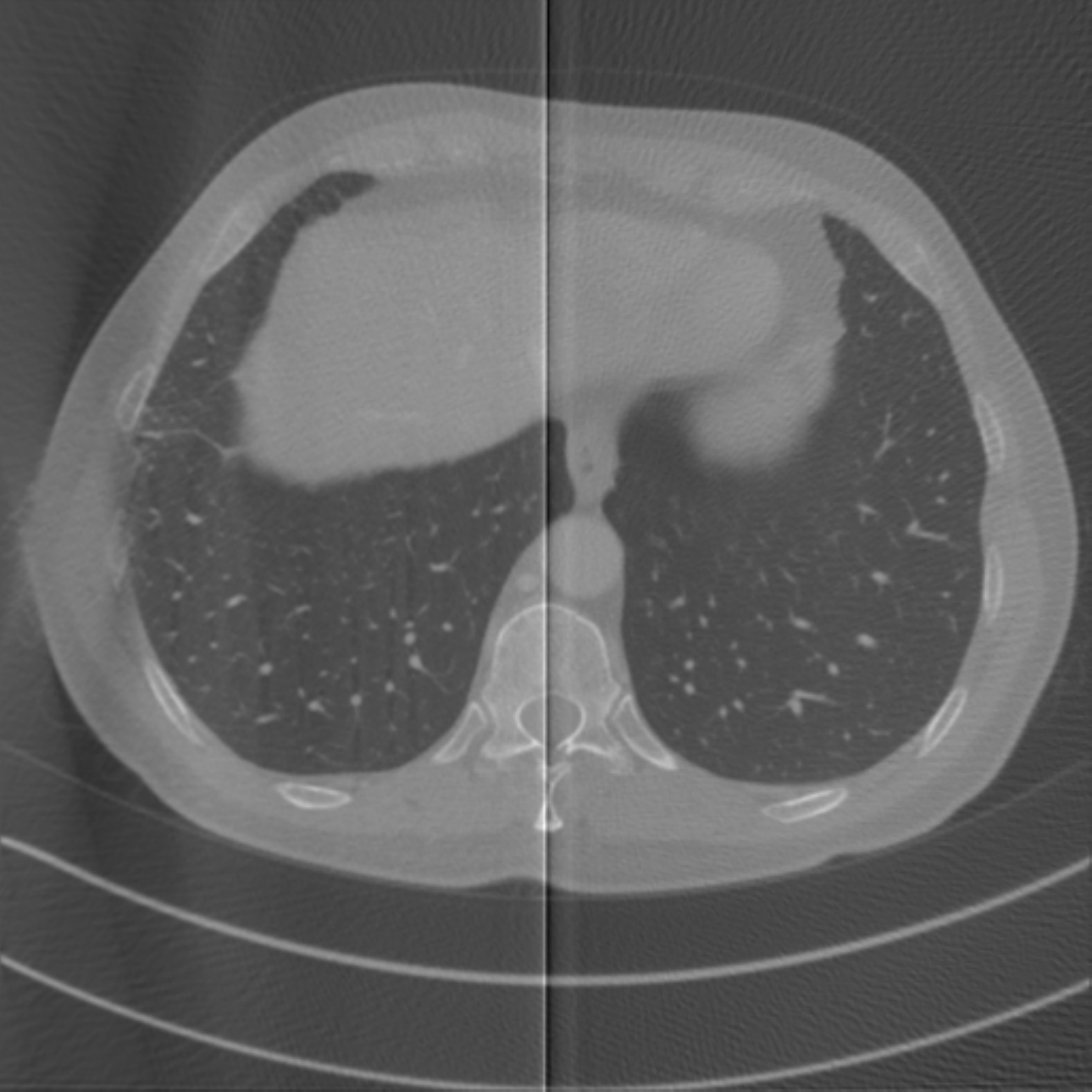}}
		\subfloat[Ours]{\includegraphics[width=0.49\columnwidth]{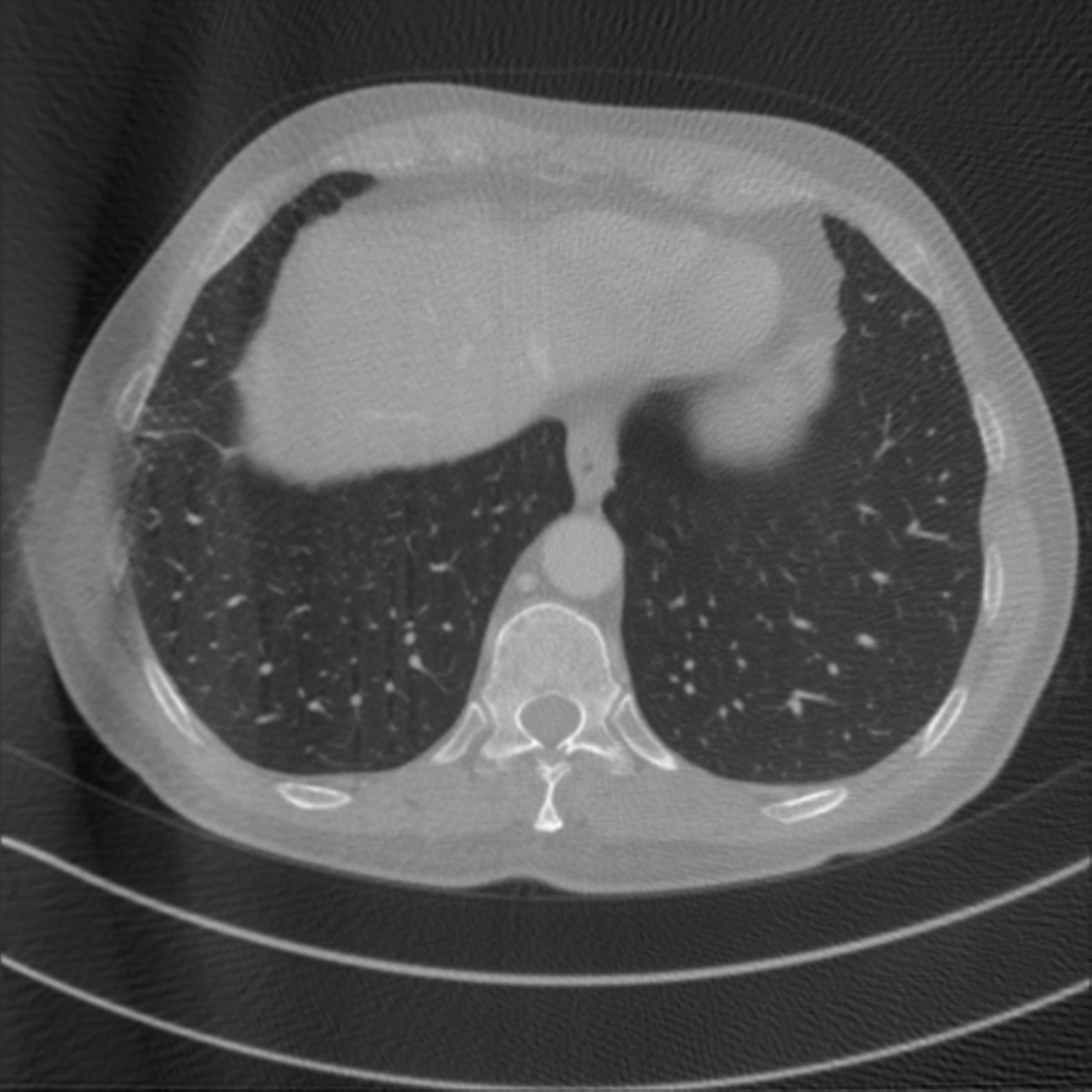}}
	}
	\caption{Sliced CT images reconstructed from super-short-scan circle cone-beam projection data.}
	\label{figcone2}
\end{figure*}

\begin{table}[!t]
	\renewcommand{\arraystretch}{1.3}
	\caption{The Averaged PSNR and SSIM of  CT Images Reconstructed by CFA, ACE and Ours from Fan-Beam Short-Scan Projection Data.}
	\label{T1}
	\centering
	\begin{tabular}{c|c|c}
		\hline
		& PSNR  & SSIM   \\ \hline
		CFA  &18.74 $\pm$0.52    &0.47$\pm$0.01\\
		ACE  &34.66$\pm$0.60  &0.83$\pm$0.01   \\
		ours	 &\textbf{34.78} $\pm$0.62 	&\textbf{0.84}$\pm$0.01 \\
		\hline
	\end{tabular}%
\end{table}%

\begin{table}[!t]
	\renewcommand{\arraystretch}{1.3}
	\caption{The Averaged PSNR and SSIM of  CT Images Reconstructed by CFA, ACE and Ours from Fan Beam Super-Short-Scan Projection Data.}
	\label{T2}
	\centering
	\begin{tabular}{c|c|c}
		\hline
		& PSNR  & SSIM   \\ \hline
		CFA  &18.51 $\pm$0.45    &0.43$\pm$0.01\\
		ACE  &25.53$\pm$0.70  &0.45$\pm$0.03   \\
		ours	 &\textbf{27.64} $\pm$0.94 	&\textbf{0.66}$\pm$0.02 \\
		\hline
	\end{tabular}%
\end{table}%

\subsection{Circle cone-beam with flat-plane detectors}
In this subsection, we give some   CT images reconstructed from the circle cone-beam projection data measured by  
flat-plane detectors to very the effectiveness of our method, and compare the results with those of the Feldkamp-Davis-Kress (FDK) algorithm  \cite{ISI:A1984SU73300005} and Noo's algorithm (ACE) \cite{ISI:000177297700011}, where we extend the weighting function of ACE such that it can be used to reconstruct circle cone-beam CT images.

To test the performances of our method and the compared algorithms, we  randomly choose 2500 full dose CT images (of size $512\times512$) from “the 2016 NIH-AAPM-Mayo Clinic Low Dose CT Grand Challenge”  \cite{data}  and use them to assemble 50  objects of size $512\times 512 \times 50$ as the original images.

 The parameters for the circle cone-beam CT with the flat-plane detector are set as follows: $u=[-494:1:494]$, $v=[-54:1:54]$, $R_o=1000$, and  $$D=\text{ceil}(R_o+\sqrt 2\times256)=1363.$$
For the sampling positions on the $\lambda$ coordinate, we set $$\lambda=[0:1:180+2\times20]\times\pi/180$$ for short-scan and $$\lambda=[0:1:180]\times\pi/180$$ for super-short-scan. The $26$th slice of the  images of the object is assumed to lie in the plane $z=0$ (i.e. the plane formed by the trajectory of the X-ray source).
The hyper-parameter $d$ in ACE \cite{ISI:000177297700011} is set as $d=10\times\pi/180$.

In Fig. \ref{figcone1}, we present some sliced CT images of one object reconstructed by FDK, ACE and ours from the short-scan circle cone-beam projection data. We can observe that there exist some stripe artifacts  in the CT images reconstructed by FDK. The visual effects of the CT images reconstructed by ACE and ours are very similar. We also use the PSNR and SSIM to measure the similarities of the reconstructed CT images and the original. From Table \ref{T3}, we can observe that the average PSNR and SSIM of our method are slightly higher than that of ACE and are  higher than that of CFA, which coincides with our observations.

In Fig. \ref{figcone2},  some sliced CT images of one object reconstructed by FDK, ACE and ours from the super-short-scan circle cone-beam projection data are shown. From Fig. \ref{figcone2}b, we can see that some stripe visual artifacts exit in the CT images reconstructed by FDK. Moreover, the left hand side parts of the CT images reconstructed by FDK suffer from severe intensity inhomogeneity. From Fig. \ref{figcone2}b, we can observe that there also exit some  undesirable vertical lines in the CT images reconstructed by ACE, which may be caused by the data incompleteness. Compared to Fig. \ref{figcone2}b and Fig. \ref{figcone2}c, the CT images reconstructed by our method suffer from less intensity inhomogeneity and their visual effects are the best as shown in Fig. \ref{figcone2}d. The PSNR and SSIM   are used to evaluate the qualities of the reconstructed images and listed in Table \ref{T4}. We can see that our method has the highest PSNR and SSIM compared to CFA and ACE.

\begin{table}
	\renewcommand{\arraystretch}{1.3}
	\caption{The Averaged PSNR and SSIM of  CT Images Reconstructed by FDK, ACE and Ours from Circle Cone-Beam Short-Scan Projection Data.}
	\label{T3}
	\centering
	\begin{tabular}{c|c|c}
		\hline
		& PSNR  & SSIM   \\ \hline
		CFA  &29.90 $\pm$1.17    &0.66$\pm$0.03\\
		ACE  &31.10$\pm$1.79  &0.67$\pm$0.04   \\
		ours	 &\textbf{31.16} $\pm$1.79 	&\textbf{0.68}$\pm$0.04 \\
		\hline
	\end{tabular}%
\end{table}%

\begin{table}
	\renewcommand{\arraystretch}{1.3}
	\caption{The Averaged PSNR and SSIM of  CT Images Reconstructed by FDK, ACE and Ours from Circle Cone-Beam Super-Short-Scan Projection Data.}
	\label{T4}
	\centering
	\begin{tabular}{c|c|c}
		\hline
		& PSNR  & SSIM   \\ \hline
		CFA  &24.69 $\pm$1.83    &0.49$\pm$0.04\\
		ACE  &26.11$\pm$1.87  &0.52$\pm$0.05   \\
		ours	 &\textbf{27.99} $\pm$1.84 	&\textbf{0.59}$\pm$0.04 \\
		\hline
	\end{tabular}%
\end{table}%

\section{Conclusion}
In this paper,  we proposed a new weighting function to deal with the redundant projection data for the arc based fan-beam CT algorithm, which was obtained via applying Katsevich’s helical CT formula \cite{ISI:000221245100004} to 2D fan-beam CT reconstruction. By extending the   arc based  algorithm to circle cone-beam geometry with the proposed weighting function, we also obtained a new FDK-similar algorithm for  circle cone-beam CT reconstruction. Experiments showed that our methods can obtained higher PSNR and SSIM compared to the related algorithms when the scanning  trajectories are super-short-scan.

% or


\appendices
\section{Algorithm for circle cone-beam CT with flat-plane detector}
\begin{figure}[h]
	\includegraphics[scale=0.5]{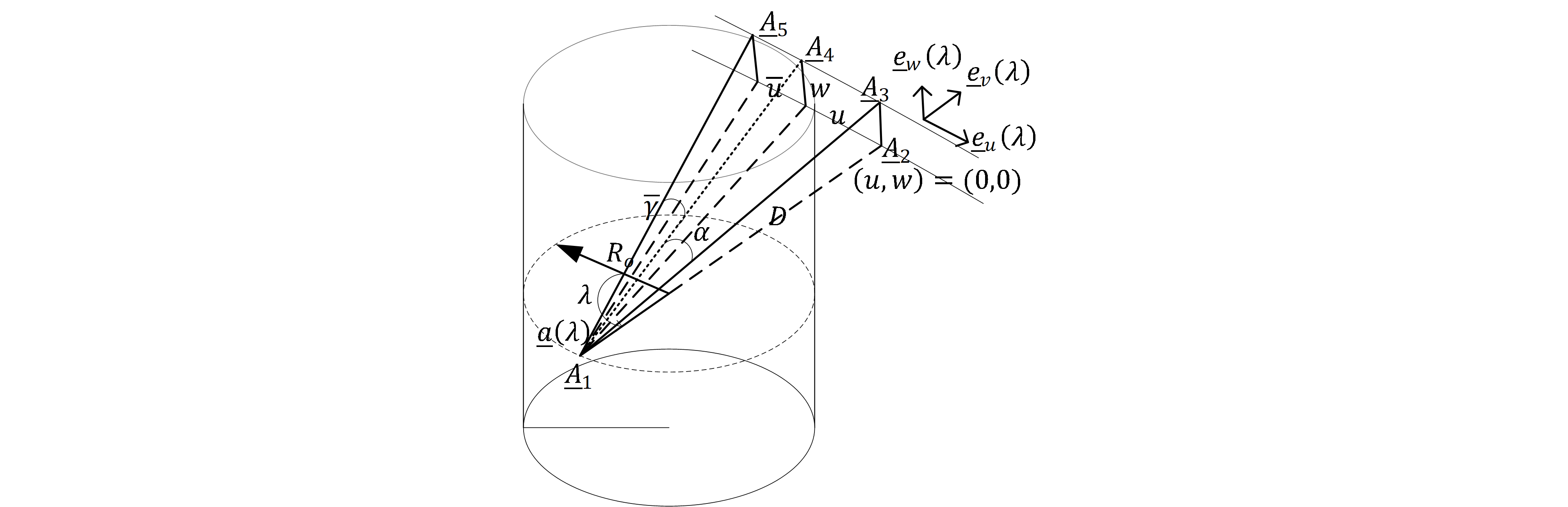}\\			
	\caption{Relation between $\gamma$ and the flat-plane detector coordinates.}
	\label{appendixA}
\end{figure}
 As can be observed from Fig. \ref{appendixA}, we have
\begin{equation}\label{key}
\begin{aligned}
\underline{e}_{u}(\lambda)=\frac{\overrightarrow{A_4A_3}}{||\overrightarrow{A_4A_3}||},~ \underline{e}_{v}(\lambda)&=\frac{\overrightarrow{A_1A_2}}{\overrightarrow{A_1A_2}},~
\underline{e}_{w}(\lambda)=\frac{\overrightarrow{A_2A_3}}{\overrightarrow{A_2A_3}},\\
||\overrightarrow{A_4A_3}||=u,~ ||\overrightarrow{A_5A_3}||=\bar u,&
~||\overrightarrow{A_2A_3}||=w,~||\overrightarrow{A_2A_3}||=D,\\
\angle A_4A_1A_3=\alpha,& ~~~~~~~~~~\angle A_5A_1A_4=\bar\gamma.
\end{aligned}
\end{equation}

We also have $\tan(\alpha+\bar\gamma)=\frac{||\overrightarrow{A_5A_3}||}{||\overrightarrow{A_1A_3}||}=\frac{\bar u}{\sqrt{D^2+w^2}}$ and so
\begin{equation}\label{eA1}
\mathrm{d} \bar\gamma=\frac{\cos^2(\alpha+\bar\gamma)}{\sqrt{D^2+w^2}}\mathrm{d} \bar u=\frac{\sqrt{D^2+w^2}}{D^2+w^2+\bar u^2}\mathrm{d} \bar u.
\end{equation}

By the Law of Sines, we have 
\begin{equation}\label{key}
\begin{aligned}
\sin\bar\gamma=&||\overrightarrow{A_5A_4}||\frac{\sin(\angle A_1A_5A_3)}{||\overrightarrow{A_1A_4}||}\\
=&\frac{\sqrt{D^2+w^2}(\bar u-u)}{\sqrt{D^2+w^2+\bar u ^2}\sqrt{D^2+w^2+u ^2}}.
\end{aligned}
\end{equation}
Therefore, 
\begin{equation}\label{eA2}
h_H(\sin\bar\gamma)=\frac{\sqrt{D^2+w^2+\bar u ^2}\sqrt{D^2+w^2+u ^2}}{\sqrt{D^2+w^2}}h_H(\bar u-u).
\end{equation}
Note that $\underline{\theta}(\lambda,u,w)=\frac{\overrightarrow{A_1A_4}}{||\overrightarrow{A_1A_4}||}$ and so $\cos \bar\gamma \underline{\theta}+\sin \bar\gamma\underline{\theta}^{\bot}=\frac{\overrightarrow{A_1A_5}}{||\overrightarrow{A_1A_5}||}$. Substituting equations (\ref{eA1}) and (\ref{eA2}) into equation (\ref{e19}), we can obtain
\begin{equation}\label{eA4}
\begin{aligned}
g^{F}(\lambda, &\underline{\theta}(\lambda,u,w))=-\frac{\sqrt{ u^2+D^2+w^2}}{D}\times\\
&\int_{-u_m}^{u_m} \mathrm{d} \bar u \frac{D}{\sqrt{\bar u^2+D^2+w^2}}h_{H}(u-\bar u) g_1(\lambda, \bar u,w),
\end{aligned}
\end{equation}

From Fig. \ref{fig4}, we can observe that
\begin{equation}\label{eA3}
\begin{aligned}
\|\underline{x}-\underline{a}(\lambda)\|=&\sqrt{(u^*)^2+(w^*)^2+D^2}\frac{(\underline{x}-\underline{a}(\lambda))\cdot\underline{e}_{v}(\lambda)}{D}\\
=&\frac{\sqrt{(u^*)^2+(w^*)^2+D^2}}{D}(R_o+\underline{x}\cdot\underline{e}_{v}(\lambda)).
\end{aligned}
\end{equation}

Substituting equation (\ref{eA3}) into equation (\ref{e18}), we can get
\begin{equation}\label{eA5}
\begin{aligned}
f(\underline{x})=&-\frac{D}{\sqrt{D^2+(w^*)^2+(u^ *)^2}}\times\\
&\frac{1}{2 \pi} \int_{\lambda_0}^{\lambda_P} \mathrm{d} \lambda \frac{\varpi_{3d}(\underline{x},\lambda)}{v^{*}(\underline x, \lambda)} g_2\left(\lambda,u^{*},w^{*} \right).
\end{aligned}
\end{equation}
 
 Canceling the factors $-\frac{\sqrt{ u^2+D^2+w^2}}{D}$ and $-\frac{D}{\sqrt{D^2+(w^*)^2+(u^*)^2}}$ in equations (\ref{eA4}) and (\ref{eA5}), we obtain the algorithm for circle cone-beam CT with the flat-plane detector.

\section {Algorithm for calculating one endpoint of a chord}
Let  $(R_o\cos\lambda_0,R_o\sin\lambda_0)$ and $(R_o\cos\lambda_1,R_o\sin\lambda_1)$ be  the two end-points of a chord on a circle with radius $r=R_o$ and $(x_1,x_2)$ be a point on the chord. Then, we have 
\begin{equation}\label{eA6}
\begin{aligned}
R_o(\cos\lambda_1-\cos\lambda_0)&=t*(x_1-R_o\cos\lambda_0),\\
R_o(\sin\lambda_1-\sin\lambda_0)&=t*(x_2-R_o\sin\lambda_0),
\end{aligned}
\end{equation}
where $t\in[0,1]$. Solving equation set (\ref{eA6}), we have
\begin{equation}\label{key}
	\begin{aligned}
	&\cos\lambda_1=\{R_o(-R_o^2 + x_1^2 + x_2^2)\cos\lambda_0 -\\
	&~~~~~~~~2x_1(-R_o^2 +x_1R_o\cos\lambda_0 + x_2R_o\sin\lambda_0)\}/\\
	&\{R_o(R_o^2 + x_1^2 + x_2^2 - 
	2x_1R_o\cos\lambda_0 - 2x_2R_o\sin\lambda_0)\},\\
	&\sin\lambda_1=\{R_o(-R_o^2 + x_1^2 + x_2^2)\sin\lambda_0-\\
	&~~~~~~~~2x_2(-R_o^2 + x_1R_o\cos\lambda_0 + x_2R_o\sin\lambda_0)\} /\\
	&\{R_o(R_o^2 + x_1^2 + x_2^2 -
	2x_1R_o\cos\lambda_0 - 2x_2R_o\sin\lambda_0)\}.
	\end{aligned}
	\end{equation}
Thus, we can get $\lambda_1$ by $\lambda_1=\arccos(\cos \lambda_1)$, where $\lambda_1$ needs to be changed by $\lambda_1=2\pi-\lambda_1$ when $\sin \lambda_1<0$.

%\appendices
%% you can choose not to have a title for an appendix
%% if you want by leaving the argument blank
%\section{}
%% use section* for acknowledgment
%\section*{Acknowledgment}

\ifCLASSOPTIONcaptionsoff
  \newpage
\fi

%\clearpage
\balance
\bibliographystyle{IEEEtran}
\bibliography{1}

\end{document}